\documentclass[10pt,twocolumn,amsmath,amssymb,aps,prb]{revtex4-1}

\usepackage{graphicx}% Include figure files
\usepackage{dcolumn}% Align table columns on decimal point
\usepackage{bm}% bold math
\usepackage{hyperref}% add hypertext capabilities
\usepackage[cp1250]{inputenc}
\usepackage[usenames,dvipsnames]{color}
\usepackage{tabularx}

\usepackage{cancel}
\usepackage{color}
\usepackage[normalem]{ulem}
\newcommand{\Tr}{\mathop{\rm Tr}}

\newcommand{\arccosh}{{\rm arcosh}}

\newcommand{\sgn}{\mathop{\rm sgn}}

\def\XXint#1#2#3{{\setbox0=\hbox{$#1{#2#3}{\int}$}
     \vcenter{\hbox{$#2#3$}}\kern-.5\wd0}}

\renewcommand{\today}{October 14, 2022}

\begin{document}

\title{Accurate localization of Kosterlitz-Thouless-type quantum phase
  transitions\\
  for one-dimensional spinless fermions}

\author{Florian Gebhard$^1$}
\email{florian.gebhard@physik.uni-marburg.de}
\author{Kevin Bauerbach$^{1}$}
\affiliation{$^1$Fachbereich Physik, Philipps-Universit\"at Marburg,
  35032 Marburg, Germany}
\author{\"Ors Legeza$^{2,3}$}
\email{legeza.ors@wigner.hu}
\affiliation{$^2$Strongly Correlated Systems Lend\"ulet Research Group, 
Institute for Solid State Physics and Optics, MTA Wigner Research Centre for
Physics, P.O.\ Box 49, 1525 Budapest, Hungary}
\affiliation{$^3$Institute for Advanced Study,
  Technical University of Munich, Lichtenbergstra\ss e 2a,
  85748 Garching, Germany}

\date{\today}

\begin{abstract}%
We investigate the charge-density wave (CDW) transition
for one-dimensional spinless fermions at half band-filling
with nearest-neighbor
electron transfer amplitude~$t$ and interaction~$V$.
The model is equivalent to the anisotropic XXZ Heisenberg model
for which the Bethe Ansatz provides an exact solution. 
For $V> V_{\rm c}= 2t$, 
the CDW order parameter and the single-particle gap are finite but
exponentially small,
as is characteristic for a Kosterlitz-Thouless transition.
It is notoriously difficult to locate 
such infinite-order phase transitions
in the phase diagram using approximate analytical and numerical approaches.
Second-order Hartree-Fock theory is qualitatively
applicable for all interaction strengths,
and predicts the CDW transition to occur at $V_{\rm c,2}^{(2)}\approx 1.5t$.
Second-order Hartree Fock theory is almost variational
because the density of quasi-particle excitations is small.
We apply the density-matrix renormalization group (DMRG)
for periodic boundary conditions for system sizes up to 514~sites
which permits a reliable extrapolation of all physical quantities
to the thermodynamic limit, apart from the critical region.
We investigate the ground-state energy, the gap,
the order parameter, the momentum distribution, the 
quasi-particle density, and the density-density correlation function
to locate $V_{\rm c}$ from the DMRG data.
Tracing the breakdown of the Luttinger liquid
and the peak in the quasi-particle density at the band edge permits us
to reproduce $V_{\rm c}$ with an accuracy of one percent.
\end{abstract}

%\pacs{72.15.Qm,75.20.Hr,75.30.Hx}

%\keywords{Impurity and Kondo physics, DMRG, Gutzwiller variational approach.}

\maketitle

\section{Introduction}
\label{sec:Intro}

The isotropic spin-1/2 Heisenberg model on a chain
was the first exactly solved many-body problem.~\cite{Bethe1931}
Three decades later, Orbach,~\cite{Orbach}
and later Yang and Yang, ~\cite{PhysRev.150.321,PhysRev.150.327}
succeeded to generalize the Bethe Ansatz to the
one-dimensional anisotropic Heisenberg (or XXZ) model,
\begin{equation}
  \hat{H}_{\rm XXZ}(\Delta)= -\frac{1}{2} \sum_{l= 1}^{L-1}
  \Bigl[
\underline{\underline{\sigma}}_l^x\cdot \underline{\underline{\sigma}}_{l+ 1}^x
  +
\underline{\underline{\sigma}}_l^y\cdot \underline{\underline{\sigma}}_{l+ 1}^y
+ \Delta \underline{\underline{\sigma}}_l^z\cdot
\underline{\underline{\sigma}}_{l+ 1}^z
\Bigr]
  \; ,
  \label{eq:XXZspinmodelll}
\end{equation}
where $\underline{\underline{\sigma}}_l^{x,y,z}$
are the Pauli matrices on site~$l$
and $\Delta\leq 0$ is the anisotropy parameter for antiferromagnetic coupling.
The XXZ model reduces to the Ising model in the limit
of strong anisotropy in the $z$-direction, $|\Delta| \to \infty$.
Since the 1960s, many exact results for the model
in the thermodynamic limit were established, e.g.,
the ground-state energy,~\cite{PhysRev.150.321,PhysRev.150.327} the 
elementary spinon
excitations,~\cite{PhysRevA.8.2526,BABELON198313,Woynarovich82,ViWo84}
and the staggered magnetization in the
Ising regime, $|\Delta| \geq 1$,~\cite{Baxter1973,IZERGIN1999679}
to name a few.
An efficient exact description of the
thermodynamic properties was achieved by Kl\"umper.~\cite{Kluemper93}

Starting from the early 1990s~\cite{JMMN92} the focus of the exact
calculations shifted to the reduced density matrix which
was fully characterized in the following
decade.~\cite{JiMi96,KMT00,GKS05,BJMST08a,JMS09,BoGo09}
This made it possible to derive a number of spin correlation functions analytically,
see, e.g., Refs.~[\onlinecite{BortzGoehmann2005,Damerauetal2007,Boosetal2008}].
Most recently, fully explicit series representations for
the exact dynamical spin correlation functions
and the spin conductivity were obtained in the
Ising regime.~\cite{PhysRevLett.126.210602,Goehmannetal2021}
Therefore, the XXZ model belongs to the best studied
and understood many-particle systems.

Via a Jordan-Wigner transformation, the antiferromagnetic
XXZ model on a chain with $\Delta \leq 0$ and zero magnetization
maps onto a model for
spinless fermions with open boundary conditions at half band-filling
with nearest-neighbor
electron transfer amplitude~$(-t)$ and repulsive
interaction~$V=-\Delta/(2t)\geq 0$.~\cite{JordanWigner1928}
While the XXZ model contains a transition to an antiferromagnetically
ordered ground state at the Heisenberg point, $\Delta =-1$,
the model for spinless fermions displays a metal-insulator transition
at $V_{\rm c}= 2t$ from a Luttinger liquid to a
charge-density wave (CDW) insulator.

A transition at finite interaction strength
is unexpected because the
nesting instability would place the transition at $V_{\rm c}^{\rm (1)}= 0^+ $,
and it requires a sophisticated renormalization group treatment to show
that the system is marginally stable against the formation of a CDW 
at weak coupling.~\cite{PhysRevLett.67.3852,RevModPhys.66.129}
The same result can be obtained using bosonization,
see Ref.~[\onlinecite{Thierrybook}] for an introduction.
To complicate matters, the single-particle gap and the order parameter
display an essential singularity at $V_{\rm c}$, i.e., they open
exponentially as a function of $1/\sqrt{V-V_{\rm c}}$,
as is characteristic for a Kosterlitz-Thouless transition.~\cite{KTtransition}
Consequently, it is notoriously difficult to determine the
critical interaction strength for the transition using
approximate analytical and numerical approaches.
Transitions of Kosterlitz-Thouless type are common in
  one-dimensional quantum systems such as in the bosonic Hubbard model,
  see e.g., Ref.~[\onlinecite{PhysRevB.61.12474}],
  or quantum spin chains, see, e.g., Ref.~[\onlinecite{PhysRevB.102.035137}].

Since spinless fermions represent a rare case of an exactly solvable model
with a metal-insulator transition at a finite interaction strength,
it is desirable to see if and how well approximate methods
are able to reproduce the
formation of CDW order at $V_{\rm c}=2t$.
As our analytical approximations we use Hartree-Fock theory
and second-order perturbation theory around
it.~\cite{PhysRevB.43.3475,PhysRevB.50.14016}
First-order Hartree-Fock theory suggests
a CDW for all finite interaction strengths, $V_{\rm c}^{\rm (1)}= 0^+ $,
corresponding to the nesting instability.
As we shall show in this work,
second-order Hartree-Fock approximation predicts a discontinuous
transition at $V_{\rm c,2}^{(2)}\approx 1.5$ with a jump in the order parameter
and concomitant discontinuities in physical quantities.

As numerical approach,
we employ the density-matrix renormalization group (DMRG)
method~\cite{White-1992b,White-1993,Schollwock-2005}
that provides
highly accurate data for finite rings with up to $L=514$ sites;
we choose periodic boundary conditions and even $L/2$
for an open-shell ground state to reduce finite-size effects.
To make contact with finite-size corrections calculated for
the XXZ model from Bethe Ansatz, we also investigate
odd $L/2$.

We identify two successful strategies
to locate accurately the CDW transition.
The first route monitors the breakdown of the metallic phase.
The properties of the Luttinger liquid
are reflected in finite-size corrections
to the ground-state energy and the gap,
and most prominently in the Luttinger parameter that determines
the momentum distribution close to the Fermi points
and the small-momentum limit of the density-density correlation function.
In the end, the accurate calculation of the Luttinger parameter
permits to locate the breakdown of the Luttinger liquid
with an accuracy of three percent.
Following an alternative route, we trace
the maximum of the quasi-particle distribution
as a function of system size and interaction. Using this independent approach,
we determine the critical
interaction strength with an accuracy of one percent.

The paper is organized as follows. In Sect.~\ref{sec:modeldef} we define
the model Hamiltonian for spinless fermions. We discuss
its relation to the XXZ~model for various boundary conditions
and particle numbers.

In Sect.~\ref{sec:exactresults} we put together exact
results from the literature
for the ground-state energy, the nearest-neighbor
single-particle density matrix,
the single-particle gap, and the charge-density wave order parameter
in the thermodynamic limit.
Since this information is often phrased for the XXZ model
and is not summarized in reviews or books,
we find it useful to compose them here for spinless fermions.
Limiting cases are derived
and discussed in the supplemental material~\cite{suppmat}
(see, also, references~[\onlinecite{BanerjeeWilkerson,GaroufalidisZagier2021}]
  therein).

In Sect.~\ref{sec:HFapproximation} we present the standard (first-order)
Har\-tree-Fock approximation
for spinless fermions. This permits us to introduce
the lower and upper Hartree-Fock bands in the reduced Brillouin zone
and the corresponding quasi-particle operators.

In Sect.~\ref{sec:2ndorderHF} we calculate the se\-cond-order
weak-coupling perturbation 
correction to the Hartree-Fock approximation, and justify its 
applicability to all interaction strengths.
Technicalities for the second-order Hartree-Fock calculations can be
found in the supplemental material.~\cite{suppmat}

In Sect.~\ref{sec:comparison} we compare approximate results with those
from the exact Bethe-Ansatz solutions. We focus on the issue how to obtain 
the critical interaction strength for the charge-density wave transition
from finite-size DMRG data.

Short conclusions, Sect.~\ref{conclusions}, close our presentation.

\section{Spinless fermions in one dimension}
\label{sec:modeldef}

We start with the introduction of the Hamiltonian for spinless fermions and
discuss its relation to the XXZ Heisenberg model.

\subsection{Hamiltonian}
The Hamiltonian for spinless fermions on a ring with $L$ lattice
sites reads
\begin{equation}
  \hat{H} = \hat{T} + \hat{V} \; .
  \label{eq:Hamiltoniandef1}
\end{equation}
The kinetic energy operator describes the transfer of fermions between
neighboring
sites with real amplitude $(-t)$ and $t>0$
\begin{equation}
  \hat{T} = (-t) \sum_{l=1}^L \left( \hat{c}_{l+1}^+\hat{c}_l^{\vphantom{+}}
  + \hat{c}_{l}^+\hat{c}_{l+1}^{\vphantom{+}}\right) \;, 
\end{equation}
where $\hat{c}_{l}^+$ ($\hat{c}_l^{\vphantom{+}}$) creates (annihilates)
a fermion on lattice site $l$,  $l=1,2,\ldots ,L$; we choose $L/2$ to be even
if not stated explicitly otherwise.
Periodic boundary conditions apply, $\hat{c}_{L+l}\equiv \hat{c}_l$.
%For comparison with literature results for the gap,~\cite{PhysRevB.84.115135}
%we apply
%open boundary conditions in appendix~\ref{app:gapbc}.
The nearest-neighbor interaction with strength~$V$ is given by
\begin{equation}
  \hat{V}= V  \sum_{l=1}^{L} \hat{n}_l \hat{n}_{l+1} \; ,
  \label{eq:definteractionnotphsymmetric}
\end{equation}
where $\hat{n}_l =\hat{c}_{l}^+\hat{c}_l^{\vphantom{+}}$
counts the number of fermions on site~$l$, and $V>0$ is repulsive.

The kinetic energy is diagonal in momentum space using the Fourier
transformation
\begin{eqnarray}
  \hat{a}_k &=&\sqrt{\frac{1}{L}} \sum_{l=1}^L e^{-{\rm i} k l} \hat{c}_l \; ,\\
  \hat{c}_l &=&\sqrt{\frac{1}{L}} \sum_k e^{{\rm i} k l} \hat{a}_k \; ,
\end{eqnarray}
where $k=(2\pi/L)m$, $m=-L/2,-L/2+1,\ldots,L/2-1$, to fulfill
periodic boundary conditions. We have
\begin{equation}
  \hat{T}= \sum_k \epsilon(k) \hat{a}_k^+\hat{a}_k^{\vphantom{+}}\; , \quad
  \epsilon(k) =-2t \cos(k)\; .
\end{equation}
In the following we shall focus on the case of half band-filling,
where the number of fermions~$N$ equals half the number of lattice sites,
$N=L/2$, and use $t\equiv 1$ as our energy unit.
The bare bandwidth is $W= 4$.

\subsection{XXZ Heisenberg model}
\label{subsc:connnectXXZandtVmodel}

The model for spinless fermions in one dimension
can be transformed into the XXZ Heisenberg
model using a Jordan-Wigner transformation.
For the moment, let us assume open boundary conditions.
On a chain, the XXZ model
is given by eq.~(\ref{eq:XXZspinmodelll}).
Using spin operators,
\begin{eqnarray}
  \hat{S}_l^{x,y,z}&=&
  \frac{1}{2} \underline{\underline{\sigma}}_l^{x,y,z} \; , \nonumber\\
    \hat{S}_l^+ &=& \hat{S}_l^x+ {\rm i}\hat{S}_l^y \; ,\nonumber \\
  \hat{S}_l^- &=& \hat{S}_l^x- {\rm i}\hat{S}_l^y \; , 
 \end{eqnarray}
the XXZ Heisenberg model reads
\begin{equation}
  \hat{H}_{\rm XXZ}(\Delta)= -\sum_{l= 1}^{L-1}\left[
  \hat{S}_l^+ \hat{S}_{l+ 1}^- +  \hat{S}_{l+ 1}^+ \hat{S}_l^-   \right]
  -2\Delta \sum_{l= 1}^{L-1}
  \hat{S}_l^z \hat{S}_{l+ 1}^z  \, .
\end{equation}
Note that spin operators on different lattice sites commute
with each other.

The Pauli particle operators
\begin{eqnarray}
  \hat{b}_l^+  &= & \hat{S}^+ _l \; ,\nonumber \\
  \hat{b}_l^{\vphantom{+ }}&= & \hat{S}_l^- \; ,\nonumber \\
  \hat{n}^b_l&= &   \hat{b}_l^+  \hat{b}_l^{\vphantom{+ } }
  =   \hat{S}_l^z + \frac{1}{2}
\end{eqnarray}
obey fermionic anticommutation relations between operators on the same site but
bosonic commutation relations between different sites. This deficiency is cured
by the Jordan-Wigner transformation,~\cite{JordanWigner1928,LIEB1961407}
\begin{eqnarray}
  \hat{c}_l^+  &= & \exp\left({\rm i}\pi \sum_{k= 1}^{l-1}\hat{n}_k^b \right)
  \hat{b}_l^+
  \; ,\nonumber \\
  \hat{c}_l^{\vphantom{+}}  &= & \exp\left(-{\rm i}\pi
  \sum_{k= 1}^{l-1}\hat{n}_k^b \right)
  \hat{b}_l^{\vphantom{+}}
  \; ,\nonumber \\
  \hat{n}_l&= &   \hat{c}_l^+  \hat{c}_l^{\vphantom{+ } }=
  \hat{b}_l^+ \hat{b}_l^{\vphantom{+}}
  = \hat{n}_l^b \; .
\end{eqnarray}
Therefore, the XXZ Heisenberg model can be written
in terms of spinless fermions as
\begin{eqnarray}
\hat{H}_{\rm XXZ}(\Delta)&=& -\sum_{l= 1}^{L-1}\left(
\hat{c}_l^+ \hat{c}_{l+ 1}^{\vphantom{+ }} +   \hat{c}_{l+ 1}^+
\hat{c}_{l}^{\vphantom{+}}  \right) \nonumber \\
&&  -2\Delta \sum_{l= 1}^{L-1}
  \left(\hat{n}_l-\frac{1}{2}\right) \left(\hat{n}_{l+ 1} -\frac{1}{2}\right)
  \label{eq:XXZtransfored}
\end{eqnarray}
when open boundary conditions are employed. 
Thus, the equivalence reads
\begin{equation}
  \hat{H}_{\rm XXZ}(-V/2)= \hat{H}(V)-V\hat{N}+ V\frac{L}{4} \; .
  \label{eqw:basicrelation}
\end{equation}
Eq.~(\ref{eqw:basicrelation}) permits to translate 
exact results for the antiferromagnetic XXZ model
$\hat{H}_{\rm XXZ}(\Delta\leq 0)$ 
to the model of spinless fermions for $V\geq 0$ for open boundary conditions.

For the case of periodic boundary conditions, 
an additional boundary term arises,~\cite{LIEB1961407}
\begin{eqnarray}
\hat{H}_{\rm XXZ}^{\rm pbc}(\Delta) &=& -\sum_{l= 1}^{L}\left(
\hat{c}_l^+ \hat{c}_{l+ 1}^{\vphantom{+ }} +   \hat{c}_{l+ 1}^+
\hat{c}_{l}^{\vphantom{+}}  \right) \nonumber \\
&&  -2\Delta \sum_{l= 1}^{L}
\left(\hat{n}_l-\frac{1}{2}\right) \left(\hat{n}_{l+ 1} -\frac{1}{2}\right)
\nonumber \\
&& + \left( \hat{c}_{L}^+ \hat{c}_1^{\vphantom{+ }}+
\hat{c}_{1}^+ \hat{c}_L^{\vphantom{+ }}\right)
\left(\exp({\rm i}\pi \hat{N})+ 1\right)\, .\;
  \label{eq:XXZtransforedpbc}
\end{eqnarray}
Therefore, a comparison of Bethe Ansatz results for the periodic XXZ~model
with those for spinless fermions on a ring are only possible in
the thermodynamic limit, or, when finite-size corrections are addressed,
for situations where the particle number~$N$ is odd.
For the ground state at half band-filling it implies that $L/2$ must be odd.
For excitations from the half-filled ground state 
we must study the sector with two particle or two hole excitations,
$N= L/2\pm 2$.

\section{Exact results}
\label{sec:exactresults}

In this section we collect exact results in the thermodynamic limit
for the ground-state energy and the nearest-neighbor
single-particle density matrix
at half band-filling, the single-particle gap, the charge-density
wave order parameter, the correlation energy, the momentum distribution,
and the density-density correlation function.

\subsection{Ground-state energy 
  and  nearest-neighbor single-particle density matrix
at half band-filling}

In the sector of half band-filling we have $N=L/2$
so that eq.~(\ref{eqw:basicrelation}) gives
\begin{equation}
e_0(V)=e_0^{\rm XXZ}(-V/2)+ \frac{V}{4} 
\end{equation}
for the energy per lattice site in the thermodynamic limit,
$N,L\to\infty$, $N/L= 1/2$,
where $e_0^{\rm XXZ}(\Delta)$ is the energy per lattice site
in the XXZ model with antiferromagnetic
anisotropy and zero magnetization.

Yang and Yang~\cite{PhysRev.150.321,PhysRev.150.327}
  give the following expressions for the ground-state energy density
  at zero magnetization ($e_0^{\rm XXZ}(\Delta)= 2 f(\Delta,0)$
  in the work of Yang and Yang),
\begin{equation}
  e_0^{\rm XXZ}(-V/2)  =
  \left\{ \begin{array}{@{}lcl@{}}
    g(\mu)  & \hbox{for} &  V = 2\cos(\mu)<2
  \; , 
  \\[6pt]
  \displaystyle 1/2-2\ln(2) & \hbox{for} &  V = 2 \; ,\\[6pt]
  h(\lambda) & \hbox{for} & V= 2\cosh(\lambda)>2\;,
  \end{array}
  \right. 
%  \label{eq:insulatingside}
\label{eq:metallicside}
\end{equation}
where
\begin{eqnarray}
  g(\mu)&= &  \frac{\cos(\mu)}{2}-
  \int_{-\infty}^{\infty}
  \frac{\sin^2(\mu){\rm d}x}{\cosh(\pi x)[\cosh(2\mu x)-\cos(\mu)]}\; ,
  \nonumber \\
  \mbox{}\\
  h(\lambda)&= &
  \frac{\cosh(\lambda)}{2}-\frac{\sinh(\lambda)}{\lambda}\left[
    \lambda+ 4\lambda \sum_{n= 1}^{\infty}\frac{1}{1+ \exp(2\lambda n)}\right]
  \nonumber \; .\\
  \label{defhlambda}
\end{eqnarray}
%\begin{equation}
%  g(\mu)=  \frac{\cos(\mu)}{2}-
%  \int_{-\infty}^{\infty}
%  \frac{\sin^2(\mu){\rm d}x}{\cosh(\pi x)[\cosh(2\mu x)-\cos(\mu)]}\; ,
%  \end{equation}
%and
%\begin{equation}
%h(\lambda)= 
%  \frac{\cosh(\lambda)}{2}-\frac{\sinh(\lambda)}{\lambda}\left[
%    \lambda+ 4\lambda \sum_{n= 1}^{\infty}\frac{1}{1+ \exp(2\lambda n)}\right]
%  \; .
%  \label{defhlambda}
%\end{equation}
The first and the third region can be continuously
extended to $V= 2$.
The above formulae can be expressed in terms of
$q$-digamma functions~\cite{Goehmannprivatecomm}
We shall not digress into the representation by special functions here.

Expansions for small and large~$V$ can be found in
the supplemental material.~\cite{suppmat}
For comparison with Hartree-Fock theory, we give
the leading-order results for weak and strong interactions,
\begin{eqnarray}
e_0(V\ll 1 )&=& -\frac{2}{\pi} + \left(\frac{1}{4}-\frac{1}{\pi^2}\right)V
 \nonumber \\[3pt]  &&
+ \left(-\frac{2}{3 \pi^3} + \frac{1}{36 \pi}\right)V^2 \; ,\nonumber \\
e_0(V\gg 1 )&=&  -\frac{1}{V} + \frac{1}{V^3} \; .
\label{eq:gsenergylimits}
\end{eqnarray}
For more details, see the supplemental material.~\cite{suppmat}

With the help of the Hellmann-Feynman theorem,~\cite{Hellmann2,Feynman}
both the potential energy and the kinetic energy can be derived
from the exact ground-state energy,
\begin{eqnarray}
  \langle  \hat{V}\rangle/L &=& V\frac{\partial e_0(V)}{\partial V} \; ,
  \nonumber \\
  \langle  \hat{T}\rangle/L &=& e_0(V)
  - V\frac{\partial e_0(V)}{\partial V} \; .
  \label{eq:Vexactgeneral}
\end{eqnarray}
Since there is no bond-order wave in the exact ground state, the
kinetic energy is just a multiple of the nearest-neighbor
single-particle density matrix,
\begin{equation}
  B_0= -\frac{1}{2L} \langle  \hat{T}\rangle =
  -\frac{1}{2}\left(e_0(V)
  - V\frac{\partial e_0(V)}{\partial V} \right)\; .
  \label{eq:exactB0general}
\end{equation}
The limiting values are
\begin{eqnarray}
  B_0(V\ll 1 )&=& \frac{1}{\pi} - \left(\frac{1}{3\pi^3}-\frac{1}{72\pi}  
  \right)V^2 \; , \nonumber \\ 
B_0(V\gg 1 )&=&  \frac{1}{V} - \frac{2}{V^3} \; .
\label{eq:B0limits}
\end{eqnarray}
For more details, see the supplemental material.~\cite{suppmat}

\subsection{Single-particle gap}

\subsubsection{Particle-hole symmetry}

The XXZ Hamiltonian in the fermionic
language~(\ref{eq:XXZtransfored}) is particle-hole symmetric
so that the chemical potential $\mu=0$ guarantees half filling
for all temperatures. To see this, we perform the particle-hole
transformation
\begin{equation}
  \tau_{\rm ph}: \quad \hat{c}_l^{\vphantom{+ }} \to (-1)^l \hat{c}_l^+
  \quad , \quad \hat{a}_k^{\vphantom{+ }} \to \hat{a}_{k+ \pi}^+
  \label{eq:phtarnsformation}
\end{equation}
that leaves the Hamiltonian~$\hat{H}_{\rm XXZ}$ in eq.~(\ref{eq:XXZtransfored})
invariant but changes the particle number operator, $\hat{N}\to L-\hat{N}$.
Therefore,
\begin{eqnarray}
  \langle \hat{N} \rangle^{\rm XXZ}(T,V,\mu= 0)
  &=&
  \frac{1}{Z}\Tr
  \left\{ e^{-\beta \hat{H}_{\rm XXZ}} \hat{N} \right\}\nonumber \\
    &= &
    \frac{1}{Z}\Tr
    \left\{ e^{-\beta \hat{H}_{\rm XXZ}}\left( L- \hat{N}\right) \right\} \nonumber \\
&= & L-       \langle \hat{N} \rangle^{\rm XXZ}(T,V,\mu= 0)
\end{eqnarray}
so that $\mu= 0$ indeed guarantees half band-filling for all
temperatures $T= 1/\beta$ and interaction strengths~$V$,
$ \langle \hat{N} \rangle^{\rm XXZ}(T,V,\mu= 0)= L/2$.

At zero temperature, the energies for adding another fermion to the half-filled
system and adding a particle to reach half filling are given by
\begin{eqnarray}
  \mu_1^{+}(V) &=& E_0  (L/2+1,V)-E_0(L/2,V)
  \; ,\nonumber \\
  \mu_1^{-}(V) &=& E_0(L/2,V)-E_0(L/2-1,V) \;.
%  \nonumber \\
\end{eqnarray}
The chemical potentials define the gap at half filling,
\begin{eqnarray}
  \Delta_1(V)&=&   \mu_1^{+}(V) - \mu_1^{-}(V)  \nonumber \\
  &=&   \mu_1^{+,{\rm XXZ}}(-V/2) - \mu_1^{-,{\rm XXZ}}(-V/2)  \nonumber \\
  &=& 2 \mu_1^{+,{\rm XXZ}}(-V/2) \nonumber \\
  &=& 2(\mu_1^+(V) -V)\;,
  \label{eq:muoneplusenough}
\end{eqnarray}
where we used particle-hole symmetry in the next to last step,
\begin{equation}
E_0^{\rm XXZ}(L-N,V)=E_0^{\rm XXZ}(N,V)
\end{equation}
for the ground-state energy with $N$ and $L-N$ fermions.
Due to eq.~(\ref{eq:muoneplusenough}),
we only need to calculate the ground-state energy at half band-filling
and with one additional fermion to calculate the single-particle
gap~$\Delta_1(V)$, or with two additional particles
when we calculate the two-particle gap $\Delta_2(V)$.

For the momentum distribution,
\begin{equation}
  n_k = \langle \hat{a}_k^+ \hat{a}_k^{\vphantom{+}}\rangle \; ,
  \label{eq:momdistdef}
\end{equation}
it is sufficient to investigate the region $|k|\leq \pi/2$
because particle-hole symmetry leads to
\begin{equation}
  n_{k} = 1- n_{k\pm \pi}
  \label{eqphforkdistribution}
\end{equation}
when $|k|>\pi/2$ and periodic boundary conditions are employed.

\subsubsection{Gap formula from the XXZ model}

In the antiferromagnetic XXZ model,
the elementary excitations are spin-1/2 objects called spinons.
For a spin-flip in the XXZ model, (at least) two spinons are required.
Adding or subtracting a particle in the model for spinless fermions
corresponds to such a spin flip. 
Since the spinon  dispersion is gapped for $\Delta <-1$,
there is a finite gap for charge excitations for $V>V_{\rm c}=2$.

The spinon dispersion for the XXZ chain is known
analytically,~\cite{PhysRevA.8.2526,BABELON198313} (recall $\Delta=-V/2$)
\begin{eqnarray}
  \epsilon_s(p,V)&=& \frac{2K(m)}{\pi}\sinh(\gamma) \sqrt{1-m\cos^2(p)} \; ,
  \nonumber \\
 \cosh(\gamma)&=&\frac{V}{2} \;, \quad 0\leq p\leq \pi \; .
\end{eqnarray}
Here, $K(m)$ is the complete elliptic integral of the first kind,
\begin{equation}
K(m)= \int_0^{\pi/2}\frac{{\rm d}\theta}{\sqrt{1-m\sin^2(\theta)}} \, 
\end{equation}
and $m$ follows from the solution of the implicit equation
\begin{equation}
  \gamma = \frac{\pi K(1-m)}{K(m)} \; .
  \label{eq:solvethisnome}
\end{equation}
Since it takes two spinons to create a spin-flip, we have
\begin{equation}
\mu_1^{+,{\rm XXZ}}(V)=2 \epsilon_s(0,V) \; .
\end{equation}
Therefore, the single-particle gap is given by
\begin{equation}
  \Delta_1(V)=\frac{8K(m)}{\pi}\sinh(\gamma) \sqrt{1-m} \; .
  \label{eq:Delta1}
\end{equation}
The single-particle gap can be expressed more compactly
in terms of Jacobi functions.~\cite{Goehmannprivatecomm}

Analytic results close to the transition and for strong coupling
are summarized in the supplemental material.~\cite{suppmat}
For comparison with approximate treatments, we list
the leading-order behavior close to the transition and
the strong-coupling result,
\begin{eqnarray}
  \Delta_1(V\gtrsim 2)&=& 16 \pi \exp\left(-\frac{\pi^2}{2\sqrt{V-2}}\right)
  \; , \\
\Delta_1(V\gg 1 )&=& 2V-8 + \frac{4}{V} \; .
\label{eq:gapexactlimits}
\end{eqnarray}
Apparently, above the transition
the gap opens exponentially in $1/\sqrt{V-V_{\rm c}}$.
A similar exponential
behavior is characteristic for the Kosterlitz-Thouless transition
in certain two-dimensional models at finite temperature.~\cite{KTtransition}
Therefore, it is said that the quantum phase transition is of
`Kos\-ter\-litz-Thouless type'.

The result for strong coupling is readily understood.
An extra fermion added to the half-filled system
leads to three fermions in a row and thus to two nearest-neighbor interactions
with an excitation energy of $2V$.
The transfer of particles between neighboring sites results in the free motion
of domain walls to the right and left. Therefore, the first correction term
to the single-particle is twice the bandwidth, namely $W= 4$
for each domain wall.

\subsection{Order parameter}

For spinless fermions, the charge density obeys
\begin{equation}
\langle \hat{n}_l\rangle = \frac{1}{2} + (-1)^l n_a(V)
\end{equation}
with $0\leq n_a(V)\leq 1/2$ when we select the CDW solution with higher
particle density on the even lattice sites.
Note that $n_a(V)$ is finite for $V>V_{\rm c}= 2$
and $n_a(V\to \infty)= 1/2$ for strong coupling.

The order parameter for the XXZ Heisenberg model in the thermodynamic limit
was calculated by Baxter using the Bethe Ansatz,\cite{Baxter1973}
and re-derived by
Izergin et al.\ using the algebraic Bethe Ansatz.\cite{IZERGIN1999679}
They give 
\begin{equation}
  s_0(q) = \left[ \prod_{m= 1}^{\infty} \frac{1-q^{2m}}{1+ q^{2m}}\right]^2 \;,
  \label{eq:s0product}
\end{equation}
where, for
\begin{equation}
  |\Delta|= \frac{V}{2} \geq 1 \; ,
\end{equation}
we have
\begin{equation}
  q(V)= |\Delta| - \sqrt{\Delta^2-1}
  = \frac{V}{2} - \sqrt{\left(\frac{V}{2}\right)^2-1} \leq 1 \; .
\end{equation}
The charge-density wave order parameter evaluated for spinless fermions
thus reads 
\begin{equation}
n_a(V)= \frac{1}{2} s_0[q(V)] \; .
\end{equation}
The order parameter can be expressed more compactly
in terms of Jacobi functions
and their derivatives.~\cite{Goehmannprivatecomm}

Analytic results close to the transition and for strong coupling
are summarized in the supplemental material.~\cite{suppmat}
For comparison with approximate treatments, we list
the leading-order behavior close to the transition and
the strong-coupling result,
\begin{eqnarray}
  n_a(V\gtrsim 2)&=&
  \frac{\pi}{\ln(q(V))} \exp\left(-\frac{\pi^2}{4\ln(q(V))}\right)\nonumber \\
  & \approx  & \frac{\pi}{\sqrt{V-2}}\exp\left(-\frac{\pi^2}{4\sqrt{V-2}}
  \right)
  \; , \nonumber \\
  n_a(V\gg 1 )&=& \frac{1}{2}-2 \left(\frac{1}{V}\right)^2
  -2 \left(\frac{1}{V}\right)^4
  \; .
  \label{eq:nalimitsexact}
\end{eqnarray}
As for the single-particle gap, we find that
the order parameter is exponentially small just above the transition.

\subsection{Correlation energy}
\label{subsec:correlationenergy}

By definition, the correlation energy is the difference between
the total interaction energy $\langle \hat{V}\rangle$ per site
and the single-particle contribution that results from a Hartree-Fock
decomposition of the four-fermion terms in $\hat{V}$,
\begin{equation}
  e_{\rm corr}(V) = \frac{1}{L} \left(
\langle \hat{V}\rangle -\langle \hat{V}^{\rm H} + \hat{V}^{\rm F}\rangle
  \right) \; ,
\end{equation}
see Sect.~\ref{sec:HFapproximation} for
the definition of
$\langle \hat{V}^{\rm H} \rangle$ and $\langle \hat{V}^{\rm F}\rangle$.
In terms of the exactly known CDW order parameter and
the nearest-neighbor single-particle density matrix we have
\begin{eqnarray}
  \langle \hat{V}^{\rm H} \rangle &=&
 VL \left(\frac{1}{4} -[n_a(V)]^2 \right)
\; , \nonumber \\
\langle \hat{V}^{\rm F}\rangle &=&
- VL [B_0(V)]^2 \; .
\end{eqnarray}
With eq.~(\ref{eq:Vexactgeneral}) and eq.~(\ref{eq:exactB0general})
we thus find for the correlation energy
\begin{eqnarray}
  e_{\rm corr}(V) &= &  V \biggl[ e_0'(V)
    -\frac{1}{4} + [n_a(V)]^2
    \nonumber \\
    && \hphantom{V \biggl[}
      + \frac{1}{4}\left( e_0(V)-Ve_0'(V)\right)^2
      \biggr] \; ,
      \label{eq:ecorrexactBA}
\end{eqnarray}
where the prime indicates the partial derivative with respect to~$V$.

\subsection{Momentum distribution}
\label{subsec:momdisexact}

The momentum distribution $n_k= \langle \hat{a}_k^+
\hat{a}_k^{\vphantom{+ }}\rangle$
has not been determined
analytically thus far, apart from some limiting cases.
It is known that the curves for $V>0$ are smooth in the thermodynamic limit
with $n_{k= \pm\pi/2}= 1/2$ due to
particle-hole symmetry, see eq.~(\ref{eqphforkdistribution}).

Below the transition, $V<V_{\rm c}= 2$,
the system describes a Luttinger liquid.~\cite{Thierrybook,PhysRevLett.64.2831}
Consequently, the momentum distribution close to the Fermi points
$k_{\pm}= \pm k_{\rm F}
= \pm \pi/2$ is known in the thermodynamic limit,
\begin{eqnarray}
  n_{k\approx k_{\rm F}}(V\leq V_{\rm c}) &=& \frac{1}{2}
  - \frac{1}{2}  \sgn(k-k_{\rm F}) |k-k_{\rm F}|^{\alpha(V)}  \; ,
  \nonumber \\
  \alpha(V) &=& \frac{1}{2}\left(K(V)+ \frac{1}{K(V)}-2\right)>0 \; ,
  \label{eq:momdisLLexact}
\end{eqnarray}
where the factor one half in front of the sign function takes into account that
the fermions are spinless.

A comparison of the elementary excitations from Bethe Ansatz
with those from a generic Luttinger liquid permits to identify
the Luttinger parameter~$K(V)$
in the metallic phase,~\cite{Thierrybook}
\begin{equation}
  K(V) = \frac{\pi}{2\arccos(-V/2)}
  \label{eq:exactLLK}
\end{equation}
for $0\leq V<V_{\rm c}= 2$.
This results in $K(V= 0)= 1$ ($\alpha(V= 0)= 0$) at the Fermi-liquid point,
and $K(V= V_{\rm c})= 1/2$ ($\alpha(V_{\rm c})= 1/4$)
at the CDW transition.
Consequently, the critical interaction can be deduced from
monitoring $K(V)$ or $\alpha(V)$
in the Luttinger-liquid phase.
The Luttinger exponent can also be extracted
from the long-range decay of the single-particle correlation function
in position space.~\cite{PhysRevB.86.155156}
The most reliable way to extract $K(V)$ is provided by
the analysis of the density-density correlation function
in the limit of long wave lengths,~\cite{Ejima2005}
see Sect.~\ref{subsec:CNNexact}.

In the insulating CDW phase, $V>V_{\rm c}$, the momentum distribution is
continuous and continuously differentiable.
For $V\gg 1$, strong-coupling perturbation theory gives for $|k|\leq \pi/2$
\begin{equation}
  n_k(V\gg 1) \approx
  \frac{1}{2} +    \frac{|\epsilon(k)|}{V}
  + {\cal O}\left(\frac{1}{V^2}\right)
\end{equation}
so that for all $|k|\leq \pi$
\begin{equation}
n_k(V\gg 1) \approx
\frac{1}{2}  +    \frac{2 \cos(k)}{V}
  + {\cal O}\left(\frac{1}{V^2}\right)\,.
  \label{eq:mklargeVexact}
\end{equation}
This relation follows from the fact that the Hartree-Fock ground state becomes
exact to leading
order in $1/V$, see Sect.~\ref{sec:HFapproximation}.

The two expressions~(\ref{eq:momdisLLexact}) and~(\ref{eq:mklargeVexact})
can be combined to
\begin{equation}
  n_{k\approx k_{\rm F}}(V) = \frac{1}{2}
  - b(V)  \sgn(k-k_{\rm F}) |k-k_{\rm F}|^{\alpha(V)}  \; ,
  \label{eq:momdisgeneralexact}
\end{equation}
where exact expressions for $\alpha(V)$ and $b(V)$ are known
in the Luttinger liquid phase, see eq.~(\ref{eq:momdisLLexact}),
and for strong coupling,
$\alpha(V\gg 1)= 1$, $b(V\gg 1)= 2/V$.

\subsection{Density-density correlation function}
\label{subsec:CNNexact}

Lastly, we list some exact results for the density-density correlation function,
\begin{equation}
  C^{\rm NN}(r,V) = \frac{1}{L} \sum_{l=1}^L
  \bigl(\langle \hat{n}_{l+r}\hat{n}_l \rangle
  - \langle \hat{n}_{l+r}\rangle \langle \hat{n}_l \rangle\bigr) \; ,
  \label{eq:CNNdef}
\end{equation}
which can be calculated analytically from Bethe Ansatz~\cite{Goehmannetal2021}
and numerically using DMRG. By inversion symmetry, we have 
$C^{\rm NN}(L-r,V)=C^{\rm NN}(r,V)$.
The limit $r\gg 1$ for $V\leq V_{\rm c}=2$
is also accessible
from field theory,~\cite{Thierrybook,PhysRevLett.64.2831,PhysRevB.39.4620}
\begin{equation}
  C^{\rm NN}(r\gg 1,V) \sim - \frac{K(V)}{2(\pi r)^2}
  +\frac{A(V) (-1)^r}{r^{1+K}[\ln(r)]^{3/2}} + \ldots \; ,
  \label{eq:CNNfieldtheory}
\end{equation}
where $A(V)$ is a constant that depends on the interaction but not
on the distance~$r$.

We extract the Luttinger exponent~$K(V)$ from the structure factor,
\begin{equation}
  \tilde{C}^{\rm NN}(q,V)
  = \sum_{r=0}^{L-1} C^{\rm NN}(r,V) e^{-{\rm i}q r}\;,
  \label{eq:CNNtildedef}
\end{equation}
where the wave numbers are from momentum space,
$q=(2\pi/L)m_q$, $m_q=-L/2,-L/2+1,\ldots,L/2-1$.
By construction, $\tilde{C}^{\rm NN}(q=0,V)=0$
because the particle number is fixed,
$N=L/2$ in the ground state.
When eq.~(\ref{eq:CNNfieldtheory}) is employed, it follows
that
\begin{equation}
  \frac{K(V)}{2}= \pi \lim_{q\to 0}\frac{\tilde{C}^{\rm NN}(q,V)}{q} \; .
  \label{eq:KfromCNN}
\end{equation}
Using this equation, the Luttinger exponent can be calculated numerically
with very good accuracy.~\cite{Ejima2005}
The limiting cases of non-interacting spinless fermions
and the limit
of strong interactions
are readily derived analytically
because the Hartree-Fock decoupling of the four-fermion term
becomes exact, see Sect.~\ref{subsec:CNNinHF}.

Eq.~(\ref{eq:CNNfieldtheory}) shows that the structure factor
diverges algebraically for $|q|\to \pi$,
with logarithmic corrections for all $V>0$ where $K(V)<1$.
In the charge-density wave insulator, $\tilde{C}^{\rm NN}(|q|= \pi,V>V_{\rm c})$
is finite.
Note that contributions from the long-range order are subtracted
in the definition of $C^{\rm NN}(r)$.
In principle, 
the CDW transition can also be inferred from the
finite-size scaling of
\begin{equation}
  S^{\rm NN}_{\pi}(L,V)
  = \tilde{C}^{\rm NN}\left(\pi - \frac{2\pi}{L},V\right)\;.
  \label{eq:defCNNpi}
\end{equation}
This quantity diverges algebraically in the Luttinger liquid and is finite
in the CDW insulator.
However, it turns out that, even for $V= 2.5$,
it requires system sizes much larger than
$L= 512$ to observe the  saturation of
$S^{\rm NN}_{\pi}(L,V=2.5)$. Therefore, we refrain from a further analysis
of this quantity.

\section{Hartree-Fock approximation}
\label{sec:HFapproximation}

In this section we derive the Hartree-Fock approximation for the
model~(\ref{eq:Hamiltoniandef1}) for spinless fermions.
We define the Hartree and Fock interactions,
diagonalize the Hartree-Fock Hamiltonian,
and optimize the Hartree-Fock ground-state energy in the
thermodynamic limit. Lastly, we calculate the density-density correlation
function in the Hartree-Fock approximation.

\subsection{Hartree and Fock interaction}

\subsubsection{Hartree interaction}

In Hartree approximation, the interaction becomes
\begin{equation}
  \hat{V}^{\rm H}= V \sum_{l=1}^L \left[
\langle \hat{n}_l \rangle \hat{n}_{l+1} +\hat{n}_{l} \langle \hat{n}_{l+1} \rangle
      -\langle \hat{n}_{l} \rangle
      \langle \hat{n}_{l+1} \rangle
      \right] \; .
\end{equation}
At half band-filling, the best Hartree solution is obtained for
a charge-density wave
\begin{equation}
\langle \hat{n}_{l} \rangle = n +(-1)^l n_a \;. 
\end{equation}
Since
\begin{equation}
  N=\sum_{l=1}^L \langle \hat{n}_{l} \rangle
  = Ln
\end{equation}
we can set $n=1/2$ from the start, irrespective of the interaction~$V$,
  whereas the alternating charge density depends on~$V$,
  \begin{equation}
    n_a(V)=\frac{1}{L}
    \sum_{l=1}^L(-1)^l \langle \hat{n}_{l} \rangle
    = \frac{1}{2}\left(
    \langle \hat{n}_{2l} \rangle - \langle \hat{n}_{2l+1}
    \rangle \right)\; .
    \label{eq:selconsistentna}
  \end{equation}
  In the following we assume $n_a(V)\geq 0$ which selects
  the symmetry-broken state with higher particle density
  on the even lattice sites.
  Since we double the unit cell, the Hartree
  Hamiltonian must be diagonalized in
  the reduced Brillouin zone (RBZ) where $-\pi/2\leq k <\pi/2$.

\subsubsection{Fock interaction}

In Hartree-Fock theory, the Hartree Hamiltonian is supplemented by
the Fock term,
\begin{eqnarray}
  \hat{V}^{\rm F}&=& V \sum_{l=1}^L \Bigl[
    \hat{c}_l^+\hat{c}_{l+1}^{\vphantom{+}}
    \langle     \hat{c}_l^{\vphantom{+}}\hat{c}_{l+1}^+ \rangle 
+  \langle\hat{c}_l^+\hat{c}_{l+1}^{\vphantom{+}}\rangle 
\hat{c}_l^{\vphantom{+}}\hat{c}_{l+1}^+ \nonumber \\
&& \hphantom{V \sum_{l=1}^L \biggl[}
-\langle\hat{c}_l^+\hat{c}_{l+1}^{\vphantom{+}}\rangle 
\langle \hat{c}_l^{\vphantom{+}}\hat{c}_{l+1}^+ \rangle
\Bigr] \; .
\end{eqnarray}
Compatible with the Hartree solution is a bond-order wave state,
\begin{equation}
  \langle \hat{c}_{l+1}^+\hat{c}_l^{\vphantom{+}}  \rangle
  =B_0 +(-1)^l B_1
\end{equation}
with complex $B_0,B_1$.

The bond-order wave loses against the charge-density wave so that
we find $B_1= 0$ and a real $B_0$ for $V\geq 0$.
We shall work with these simplifications right from the start.

\subsection{Diagonalization of the Hartree-Fock Hamiltonian}
\label{subsec:HFDiagonlaization}

To leading order in the Hartree-Fock approximation,
the Hartree-Fock Hamiltonian,
\begin{equation}
\hat{H}^{\rm HF}= \hat{T}+ \hat{V}^{\rm H}+ \hat{V}^{\rm F}\;, 
  \end{equation}
must be diagonalized.

\subsubsection{Operators in the reduced Brillouin zone}

We have
  \begin{equation}
    \hat{T}= \sum_{k\in{\rm RBZ}} \epsilon(k)
    \left(
    \hat{a}_k^+\hat{a}_k^{\vphantom{+}} -
    \hat{a}_{k+\pi}^+\hat{a}_{k+\pi}^{\vphantom{+}}\right)
    \end{equation}
  because of the nesting property of the dispersion relation,
  $\epsilon(k+\pi)=-\epsilon(k)$.

  For the Hartree interaction we find
  \begin{eqnarray}
    \hat{V}^{\rm H} &=& V \sum_{l=1}^L (n+(-1)^l n_a)\hat{n}_{l+1}
      +\hat{n}_l (n+(-1)^{l+1}n_a)
    \nonumber \\    &&
      - V \sum_{l=1}^L
      (n+(-1)^ln_a)  (n+(-1)^{l+1}n_a)\nonumber \\
%    &=& -VL(n^2-n_a^2)+Vn \sum_1^{L}\left(\hat{n}_l+\hat{n}_{l+1}\right)
%    \nonumber \\    &&
%    +Vn_a \sum_{l=1}^L \left((-1)^l \hat{n}_{l+1}+(-1)^{l+1}\hat{n}_l\right)
%    \nonumber \\
%    &=& VL(n^2+n_a^2)-2Vn_a \sum_l (-1)^l \hat{c}_l^+\hat{c}_l^{\vphantom{+}}
%    \nonumber \\
%    &=& VL(n^2+n_a^2)-2Vn_a \sum_k \hat{a}_k^+\hat{a}_{k+\pi}^{\vphantom{+}}
%    \nonumber \\
    &=& VL(n^2+n_a^2)-2Vn_a \sum_{k\in {\rm RBZ}}
    \left(\hat{a}_k^+\hat{a}_{k+\pi}^{\vphantom{+}}
    +\hat{a}_{k+\pi}^+\hat{a}_{k}^{\vphantom{+}}\right) \; ,\nonumber \\
  \end{eqnarray}
  where we used that $\hat{N}=N$ in the sector of fixed particle number~$N$.

For the Fock interaction we find
\begin{eqnarray}
  \hat{V}^{\rm F}&=& V L B_0^2
+ V \sum_{k\in{\rm RBZ}} b_0(k)  \left(   \hat{a}_k^+\hat{a}_k^{\vphantom{+}} -
  \hat{a}_{k+\pi}^+\hat{a}_{k+\pi}^{\vphantom{+}}\right) \nonumber \; ,\\
  b_0(k)&=&-2B_0\cos(k) \; .
\end{eqnarray}
The Hartree-Fock Hamiltonian in the reduced Brillouin zone
${\rm RBZ}= \{-\pi/2 \leq k <\pi/2\}$ reads 
\begin{eqnarray}
  \hat{H}^{\rm HF} &=& VL\left(n^2+n_a^2+B_0^2\right) \nonumber \\
  && + \sum_{k\in{\rm RBZ}} \widetilde{\epsilon}(k)
   \left(
    \hat{a}_k^+\hat{a}_k^{\vphantom{+}} -
    \hat{a}_{k+\pi}^+\hat{a}_{k+\pi}^{\vphantom{+}}\right)
    \nonumber \\
    && 
    - \sum_{k\in{\rm RBZ}}
    2Vn_a \left(\hat{a}_k^+\hat{a}_{k+\pi}^{\vphantom{+}}
    +\hat{a}_{k+\pi}^+\hat{a}_{k}^{\vphantom{+}}\right)
    \label{eq:HFbeforediagonalization}
\end{eqnarray}
with
\begin{equation}
  \widetilde{\epsilon}(k)  =\epsilon(k) + V b_0(k)= -2(1+ V B_0) \cos(k) 
  \; .
\end{equation}

\subsubsection{Diagonalization}

For the diagonalization of the Hartree Hamiltonian we introduce for
each $k\in{\rm RBZ}$
\begin{eqnarray}
  \hat{a}_k &=& \cos(\varphi_k) \hat{\alpha}_k-\sin(\varphi_k)\hat{\beta}_k \;,
  \nonumber \\
  \hat{a}_{k+\pi} &=& 
  \sin(\varphi_k) \hat{\alpha}_k+\cos(\varphi_k)\hat{\beta}_k\;.
  \label{eq:introducephik}
\end{eqnarray}
The operators $\hat{\alpha}_k$ and $\hat{\beta}_k$
obey fermionic commutation relations for real $0\leq \varphi_k<2\pi$.

For each $k\in{\rm RBZ}$ we thus have to diagonalize
%{\arraycolsep=2pt
  \begin{eqnarray}
  \hat{h}_k^{\rm HF}&=& \widetilde{\epsilon}(k) 
    \left(u_k \hat{\alpha}_k^+ -v_k \hat{\beta}_k^+ \right)
    \left(u_k \hat{\alpha}_k^{\vphantom{+}} -v_k \hat{\beta}_k^{\vphantom{+}} \right)
    \nonumber \\
   && -  \widetilde{\epsilon}(k)
    \left(v_k \hat{\alpha}_k^+ +u_k \hat{\beta}_k^+ \right)
    \left(v_k \hat{\alpha}_k^{\vphantom{+}} +u_k \hat{\beta}_k^{\vphantom{+}} \right)
    \nonumber \\
  && -2Vn_a 
    \left(u_k \hat{\alpha}_k^+ -v_k \hat{\beta}_k^+ \right) 
    \left(v_k \hat{\alpha}_k^{\vphantom{+}} +u_k \hat{\beta}_k^{\vphantom{+}} \right)
\nonumber \\
  && -2Vn_a  \left(v_k \hat{\alpha}_k^+ +u_k \hat{\beta}_k^+ \right)
    \left(u_k \hat{\alpha}_k^{\vphantom{+}} -v_k \hat{\beta}_k^{\vphantom{+}} \right)
     , \nonumber \\
\end{eqnarray}
  where we abbreviated $u_k=\cos(\varphi_k)$ and $v_k=\sin(\varphi_k)$.

  The non-diagonal terms proportional to $\hat{\alpha}_k^+
  \hat{\beta}_k^{\vphantom{+}}$
must vanish. This leads to the condition
\begin{equation}
\tan(2\varphi_k)= -\frac{2Vn_a}{\widetilde{\epsilon}(k)}\geq 0  \; ,
\end{equation}
and 
\begin{eqnarray}
  \cos(2\varphi_k) &= & \frac{|\widetilde{\epsilon}(k)|}{E(k)} \; ,
  \nonumber \\
  2u_kv_k= \sin(2\varphi_k) &= & \frac{2 V n_a}{E(k)} \; ,
          \label{eq:cossinandallthat}
      \label{eq:defineEofk}\\
 u_k^2=  \cos^2(\varphi(k))&= & \frac{1}{2}\left( 1+
    \frac{|\widetilde{\epsilon}(k)|}{E(k)}    \right) \; ,
    \nonumber \\
    v_k^2=   \sin^2(\varphi(k))&= & \frac{1}{2}\left( 1-
    \frac{|\widetilde{\epsilon}(k)|}{E(k)}    \right) \; ,\nonumber\\
      E(k)&=& \sqrt{(\epsilon(k)+Vb_0(k))^2+(2Vn_a)^2} \; .\nonumber
\end{eqnarray}
The Hartree-Fock Hamiltonian becomes diagonal in the new basis,
\begin{equation}
  \hat{H}^{\rm HF}=
VL\left(n^2+n_a^2+B_0^2\right) 
  +\sum_{k\in{\rm RBZ}} E(k)\biggl(
      \hat{\beta}_k^+\hat{\beta}_k^{\vphantom{+}}
      -     \hat{\alpha}_k^+\hat{\alpha}_k^{\vphantom{+}}\biggr)
      \label{eq:HFdiagonal}
\end{equation}
with the dispersion relation $E(k)$.
%\begin{equation}
%  \label{eq:defineEofk}
%\end{equation}
The Hamiltonian parametrically depends on $B_0(V)$ and $n_a(V)$.

\subsection{Minimization of the Hartree-Fock ground-state energy
in the thermodynamic limit}

The optimal Hartree-Fock energy can thus be found from the minimization of
the simplified Hartree-Fock energy functional ($n= 1/2$)
\begin{equation}
E_0^{\rm HF }(B_0,n_a,V)=
VL\left(n^2+n_a^2+B_0^2\right) 
-\sum_{k\in{\rm RBZ}} E(k) 
\label{eq:HFsimplified}
\end{equation}
for real $B_0$, $n_a$. 
In the thermodynamic limit, eq.~(\ref{eq:HFsimplified}) can be expressed as
\begin{eqnarray}
  e_0^{\rm HF}(B_0,n_a,V)
  &=&\lim_{L\to\infty} \frac{E_0^{\rm HF }(B_0,n_a,V)}{L} \nonumber \\
  &=&  V\left(n^2+n_a^2+B_0^2\right) \nonumber \\
  &&  - \frac{1}{\pi} \sqrt{a^2+b^2} E\left[a^2/(a^2+b^2)\right] \, ,\;
\end{eqnarray}
where $E[m]$ ($0\leq m\leq 1$) is the complete elliptic integral
of the second kind,
\begin{equation}
  E[m] = \int_0^{\pi/2} {\rm d}\varphi\sqrt{1-m \sin^2(\varphi)}   \; ,
  \label{eq:ellipticEdef}
\end{equation}
and we defined the abbreviations
\begin{eqnarray}
  a&=& 2(1+VB_0)\; , \nonumber \\
  b &=& 2Vn_a \; .
\end{eqnarray}
For general interactions and system sizes, the optimization of the
Hartree-Fock ground-state energy has to be done numerically.

\subsubsection{Small interactions}

The minimization of $e_0^{\rm HF}(B_0,n_a,V)$ %\equiv e_0^{\rm HF}$
can be carried out analytically
for small~$V$. The Taylor series up to third order in~$V$
reads
\begin{eqnarray}
  e_0^{\rm HF}(B_0,n_a,V)&\approx & -\frac{2}{\pi}
  +\left( \frac{1}{4} -\frac{2B_0}{\pi}+B_0^2+n_a^2\right)V\nonumber  \\
&&  +\frac{V^2n_a^2}{2\pi}\left(
  -1 - 4 \ln(2) + 2 \ln(n_aV) \right)
  \nonumber  \\
  && +\frac{V^3B_0n_a^2}{2\pi}\left(
  -1+ 4 \ln(2) - 2 \ln(n_aV) \right)
  \nonumber  \\
  && +{\cal O}\left(V^4\right) \;.
\label{eq:e0HFsmallVTaylor}
\end{eqnarray}
Its minimization leads to two coupled equations for
$B_0$ and $n_a$.
The solution for $B_0$ is given by
\begin{equation}
  B_0 = \frac{1}{4\pi} \left(4 + n_a^2 V^2 (1- 4\ln(2))
  + 2 n_a^2 V^2 \ln(n_aV)\right) \; .
\end{equation}
We insert this result into the minimization equation for $n_a$ 
and expand to third order in~$V$ to find the solution
\begin{equation}
  n_a(V\to 0)=\frac{4}{V}\exp\left(-\frac{\pi}{V}-1\right) \;.
  \label{eq:nasmallVanalyt}
\end{equation}
In Hartree-Fock theory, the order parameter is finite for all $V>0$
and displays an
essential singularity at $V=0$.
The Fock parameter  deviates exponentially from its bare value $B_0(V=0)=1/\pi$,
\begin{equation}
  B_0(V\to 0)=\frac{1}{\pi} -\frac{4(2\pi+V)}{\pi V}
  \exp\left(-\frac{2\pi}{V}-2\right) \;.
\end{equation}
Consequently,
the optimized Hartree-Fock ground-state energy
per site for small interactions becomes
\begin{eqnarray}
  e_0^{\rm HF, min}(V\to 0) &=& -\frac{2}{\pi}
  + \left(\frac{1}{4}-\frac{1}{\pi^2}\right)V \nonumber \\
  &&  -\frac{8(\pi-V)}{\pi^2} \exp\left(-\frac{2\pi}{V}-2\right)
  \label{eq:HFenergysmallV}\\
  &&
-\frac{16(2\pi+V)^2}{\pi^2V} \exp\left(-\frac{4\pi}{V}-4\right) \nonumber
\;.
\end{eqnarray}
This formula agrees with the numerically
determined value with an accuracy of better
than $10^{-3}$ for $V\leq 1$. The error is only 5\% at $V=2$.

The Hartree-Fock theory reproduces 
the exact ground-state energy and nearest-neighbor single-particle
density matrix
for small interactions~(\ref{eq:gsenergylimits}) to first order
but lacks the correct second-order terms.

\subsubsection{Large interactions}
\label{subsubsec:largeinteractions}

For large interactions, the energy can be expanded in a power series in $1/V$.
To find the series, we also expand the variational parameters $B_0$ and $n_a$
in inverse powers of~$V$. It turns out that $n_a$ ($B_0$) contains
only even (odd) powers,
\begin{eqnarray}
  n_a &=& \frac{1}{2} +\frac{\delta_2}{V^2} +\frac{\delta_4}{V^4}
  +\frac{\delta_6}{V^6}
  +\frac{\delta_8}{V^8}
  +   {\cal O}(V^{-10})
  \;, \nonumber \\
  B_0&=&\frac{b_1}{V} +\frac{b_3}{V^3} +\frac{b_5}{V^5} +\frac{b_7}{V^7}
  +   {\cal O}(V^{-9})\; .
\end{eqnarray}
Up to the given order we find from the minimization
of the Hartree-Fock ground-state energy
\begin{eqnarray}
  n_a &=& \frac{1}{2} -\frac{2}{V^2} +\frac{10}{V^4}-\frac{64}{V^6}
  +\frac{466}{V^8}+    {\cal O}(V^{-10}) \;, \label{eq:nalargeVanalyt} \\[6pt]
  B_0&=&\frac{1}{V} -\frac{4}{V^3} +\frac{24}{V^5} -\frac{168}{V^7}
  +   {\cal O}(V^{-9})\; . 
\end{eqnarray}
The expansion reproduces the Hartree-Fock result for the order parameter $n_a$
with an accuracy
of at least $6\cdot 10^{-3}$ ($6\cdot 10^{-4}$) for $V/t\geq 4$ ($V/t\geq 5$),
and for $B_0$ with an accuracy
of better than $2\cdot 10^{-2}$ ($3\cdot 10^{-3}$) for $V/t\geq 4$ ($V/t\geq 5$).

Hartree-Fock theory reproduces the exact order para\-me\-ter
in the strong-coupling limit
to second order in $1/V$, see eq.~(\ref{eq:nalimitsexact}).
Corrections are of the order $1/V^4$.
Moreover, it gives the correct leading order for $B_0$, see
eq.~(\ref{eq:B0limits}), with corrections of the order $1/V^3$.

With these parameters, the Hartree-Fock ground-state energy can be calculated
up to 15th order in $t/V$,
\begin{eqnarray}
  e_0^{\rm HF}(V\gg 1) &= & -\frac{1}{V}+\frac{2}{V^3}-\frac{8}{V^5}
  +\frac{42}{V^7}-\frac{256}{V^9}+\frac{1712}{V^{11}}
  \nonumber \\
&&  -\frac{12192}{V^{13}}+\frac{90858}{V^{15}} + {\cal O}(V^{-17}) \; .
\end{eqnarray}
The expansion reproduces the Hartree-Fock result for the ground-state energy
with an accuracy
of better than $10^{-2}$ ($10^{-4}$) for $V/t\geq 3$ ($V/t\geq 4$).

In the strong-coupling limit, the Hartree-Fock approximation
reproduces the exact ground-state energy to leading order in $1/V$,
see eq.~(\ref{eq:gsenergylimits}), with corrections of the order $1/V^3$.

\subsubsection{Hartree-Fock single-particle gap}

When we add a particle or hole to the half-filled state, the variational
parameters do not have to be re-adjusted because the Hartree-Fock energy
is minimal at $n_a^{(0)}$ and $B_0^{(0)}$.
Corrections of the form
$x^{(0)}\to x^{(0)}+p/L$ thus lead to corrections of the order $1/L$ in
$E_0^{\rm HF, min}$ whereas the dominant correction of order unity results from
the additional particle or hole.
Therefore, we obtain the Hartree-Fock chemical potentials from the Hartree-Fock
band structure
\begin{eqnarray}
  \mu_1^+&=& E_0^{\rm HF, min}(L+1)-E_0^{\rm HF, min}(L)
  %\nonumber \\  &=&
  = V+\bar{E}(\pi/2) %=V(1+2n_a)
  \; , \nonumber \\
  \mu_1^- &=& E_0^{\rm HF, min}(L)-E_0^{\rm HF, min}(L-1)
  %\nonumber \\&=&
  = V-\bar{E}(\pi/2) %=V(1-2n_a)
  \nonumber \\
\end{eqnarray}
with $\bar{E}(\pi/2)= 2Vn_a(V)$,
and the gap for single-particle excitations becomes
\begin{equation}
  \Delta_1^{\rm HF}(V)=4Vn_a(V) \;,
  \label{eq:HForderandgap}
\end{equation}
where $n_a(V)$ is the Hartree-Fock order parameter.
For explicit expressions for $n_a(V)$
for small and large interactions,
see eqs.~(\ref{eq:nasmallVanalyt}) and~(\ref{eq:nalargeVanalyt}), respectively.

In particular, the leading orders in the strong-coupling expansion read
\begin{equation}
  \Delta_1^{\rm HF}(V\gg t)=4Vn_a(V)\approx
  2V -\frac{8t^2}{V} \; .
\end{equation}
In strong coupling, Hartree-Fock theory reproduces only the leading order
of the exact single-particle gap, see eq.~(\ref{eq:gapexactlimits}).
The domain walls in the charge-density wave are mobile
in the exact solution whereas they are localized in the Hartree-Fock
description. Therefore, at strong coupling,  the Hartree-Fock approximation
lacks a gap contribution of the order unity.

Since this basic problem is not cured by second-order perturbation theory,
we refrain from a comparison of the Hartree-Fock and exact single-particle
gaps.

\subsection{Density-density correlation function}
\label{subsec:CNNinHF}

In Hartree-Fock theory, the four-fermion term
in the density-density correlation function in eq.~(\ref{eq:CNNdef})
factorizes,
\begin{equation}
  C^{\rm NN}_{\rm HF}(r)= \frac{\delta_{r,0}}{L} \sum_l \langle \hat{n}_l \rangle
  (1-\langle \hat{n}_l \rangle)  -\frac{(1-\delta_{r,0})}{L} \sum_l |P_{l+r,l}|^2
  \; ,
\end{equation}
where $P_{l,m}$ is the single-particle density matrix,
\begin{equation}
  P_{l,m}= \langle \hat{c}_{l}^+ \hat{c}_m^{\vphantom{+}} \rangle
  \; .
\end{equation}
For $V=0$ the single-particle density matrix is the Fourier transform
of the momentum distribution,
\begin{equation}
  P_{l,m}^{(0)}= \frac{1}{L} \sum_k e^{-{\rm i}k(l-m)} n_k=P_{m,l}^{(0)}=
  \Bigl[P_{l,m}^{(0)}\Bigr]^*
  \; .
\end{equation}
Upon Fourier transformation we thus find
\begin{equation}
  \tilde{C}_0^{\rm NN}(q)= \frac{1}{2}-\frac{1}{L} \sum_k n_k n_{k+q}
  = \frac{|q|}{2\pi}
  \label{eq:CNNVequalzero}
\end{equation}
and thus $K(V=0)=1$ for the Luttinger parameter, as expected.

For $V\geq 0$ and $r\neq 0$, Hartree-Fock theory gives
\begin{eqnarray}
  P_{r+ l,l} &=& \frac{1}{L} \sum_{k\in {\rm RBZ}} e^{-{\rm i} kr}  \biggl[
    \langle      \hat{a}_{k}^+ \hat{a}_{k}^{\vphantom{+ }}      \rangle
    + (-1)^r \langle      \hat{a}_{k+ \pi}^+ \hat{a}_{k+ \pi}^{\vphantom{+ }}
    \rangle
    \nonumber \\
    && \hphantom{\frac{1}{L} \sum_{k\in {\rm RBZ}} e^{-{\rm i} kr}  \biggl[}
      (-1)^l\Bigl[ 1+ (-1)^{r}\Bigr]
      \langle      \hat{a}_{k+ \pi}^+ \hat{a}_{k}^{\vphantom{+ }}      \rangle
      \biggr]\,,\nonumber \\
\end{eqnarray}
where $\langle      \hat{a}_{k+ \pi}^+ \hat{a}_{k}^{\vphantom{+ }}      \rangle
= 
\langle      \hat{a}_{k}^+ \hat{a}_{k+ \pi}^{\vphantom{+ }}      \rangle
=u_kv_k$ is real,
$\langle      \hat{a}_{k}^+ \hat{a}_{k}^{\vphantom{+ }}      \rangle=u_k^2$,
and 
$\langle      \hat{a}_{k+ \pi}^+ \hat{a}_{k+ \pi}^{\vphantom{+ }}      \rangle=v_k^2
= 1-u_k^2= 1-\langle      \hat{a}_{k}^+ \hat{a}_{k}^{\vphantom{+ }}      \rangle$,
in agreement with eq.~(\ref{eqphforkdistribution}).
Then,  for $r \neq 0$,
\begin{eqnarray}
  \frac{1}{L}\sum_l |P_{l+ r,l}|^2 &= & |P_1(r)|^2+ |P_2(r)|^2 \; , \nonumber \\
  P_1(r) &= & (-1)^r \frac{1}{L} \sum_{k\in{\rm RBZ}}   e^{-{\rm i}k r} \nonumber \\
  && 
  + 
  \frac{r_{\rm o}}{L} \sum_{k\in{\rm RBZ}}   e^{-{\rm i}k r}
  \left(1+ \frac{|\tilde{\epsilon}(k)|}{E(k)} \right)
  \nonumber\\
  &= & 
 \frac{r_{\rm o}}{L} \sum_{k\in{\rm RBZ}}   e^{-{\rm i}k r}
  \frac{|\tilde{\epsilon}(k)|}{E(k)}   
  \; , \nonumber \\
  P_2(r) &= &
  \frac{  r_{\rm e}}{L} \sum_{k\in{\rm RBZ}}   e^{-{\rm i}k r}
  \frac{2Vn_a}{E(k)} \; , 
  \end{eqnarray}
where
$r_{\rm o}= (1-(-1)^r)/2$ and
$r_{\rm e}= (1+ (-1)^r)/2$ are unity when $r$ is odd or even, respectively,
and zero else.
Performing the Fourier transformation, the density-density correlation function
in the Hartree-Fock approximation becomes
\begin{eqnarray}
  \tilde{C}_{\rm HF}^{\rm NN}(q)
  &= & \frac{1}{4}
  - \frac{1}{2}  \int_{-\pi/2}^{\pi/2}\frac{{\rm d}k}{2\pi} F(k,q)
  \; , \nonumber \\
  F(k,q) &= &   \frac{\tilde{\epsilon}(k)}{E(k)}
    \frac{\tilde{\epsilon}(k+ q)}{E(k+ q)}   
    +   \frac{2Vn_a}{E(k)}\frac{2Vn_a}{E(k+ q)}
\end{eqnarray}
in the thermodynamic limit. For $V= 0$, the result~(\ref{eq:CNNVequalzero})
is recovered; note that $E(k+ q)$ is always positive but
$\tilde{\epsilon}(k+ q)$ changes its sign at $k= \pi/2-q$ ($k= -\pi/2-q$)
when $q>0$ ($q<0$).

In the limit of strong coupling, the Hartree-Fock ground-state becomes exact to
leading order in $1/V$. Therefore,
the strong-coupling result for the spin-spin correlation functions becomes
\begin{equation}
\tilde{C}^{\rm NN}(q,V\gg  1) = \frac{2(1-\cos(q))}{V^2} \; .
\label{eq:CNNtildelargeq}
\end{equation}
This corresponds to the fact that, to leading order in $1/V$, the
single-particle density matrix is finite only for nearest neighbors.

\section{Second-order Hartree-Fock approximation}
\label{sec:2ndorderHF}

In this section, we calculate the second-order correction in the interaction
around the Hartree-Fock solution presented in the previous section.
This concept was applied earlier to the extended Hubbard model
around the limit of high dimensions.~\cite{PhysRevB.50.14016}

First, we formally expand the ground-state energy and the momentum distribution
to second order, and identify the required excited states.
Next, we argue that second-order Hartree-Fock theory is applicable
for spinless fermions for all interaction strengths, 
and calculate the second-order corrections to the ground-state energy and the
momentum distribution. Finally, we discuss the
metal-insulator transition in second-order Hartree-Fock theory.

\subsection{Formal expansion}

For the derivation of the formal second-order expansion,
we assume that $n_a$ and $B_0$ are fixed.~\cite{PhysRevB.43.3475}

\subsubsection{Perturbation operator}

We write
\begin{equation}
\hat{H}= \hat{T}+ \hat{V}= \hat{H}_{\rm HF}+ \hat{V}_{\perp}
\end{equation}
with the perturbation operator
\begin{equation}
\hat{V}_{\perp} = \hat{V}-\hat{V}^{\rm H}-\hat{V}^{\rm F} \; .
\end{equation}

\subsubsection{Ground state to first order}

The ground state to first order in the perturbation reads
\begin{equation}
  |\psi_0\rangle^{(1)} = |0\rangle + \sum_{|n\rangle \neq |0\rangle}
  |n\rangle \frac{\langle n | \hat{V}_{\perp} | 0\rangle}{E_0^{(0)}-E_n^{(0)}}
  \; .
  \label{eq:defexactgs1storder}
\end{equation}
Here,
\begin{equation}
|0\rangle = \prod_{k\in {\rm RBZ}}\hat{\alpha}_k^+ |{\rm vac}\rangle 
\end{equation}
is the Hartree-Fock ground state for given parameters $B_0$ and $n_a$.
Moreover, $|n\rangle$ are exact excited states of
the Hartree-Fock Hamiltonian $\hat{H}^{\rm HF}$, see eq.~(\ref{eq:HFdiagonal})
for its diagonalized form.

\subsubsection{Ground-state energy to second order}
\label{subsubsec:gsenergy2nd}

To second order in $V$, the ground-state energy reads
\begin{equation}
E_0^{(2)}(V) =E_0^{\rm HF}(V) + 
\sum_{|n\rangle \neq |0\rangle}
\frac{|\langle n | \hat{V}_{\perp} | 0\rangle|^2}{E_0^{(0)}-E_n^{(0)}} \; .
\label{eq:2ndorderformal}
\end{equation}
All first-order contributions are contained in the Hartree-Fock energy, i.e.,
\begin{equation}
  \langle 0 | \hat{V}_{\perp} |0 \rangle =
  \langle 0 | \hat{V}-\hat{V}^{\rm H}-\hat{V}^{\rm F} |0 \rangle =0
\end{equation}
by construction.

\subsubsection{Quasi-particle occupation numbers}

We are interested in the expectation values of the
occupation number operators in the Hartree-Fock basis,
$  \hat{n}_{p,\alpha}=
\hat{\alpha}_p^+  \hat{\alpha}_p^{\vphantom{+}} $
and
$  \hat{n}_{p,\beta}
= \hat{\beta}_p^+  \hat{\beta}_p^{\vphantom{+}} $,
\begin{eqnarray}
  n_{p,\alpha}
  &=& {}^{(1)}\langle \psi_0 |
  \hat{n}_{p,\alpha}
%  \hat{\alpha}_p^+  \hat{\alpha}_p^{\vphantom{+}}
  | \psi_0 \rangle^{(1)}
  \; , \nonumber \\
  n_{p,\beta}
  &=& {}^{(1)}\langle \psi_0 |
  \hat{n}_{p,\beta}
  %\hat{\beta}_p^+  \hat{\beta}_p^{\vphantom{+}}
  | \psi_0 \rangle^{(1)} \; .
\end{eqnarray}
We know that ($p\in {\rm RBZ}$)
\begin{equation}
  \hat{\alpha}_p^+  \hat{\alpha}_p^{\vphantom{+}}+
  \hat{\beta}_p^+  \hat{\beta}_p^{\vphantom{+}} =
  \hat{a}_p^+  \hat{a}_p^{\vphantom{+}}+
    \hat{a}_{p+ \pi}^+  \hat{a}_{p+ \pi}^{\vphantom{+}}\; .
\end{equation}
We can use particle-hole symmetry at half band-filling,
see eq.~(\ref{eqphforkdistribution}), to show that
\begin{equation}
%  \langle
  \hat{a}_p^+  \hat{a}_p^{\vphantom{+}}+
  \hat{a}_{p+ \pi}^+  \hat{a}_{p+ \pi}^{\vphantom{+}} %\rangle
  = 1 \; .
\end{equation}
Therefore,
\begin{equation}
  n_{p,\alpha} = 1- n_{p,\beta}
  \label{eq:HFqpoccoanciesrelation}
\end{equation}
for all interactions
so that it is sufficient to calculate $n_{p,\beta}$.

Since the excited states $|n\rangle$ in eq.~(\ref{eq:defexactgs1storder})
are eigenstates of the occupation number operators
 we have $\langle 0 | n \rangle= 0$ and $n_{p,\beta}^{(0)}= 0$. Thus,
we readily find
\begin{equation}
n_{p,\beta}(V)= 
\sum_{|n\rangle \neq |0\rangle}
\frac{|\langle n | \hat{V}_{\perp} | 0\rangle|^2}{\bigl(E_0^{(0)}-E_n^{(0)}\bigr)^2}
\langle n | \hat{n}_{p,\beta}| n \rangle \; .
\label{eq:occofbetap}
\end{equation}
An important quantity is the density of quasi-particle
excitations of the bare Hartree-Fock ground state,
\begin{equation}
  n_{\beta}(V) = \frac{1}{L} \sum_{p\in {\rm RBZ}} n_{p,\beta}(V)
  \label{eq:deftotalqpdensity}
\end{equation}
with $0\leq n_{\beta} \leq 1/2$. Second-order perturbation theory
remains meaningful for all interaction strengths if $n_{\beta}(V) \ll 1/2$
for all~$V$, see Sect.~\ref{subsec:almostvariational}.

\subsubsection{Excited states}

Since $\hat{V}$ contains two creation and two annihilation operators,
the intermediate excited states~$|n\rangle$ can contain one or at most two
particle-hole excitations,
\begin{eqnarray}
  |n_1 \rangle \equiv |k;p\rangle &=& \hat{\beta}_k^+ \hat{\alpha}_p^{\vphantom{+}}
  |0\rangle \; , \nonumber \\
  |n_2 \rangle \equiv |k_1,k_2;p_1,p_2\rangle &=&
  \hat{\beta}_{k_1}^+ \hat{\alpha}_{p_1}^{\vphantom{+}}
  \hat{\beta}_{k_2}^+  \hat{\alpha}_{p_2}^{\vphantom{+}}
  |0\rangle
  \label{eq:SDterms}
\end{eqnarray}
with $k_1<k_2$ and  $p_1<p_2$.
The excitation energies are
\begin{eqnarray}
  E_0^{(0)}-E_{n_1}^{(0)}&=& -\left(E(k)+ E(p)\right) \; , \nonumber \\
  E_0^{(0)}-E_{n_2}^{(0)}&=& -\left(E(k_1)+ E(k_2)+  E(p_1)+ E(p_2)\right)
  \; .\nonumber \\
  \end{eqnarray}
The matrix elements are calculated
in the supplemental material.~\cite{suppmat}
In particular, we have
\begin{equation}
  \langle 0 | \hat{V}_{\perp}| n_1\rangle = 0 
  \label{eq:n1givesnothing}
\end{equation}
so that only two-particle excitations need to be taken into account.

\subsubsection{Momentum distribution}

It is sufficient to calculate the momentum distribution $n_k$
for $|k|\leq \pi/2$ because particle-hole symmetry leads to
$n_{k} = 1- n_{k\pm \pi}$, see eq.~(\ref{eqphforkdistribution}).
Using eq.~(\ref{eq:introducephik}) we find in second-order Hartree-Fock 
theory
\begin{eqnarray}
  n_k &=& u_k^2 \langle \hat{\alpha}_k^+  \hat{\alpha}_k^{\vphantom{+}}
  \rangle + v_k^2 \langle \hat{\beta}_k^+  \hat{\beta}_k^{\vphantom{+}}\rangle
  \nonumber \\[3pt]
  &= &
  \frac{1}{2}\left( 1+
  \frac{|\widetilde{\epsilon}(k)|}{E(k)}    \right)
  - \frac{|\widetilde{\epsilon}(k)|}{E(k)}   n_{k,\beta}\; ,
  \label{eq:nkinHF2nd}
\end{eqnarray}
where we employed
eqs.~(\ref{eq:cossinandallthat}) and~(\ref{eq:n1givesnothing}).
Therefore, it is sufficient to calculate the quasi-particle density $n_{k,\beta}$
to derive the
Hartree-Fock momentum distribution.

We can use this relation to prove eq.~(\ref{eq:mklargeVexact}) for the
momentum distribution in the strong-coupling limit.
Since the Hartree-Fock ground state becomes
exact to leading order in $1/V$, we use in eq.~(\ref{eq:nkinHF2nd}) that
$n_{k,\beta}= {\cal O}(1/V^2)$,
$E(k)\approx V$, and $\widetilde{\epsilon}(k)\approx 2\epsilon(k)$
because $VB_0\approx 1$.

Note that weak-coupling perturbation theory
in the absence of CDW order leads to
a logarithmically divergent momentum distribution in the thermodynamic limit
for $|k|\to \pi/2$. This divergence signals that the Fermi gas breaks down
and must be replaced by a Luttinger liquid.~\cite{Thierrybook}
To circumvent this singularity, we later show the
second-order Hartree-Fock momentum distribution for a small but finite
CDW order parameter, $n_a^{\rm inf}=10^{-6}$, even though the minimization
leads to $n_a= 0$ in the thermodynamic limit.

\subsection{Almost-variational property}
\label{subsec:almostvariational}

The Hartree-Fock approximation is a variational theory that gives an upper bound
to the exact ground-state energy for all interaction strengths.
For fixed $n_a$ and $B_0$, the second-order Hartree-Fock energy
provides a systematic energy correction for weak interactions.
Apparently, one would rather minimize the full energy expression including
the second-order term to optimize the parameters $B_0$ and $n_a$
(`second-order Hartree-Fock approximation').
Before we shall follow this route, we give some arguments how this approach
can be justified.
In fact, the optimal second-order Hartree-Fock energy does not
necessarily provide a true variational bound for all interaction strengths
but corrections are small in the limit $n_{\beta}(V)\ll 1/2$
which is the case for spinless fermions in one dimension for all $V$
where $n_{\beta}(V_{\rm max})\approx 0.01$, see Sect.~\ref{sec:comparison}.

As in quantum chemistry, we make the variational Ansatz
for the exact ground state
\begin{equation}
  |\psi_0\rangle = |0\rangle + \sum_{n\neq 0} \Phi_n |n\rangle \; ,
  \label{eq:defvariationalstate}
\end{equation}
where $\Phi_n$ are complex coefficients and $|n\rangle$ are the
Har\-tree-Fock eigenstates.
Since the Hartree-Fock states form a complete set, the exact ground state
can be written in this form.
If we restrict ourselves to the states in eq.~(\ref{eq:SDterms}),
we recover the singlet-doublet (SD) approximation
where up to two particle-hole excitations
of the Hartree-Fock ground state $|0\rangle$ are included in
$|\psi_0^{\rm SD}\rangle$.

The expectation value for the Hamiltonian reads
\begin{eqnarray}
  H(\psi_0) &= & \langle \psi_0 | \hat{H} | \psi_0\rangle \nonumber \\
  &= &
  E_0^{\rm HF} + \sum_{n\neq 0}E_n^{\rm HF} |\Phi_n|^2  \nonumber \\
&& + \sum_{n\neq 0}
\left(\Phi_n^*\langle n | \hat{V}_{\perp} | 0\rangle
  + \Phi_n \langle 0 | \hat{V}_{\perp} | n\rangle \right)\nonumber \\
&&  + \sum_{m,n\neq 0} \Phi_n^* \Phi_m \langle n | \hat{V}_{\perp} | m\rangle\; .
\end{eqnarray}
The norm of the state $|\psi_0\rangle$ is given by
\begin{equation}
N(\psi_0) = \langle \psi_0 | \psi_0\rangle = 1 + \sum_{n\neq 0} |\Phi_n|^2 \; .
\end{equation}
Next, we optimize the variational ground-state energy 
\begin{equation}
E_0= \frac{H(\psi_0)}{N(\psi_0)}
\end{equation}
with respect to $\Phi_n^*$ to find
\begin{equation}
  \left(E_0-E_n^{\rm HF}\right)\Phi_n = \langle n |\hat{V}_{\perp}| 0\rangle
  + \sum_{m\neq 0} \langle n | \hat{V}_{\perp} | m\rangle \Phi_m \; ,
  \label{eq:Sequatoonrephrased}
\end{equation}
which is nothing but the Schr\"odinger equation expressed
in the Hartree-Fock basis.

We now assume that the last term in eq.~(\ref{eq:Sequatoonrephrased})
is small. This is justified
in weak coupling when the amplitudes $\Phi_m\propto V$ are small,
or when the density of excitations is small for all~$V$, as is the case
for spinless fermions in one dimension.
At the same level of approximation, we must replace $E_0$ by $E_0^{\rm HF}$
to find
\begin{equation}
  \left(E_0^{\rm HF}-E_n^{\rm HF}\right)\tilde{\Phi}_n
  = \langle n |\hat{V}_{\perp}| 0\rangle \; ,
\end{equation}
which gives $\tilde{\Phi}_n$ from second-order perturbation theory
with respect to the Hartree-Fock approximation,
\begin{equation}
  \tilde{\Phi}_n = \frac{\langle n | \hat{V}_{\perp} | 0\rangle}{
    E_0^{\rm HF}-E_n^{\rm HF}} \; ,
\end{equation}
so that we recover eq.~(\ref{eq:2ndorderformal})
that was the basis of our considerations. To be consistent,
we had to approximate $N(\psi_0)\approx 1$.

While $N(\psi_0)\approx 1$ is guaranteed for small interaction strengths,
this is not obvious for large interactions.
In the SD approximation, we have
\begin{equation}
N(\psi_0^{\rm SD}) = 1+ \frac{1}{2} n_{\beta}(V) \;.
\end{equation}
Now that $n_{\beta}(V) \ll 1$ for all interactions, corrections due to
the norm term are small. For the same reason, the last term
in eq.~(\ref{eq:Sequatoonrephrased}) is small because
it describes the scattering between dilute quasi-particle excitations.

In sum, a meaningful second-order perturbation theory
around the Hartree-Fock solution requires
dilute quasi-particle excitations above the Hartree-Fock ground-state.
For spinless fermions in one dimension, the condition $n_{\beta}(V) \ll 1 $
is fulfilled for all interaction strengths, and the ground-state energy
obeys an `almost-variational' property.

\subsection{Ground-state energy and order parameter}

The optimization of the ground-state energy must be done numerically.
The corresponding formulae are derived
in the supplemental material
for finite system sizes and in the thermodynamic limit.~\cite{suppmat}

\subsubsection{Hartree-Fock energy functional to second order}

For our further analysis of the equations in the
thermodynamic limit, we introduce the variable
\begin{equation}
u= \frac{n_a V}{1+ B_0V}
\end{equation}
and use $u$ instead of $n_a$ as variational parameter.
The energy functional in the thermodynamic limit can be written as
\begin{eqnarray}
  e_0^{(2)}(B_0,u,V) &= & -\frac{2}{\pi} \left(1+B_0V \right)\sqrt{1+u^2}
    E\left[\frac{1}{1+u^2}\right] \nonumber \\
&&    + V\left[\frac{1}{4}+ B_0^2+ \left(\frac{u(1+ B_0V)}{V}\right)^2\right]
    \nonumber \\
    &&   + \frac{V^2}{1+ B_0V} \bar{e}(u) \label{eq:tripleintegralofu} \; ,
\end{eqnarray}
where
\begin{eqnarray}
 \bar{e}(u)&= &
      -\frac{1}{2} \int_{-\pi/2}^{\pi/2}
  \frac{{\rm d} k_1}{2\pi}
  \int_{-\pi/2}^{k_1}   \frac{{\rm d} p_1}{2\pi}
  \int_{-\pi/2+ k_1-p_1}^{\pi/2}   \frac{{\rm d} p_2}{2\pi}\nonumber \\
  && %\hphantom{-\frac{1}{2}}
\frac{
  \left|\bar{A}(k_1,p_1+ p_2-k_1;p_1,p_2)\right|^2}{
  \bar{E}(k_1)+\bar{E}(p_1+ p_2-k_1)+\bar{E}(p_1)+\bar{E}(p_2)}
\nonumber \\[6pt]
&& -\frac{1}{2} \int_{-\pi/2}^{\pi/2}
  \frac{{\rm d} k_1}{2\pi}
  \int_{k_1}^{\pi/2}   \frac{{\rm d} p_1}{2\pi}
  \int_{\pi/2+ k_1-p_1}^{\pi/2}   \frac{{\rm d} p_2}{2\pi}\nonumber \\
  && %\hphantom{-\frac{1}{2} }
\frac{
  \left|\bar{B}(k_1,p_1+ p_2-k_1-\pi;p_1,p_2)\right|^2}{
  \bar{E}(k_1)+\bar{E}(p_1+ p_2-k_1-\pi)+\bar{E}(p_1)+\bar{E}(p_2)} 
\nonumber 
\end{eqnarray}
with
\begin{equation}
\bar{E}(k) = \sqrt{\epsilon(k)^2+ (2u)^2}
\end{equation}
and $\epsilon(k)= -2\cos(k)$ as before. Again,
$E[x]$ in eq.~(\ref{eq:tripleintegralofu})
is the complete elliptic integral of the second kind, see
eq.~(\ref{eq:ellipticEdef}).
In addition,
\begin{eqnarray}
  \left|\bar{A}(k_1,k_2;p_1,p_2)\right|^2 &=&
  \bar{Q}_1(u;k_1,k_2)\bar{Q}_1(u;p_1,p_2)\nonumber \\
  &&+
  \bar{Q}_2(u;k_1,k_2)\bar{Q}_2(u;p_1,p_2)
  \nonumber \\
  && -2\bar{Q}_3(u;k_1,k_2)\bar{Q}_3(u;p_1,p_2)
  \; ,    \label{eq:finalAsqofu}\nonumber \\
    \left|\bar{B}(k_1,k_2;p_1,p_2)\right|^2 &=&
    \bar{Q}_1(u;k_1,k_2)\bar{Q}_2(u;p_1,p_2)\nonumber \\
    && +
  \bar{Q}_2(u;k_1,k_2)\bar{Q}_1(u;p_1,p_2)
  \nonumber \\
  && + 2\bar{Q}_3(u;k_1,k_2)\bar{Q}_3(u;p_1,p_2)\nonumber \\
   \label{eq:finalBsqofu}
\end{eqnarray}
with
\begin{eqnarray}
\bar{Q}_1(u;k_1,k_2) &= & 
  2\sin^2[(k_2-k_1)/2]\nonumber \\
&&\times \left( 1 +
  \frac{\epsilon(k_1)\epsilon(k_2)}{\bar{E}(k_1)\bar{E}(k_2)}
  -\frac{(2u)^2}{\bar{E}(k_1)\bar{E}(k_2)}\right) , \nonumber \\[3pt]
\bar{Q}_2(u;k_1,k_2) &= &   2\cos^2[(k_2-k_1)/2]\nonumber \\
&&\times \left( 1 -
  \frac{\epsilon(k_1)\epsilon(k_2)}{\bar{E}(k_1)\bar{E}(k_2)}
  -\frac{(2u)^2}{\bar{E}(k_1)\bar{E}(k_2)}\right) , \nonumber \\[3pt]
%
  %
  %
% \bar{Q}_3(u;k_1,k_2)  &= &  \sin(k_2-k_1)\nonumber \\
% &&\times   \left(\frac{2u}{\bar{E}(k_2)}-\frac{2u}{\bar{E}(k_1)}\right) \; .
  \bar{Q}_3(u;k_1,k_2)  &= &  \sin(k_2-k_1)
\left(\frac{2u}{\bar{E}(k_2)}-\frac{2u}{\bar{E}(k_1)}\right) \; .
\end{eqnarray}

\subsubsection{Limiting cases}

In the absence of a charge-density wave order, $n_a=u= 0$,
the energy function reads
\begin{eqnarray}
  e_0^{(2)}(B_0,0,V)&=& -\frac{2}{\pi}\left(1+B_0 V\right) +
  \left(\frac{1}{4}+B_0^2\right)V \nonumber \\
  &&    + \left(-\frac{2}{3 \pi^3} + \frac{1}{36 \pi}\right)\frac{V^2}{1+VB_0}
  \; ,
  \label{eq:noCDWenergy}
\end{eqnarray}
see the supplemental material,~\cite{suppmat}
with the correct second-order
coefficient, see eq.~(\ref{eq:gsenergylimits}), and $B_0\approx 1/\pi$
for $V\lesssim 1$.
Since the expression~(\ref{eq:noCDWenergy}) leads to a diverging energy
for $V\gg 1$,
the CDW order must be present above some critical interaction strength.

For large interactions, the second-order correction does not change
the leading-order terms for the order parameter~$n_a$,
nor for $B_0$, see eq.~(\ref{eq:nalargeVanalyt}). However, the
Hartree-Fock energy to third order in $1/V$ is shifted towards
the exact values,
\begin{eqnarray}
  e_0^{\rm HF}(V\gg t) &\approx & -\frac{1}{V}+ 2
  \frac{1}{V^3} \; ,\nonumber \\
    e_0^{\rm HF,2nd}(V\gg t) &\approx & -\frac{1}{V}+ \left(2-\frac{1}{4}\right)
  \frac{1}{V^3} \; , \nonumber\\
e_0^{\rm exact}(V\gg t) &\approx & -\frac{1}{V}+ \frac{1}{V^3} \; ,
\end{eqnarray}
see the supplemental material.~\cite{suppmat}

\subsection{Occupation numbers}

As shown in the supplemental material,~\cite{suppmat}
the occupancies in second-order perturbation theory are given by
\begin{equation}
  n_{s,\beta}= n_{s,\beta}^{(1)}+ n_{-s,\beta}^{(1)}  + n_{s,\beta}^{(4)}
  +   n_{-s,\beta}^{(4)}
\end{equation}
with
\begin{eqnarray}
n_{s,\beta}^{(1)}&= & \frac{V^2}{2(1+ VB_0)^2}
\int_s^{\pi/2} \frac{{\rm d}p_1}{2\pi}
  \int_{-\pi/2}^{\pi/2-p_1+ s} \frac{{\rm d}p_2}{2\pi}\nonumber\\
&&\frac{\left|\bar{A}(s,p_1+p_2-s;p_1,p_2)\right|^2}{
    [\bar{E}(s)+\bar{E}(p_1+p_2-s)+\bar{E}(p_1)+\bar{E}(p_2)]^2}
  \nonumber \\
\end{eqnarray}
and
\begin{eqnarray}
  n_{s,\beta}^{(4)}&= & \frac{V^2}{2(1+ VB_0)^2}
  \int_s^{\pi/2} \frac{{\rm d}p_1}{2\pi}
  \int_{\pi/2-p_1+ s}^{\pi/2} \frac{{\rm d}p_2}{2\pi}\nonumber \\
 && \frac{\left|\bar{B}(s,p_1+p_2-s-\pi;p_1,p_2)\right|^2}{
    [\bar{E}(s)+\bar{E}(p_1+p_2-s-\pi)+\bar{E}(p_1)+\bar{E}(p_2)]^2}
  \nonumber\\
\end{eqnarray}
in the thermodynamic limit. Apparently,
the momentum distribution is inversion symmetric,
$n_{s,\beta}= n_{-s,\beta}$.

For small interactions,
the occupations of the upper Hartree-Fock bands are small,
of the order~$V^2$. For large interactions, they are equally small,
of the order $1/V^2$, because
Hartree-Fock theory for the ground state becomes exact to leading
order in $1/V$. The maximum number of excited quasi-particles
can be expected to occur around the metal-insulator transition.

\subsection{Metal-insulator transition in second-order perturbation theory}
\label{sec:MITin2ndorder}

Here, we shall show that the order parameter is finite for
$0<V<V_{\rm c,1}^{(2)}\approx 0.21$, where it is exponentially small.
It exactly vanishes in the region $V_{\rm c,1}^{(2)}<V<V_{\rm c,2}^{(2)}\approx 1.51$
where it jumps to a finite value with discontinuities in all observables,
including the ground-state energy.

\subsubsection{Energy functional for small order parameter}

For small~$u$, we expand the energy functional,
\begin{eqnarray}
  e_0^{(2)}(B_0,u,V) &\approx & -\frac{2}{\pi} \left(1+B_0V \right)\nonumber \\
&&      + V\left[\frac{1}{4}+ B_0^2+ \left(\frac{u(1+ B_0V)}{V}\right)^2\right]
      \nonumber \\
      && + \frac{(1+B_0V)u^2 }{2\pi} \left(2\ln(u) -1- 4\ln(2)\right)
      \nonumber \\
&&      + \frac{V^2}{1+ B_0 V}\bar{e}(u)
      \label{eq:minimizethisforsmallu}
\end{eqnarray}
with
\begin{equation}
  \bar{e}(u\ll 1)=e_0^{(2) }
  + u^2\left(\alpha [\ln(u)]^2+ \beta \ln(u) +  \gamma\right) \; ,
\end{equation}
where $e_0^{(2)}= -2/(3\pi^3)+ 1/(36\pi)$
  from eq.~(\ref{eq:gsenergylimits}).
Corrections are of the order $u^4[\ln(u)]^2$.

The coefficients are determined from a numerical fit for
$(\bar{e}(u)-e_0^{(2)})/u^2$ in the
interval $I= [0.01,0.1]$ where the energy can be calculated with
a relative accuracy of $10^{-10}$ using {\sc Mathematica}.~\cite{Mathematica12}
We find
\begin{equation}
\alpha = 0.1573 \;, \quad \beta = 0.3726\; , \quad \gamma = 0.4121\; .
\end{equation}
Note that the three-parameter fit is fairly sensitive.

\subsubsection{Nearest-neighbor transfer amplitude}

The minimization of the energy expression
in eq.~(\ref{eq:minimizethisforsmallu}) at $u= 0$
with respect to $B_0$ leads to
the third-order equation for $B_0\equiv B_0(0,V)$,
\begin{equation}
  -\frac{2 V}{\pi} (1 + B_0 V)^2
  + 2 B_0 V + 4 B_0^2 V^2 - e_0^{(2)}V^3 +  2 B_0^3 V^3
  = 0 \; .
\end{equation}
$B_0(V)$ decreases from its value $B_0(0,0)= 1/\pi\approx 0.318$ to
$B_0(0,V= 1.6)\approx 0.311$, i.e., it remains essentially constant
up to moderate interactions.

When the order parameter for the charge-density wave is finite, $u>0$,
and $V<V_{\rm c,1}^{(2)}$, the corrections to the value at $u= 0$ are exponentially
small as in Hartree-Fock theory, see Sect.~\ref{sec:HFapproximation},
and we may use $B_0(u,V)\approx B_0(0,V)\equiv B_0$ in the following.

\subsubsection{Order parameter}

When $u\neq 0$, the minimization equation for $u$ reduces to
a quadratic equation in $y= -\ln(u)>0$,
\begin{eqnarray}
2(1+ B_0 V)^3 + 
V^3 (\beta + 2 \gamma - 2 (\alpha + \beta) y + 2 \alpha y^2) =&&
\nonumber \\
    \frac{ 2 V}{\pi} (1 + B_0 V)^2 (y + 2\ln(2))  \; .&&\nonumber \\
    \label{eq:parabola}
\end{eqnarray}
The discriminant of the equation is {\sl negative\/} in the range
$0.231\approx V_{\rm c,1}^{(2)}<V<V_{\rm c,2}^{(2)}\approx 1.54$.
Therefore, there is no charge-density wave order between $V_{\rm c,1}^{(2)}$ and
$V_{\rm c,2}^{(2)}$.

The region $0<V<V_{\rm c,1}^{(2)}$ cannot be studied numerically because the order
parameter is exponentially small.
Indeed, for $V\to 0$ we have
\begin{equation}
  n_a(V\ll 1) \approx \left(1+ \frac{V}{\pi}\right)\frac{4}{V}
  \exp\left(-\frac{\pi}{V}-(1+ \alpha \pi^3)\right)
\end{equation}
using $B_0\approx 1/\pi$. Corrections in the exponent are of the order of $6V$.
In comparison with the Hartree-Fock result to leading order,
see eq.~(\ref{eq:nasmallVanalyt}), the order parameter is smaller by
the factor $\exp(-\alpha\pi^3)\approx 0.008$
so that the already exponentially small
Hartree-Fock order parameter is reduced in second-order perturbation theory
by additional two orders of magnitude. Numerically,
$n_a(V<V_{\rm c,1}^{(2)})< 10^{-8}$.

While $V_{\rm c,1}^{(2)}$ cannot be identified numerically, we find that
\begin{equation}
  V_{\rm c,2}^{(2)}\approx 1.515
\end{equation}
from the numerical minimization of the full energy functional.
This value agrees very well with the value where the discriminant of the
quadratic equation~(\ref{eq:parabola}) becomes positive.
At $V_{\rm c,2}^{(2)}$, the order parameter jumps to a finite value,
$n_a(V= V_{\rm c,2}^{(2)})\approx 0.085$, in good agreement with the result
from the calculation for small~$u$,
$n_a^{{\rm small}\, u}(V= V_{\rm c,2}^{(2)})\approx 0.07$.

%The findings indicate that the order parameter is actually zero
%for $V<V_{\rm c}$ with $V_{\rm c}$ of the order unity.
%However, this exact result cannot be obtained from
%perturbation theory to finite order.

\section{Comparison}
\label{sec:comparison}

We start this section with some technical information about the
DMRG implementation.
Second, we show the ground-state energy and
the single-particle density matrix
for nearest neighbors that do not signal the charge-density wave
transition.
It requires detailed information from Bethe Ansatz and field theory
on the finite-size corrections to the ground-state energy
to estimate the critical interaction from the ground-state energy.

The metal-to-insulator transition
is seen in the single-particle gap and in the
CDW order parameter that we discuss next.
Since both quantities display a Kosterlitz-Thouless behavior
with an essential singularity at the critical interaction,
it is not possible to extract 
the critical interaction from finite-size extrapolations reliably
for any choice of boundary conditions.
The DMRG gap data for periodic boundary conditions
and odd particle numbers permit to reproduce the
Bethe Ansatz results for the leading-order finite-size corrections in the
metallic regime from which one can estimate the critical interaction
strength.

The correlation energy displays a maximum as a function
of the interaction strength. However, its position is not identical to
the critical interaction. The momentum
and quasi-particle distributions and, finally, the
density-density correlation function provide the necessary information 
to extrapolate reliably 
the critical interaction strength from the Luttinger parameter
and from the quasi-particle density.

%We defer a detailed discussion of the gap and its dependence
%on system size and boundary conditions to appendix~\ref{app:gapbc}.

\subsection{DMRG technicalities}

Before we start the comparison of analytic and numerical results, we
compile some technical remarks on the implementation
of our DMRG code. Moreover, we introduce the notion of natural orbitals
and discuss their relation to the Hartree-Fock levels.

\subsubsection{Coding}

We apply the real-space DMRG
algorithm~\cite{White-1992b,White-1993,Schollwock-2005}
to the Hamiltonian~(\ref{eq:Hamiltoniandef1}).
Since the model has a gapless energy spectrum up to the critical Coulomb
coupling $V_{\rm c}=2$ in the thermodynamic limit,
its numerical analysis requires relatively high numerical accuracy
for a reliable finite-size scaling.
Therefore, we keep the truncation error
below $\delta\varepsilon_{\rm Tr}=10^{-8}$
for the whole range $0\le V \le 8$, and use a minimum bond
dimension $D=1024$.~\cite{legeza2003,Legeza2004}
For $V>2.5$, the latter condition results
in a much lower truncation error, i.e.,
we find $\delta\varepsilon_{\rm Tr}=10^{-14}\ldots 10^{-10}$.

We run between seven to eleven sweeps to acquire symmetric data
sets in position space when expectation
values of zero-point and one-point correlation functions are calculated.
We use Davidson and/or Lanczos methods for the diagonalization
of the effective Hamiltonian and enforce a very tight error threshold, i.e.,
the residual error is set to $10^{-10}$.

We apply periodic boundary conditions for system sizes corresponding to
an open-shell ground-state configuration. To lift the ground-state degeneracy,
we employ a very small pinning field in the range of $\Delta_{\rm pin}=10^{-4}$.
In order to check boundary effects, we also perform calculations
for closed-shell configurations,
and occasionally for open boundary conditions.
The finite-size scaling analysis is carried out for systems with
up to $L=514$ sites.

\subsubsection{Single-particle density matrix and natural orbitals}
\label{subsubsec:definenkalphankbeta}

DMRG provides the single-particle density matrix in position space,
\begin{equation}
  P_{l,m}=\langle \hat{c}_l^+\hat{c}_{m}^{\vphantom{+}} \rangle \; .
  \label{eq:SPDMorOBDM}
\end{equation}
Upon Fourier transformation, we have
\begin{equation}
  \widetilde{P}_{k,p}=\langle \hat{a}_k^+\hat{a}_{p}^{\vphantom{+}} \rangle\; .
  \label{eq:SPDMorOBDMkspace}
\end{equation}
In the presence of a charge-density wave,
the unit cell doubles, and we thus find for $|k|, |p| \leq \pi$
\begin{equation}
  \widetilde{P}_{k,p}=
  \langle \hat{a}_k^+\hat{a}_{k}^{\vphantom{+}} \rangle
  \delta_{p,k} + 
  \langle \hat{a}_k^+\hat{a}_{k\pm \pi}^{\vphantom{+}} \rangle
  \delta_{p,k\pm \pi} 
  \; .
  \label{eq:SPDMorOBDMkspaceagain}
\end{equation}
Numerically, deviations are of the order $10^{-4}$.

To find the `natural orbitals', we have to diagonalize
the $2\times 2$-matrices
\begin{equation}
  \underline{\underline{M}}_k =
\left(
  \begin{array}{@{}cc@{}}
    1/2 & 0\\
    0 & 1/2
    \end{array}
  \right)+
  \left(
  \begin{array}{@{}cc@{}}
    n_k-1/2 & d_k\\
    d_k & -(n_k-1/2)
    \end{array}
  \right)
\end{equation}
in the reduced Brillouin zone, $|k|\leq \pi/2$,
where we used particle-hole symmetry,
$n_{k\pm \pi}=1-n_k$, and abbreviated
\begin{equation}
d_k= \langle \hat{a}_k^+\hat{a}_{k\pm \pi}^{\vphantom{+}} \rangle=d_k^* \; .
\label{eq:defdk}
\end{equation}
Note that the order parameter is the sum over the non-diagonal
matrix elements,
\begin{equation}
n_a= \frac{1}{L} \sum_{k\in {\rm RBZ}} (d_k+d_k^*) \; .
\end{equation}
The same type of diagonalization is carried out in Hartree-Fock theory,
see Sect.~\ref{subsec:HFDiagonlaization},
where $n_k-1/2$ is replaced by $\widetilde{\epsilon}(k)$ and $d_k$ by
$(-2Vn_a)$, see eq.~(\ref{eq:HFbeforediagonalization}).
Due to this similarity, we call the natural orbitals
as the states in the upper and lower Hartree-Fock band.

The eigenvalues of the matrix $\underline{\underline{M}}_k$ are the
level occupancies $n_{k,\alpha/\beta}$.
They obey $n_{k,\alpha}=1-n_{k,\beta}$ due to particle-hole symmetry,
see eq.~(\ref{eq:HFqpoccoanciesrelation}).
Therefore, we shall only address the occupation density
$n_{k,\beta}$ of the upper Hartree-Fock band.

\subsection{Ground-state energy at half band-filling and
  nearest-neighbor single-particle density matrix}

\subsubsection{Ground-state energy}

In table~\ref{tab:DMRGexactenergy} we give the DMRG ground-state energy
per lattice site for
systems with $L= 8,16,32,64,128,256,512$ sites at half band-filling,
and compare it to the exact Bethe-Ansatz
results~\cite{PhysRev.150.321,PhysRev.150.327}
at $V= 0,0.8,1.4,2,4$. Apparently, the convergence to the thermodynamic limit
is very fast,
and the DMRG data are accurate to five (four) digits for $V\leq 2$
($V \leq 4$).

In table~\ref{tab:L64andTDL-ge-table}
we compare the ground-state energy per lattice site
from (second-order) Hartree-Fock approximation with those  from DMRG
for $L= 64$ and to those from Bethe Ansatz 
in the thermodynamic limit. It is seen that the second-order Hartree-Fock
theory provides very accurate
results for $V\lesssim 1.4$, with errors of about one percent.
Even for large interactions, $V= 4$, the errors are below five percent.
Although unwarranted by a variational principle, the second-order
Hartree-Fock energies are upper bounds to the exact energies. 

\begin{table}[t]
\begin{center}
\begin{tabular}{|c|c|c|c|c|c|}\hline
$L\backslash V$   & $0$        & $0.8$      & $1.4$ & $ 2$ & $ 4 $\\ \hline
$8$  & $-0.60357$ & $-0.49729$ & $-0.42832$ & $-0.36857$  & $-0.23184$ \\ 
$16$ & $-0.62842$ & $-0.51846$ & $-0.44599$ & $ -0.38208$& $-0.23435$ \\ 
$32$ & $-0.63458$ & $-0.52371$ & $-0.45036$ & $-0.38529$ & $-0.23448$ \\ 
$64$ & $-0.63611$ & $-0.52502$ & $-0.45145$ & $-0.38606$ & $-0.23448$ \\ 
$128$& $-0.63649$ & $-0.52535$ & $-0.45172$ & $-0.38624$ & $-0.23448$ \\ 
$256$& $-0.63659$ & $-0.52543$ & $-0.45178$ & $-0.38628$ & $-0.23448$ \\ 
  $512$& $-0.63661$ & $-0.52545$ & $-0.45180$ & $-0.38629$ &  \\ \hline
&&&&&\\[-7pt]
  BA
  & $-0.63662$ & $-0.52545$ & $-0.4518$0 & $-0.38629$ & $-0.23444$ \\ \hline
\end{tabular}
\caption{Ground-state energy per lattice site for spinless fermions
  for systems with $L$ sites and $V= 0,0.8,1.4,2,4$ from DMRG
  with a small symmetry-breaking pinning field.
  The last line contains the exact ground-state energy obtained
  in the thermodynamic limit from Bethe Ansatz.\label{tab:DMRGexactenergy}}
\end{center}
\end{table}

\begin{table}[hb]
  \begin{tabular}[t]{@{}ll@{}}
    (a) &
%  \end{flushleft}
%  \begin{center}
\begin{tabular}[t]{|c|c|c|c|c|}\hline
Method$\backslash V$   & $0.8$        & $1.4$      & $2$ & $ 4 $ \\ \hline
HF & $-0.51774$ & $-0.43415$ & $-0.36458$ & $ -0.22469$ \\
2nd HF & $-0.52368$ & $-0.44502$ & $-0.37339$ & $-0.22729$ \\
DMRG & $-0.52502$ & $-0.45145$ & $-0.38606$ & $-0.23448$ \\ \hline
\end{tabular}\\
\mbox{}& \\%[12pt]
%  \end{center}
%  \begin{flushleft}
    (b) &
%  \end{flushleft}
%  \begin{center}
\begin{tabular}[t]{|c|c|c|c|c|}\hline
Method$\backslash V$   & $0.8$        & $1.4$      & $2$ & $ 4 $ \\ \hline
HF & $-0.51784$ & $-0.43415$ & $-0.36458$ & $ -0.22469$ \\
2nd HF & $-0.52414$ & $-0.44568$ & $-0.37339$ & $-0.22729$ \\
BA & $-0.52545$ & $-0.45180$ & $-0.38629$ & $-0.23444$ \\ \hline
\end{tabular}
\end{tabular}
\caption{(a) Ground-state energy per lattice site for spinless fermions
  for $L=64$ sites from
  Hartree-Fock (HF) and second-order Hartree-Fock (HF~2nd)
  approximation and DMRG for $V= 0.8,1.4,2,4$. %\newline
  (b) As in (a) but for the thermodynamic limit; exact results are from
  Bethe Ansatz (BA).\label{tab:L64andTDL-ge-table}}
%\end{center}
%\end{flushleft}
\end{table}

\begin{figure}[t]
\begin{center}
\includegraphics[width=8cm]{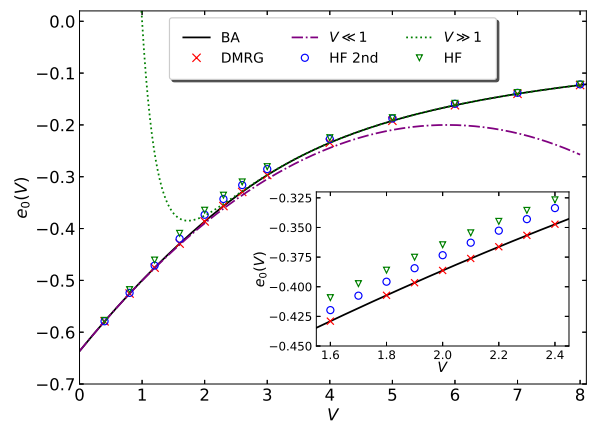}
\end{center}
\caption{Ground-state energy density for spinless fermions
  at half band-filling
  as a function of the nearest-neighbor interaction~$V$ from Bethe Ansatz (BA),
  DMRG for $L= 512$ sites,
  and (second-order) Hartree-Fock  (HF, HF~2nd).
  Dashed and dotted lines correspond to the small-$V$ and large-$V$ expansions in
  eq.~(\protect\ref{eq:gsenergylimits}).\label{fig:gsenergyTDL}}
\end{figure}

Figure~\ref{fig:gsenergyTDL} shows the ground-state energy per lattice site
in the thermodynamic limit as a function of the interaction strength.
On the scale of the figure, the DMRG data for $L= 512$ sites
lie on top of the exact results.
The Hartree-Fock approximation
becomes exact for small and large interactions, and provides
a very good estimate for the ground-state energy even for intermediate
interactions, see inset. The inclusion of the second-order corrections
improves the energy estimate systematically for all interaction strengths.

\begin{figure}[hb]
\begin{center}
\includegraphics[width=8cm]{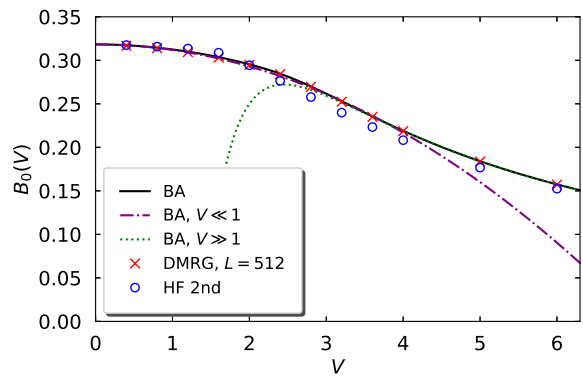}
\end{center}
\caption{Single-particle density matrix $B_0(V)$
  between near\-est neighbors 
  as a function of the nearest-neighbor interaction~$V$ from Bethe Ansatz (BA),
  including the small and large coupling asymptotes,
  from DMRG for $L=512$ sites, and from second-order
  Hartree-Fock (HF 2nd).
Dashed and dotted lines correspond to the small-$V$ and large-$V$ expansions in
  eq.~(\protect\ref{eq:B0limits}).\label{fig:B0TDL}}
\end{figure}

\subsubsection{Nearest-neighbor single-particle density matrix}

In Fig.~\ref{fig:B0TDL} we show the nearest-neighbor
single-particle density matrix~$B_0(V)$ from Bethe Ansatz, see
eq.~(\ref{eq:exactB0general}), and its limiting behavior for small and large
interactions, see eq.~(\ref{eq:B0limits}), together with the results
from second-order Hartree-Fock theory and DMRG data for $L=512$ sites.
As for the ground-state energy, the DMRG data lie on top the Bethe-Ansatz result
on the scale of the figure. Second-order Hartree-Fock theory
is exact for small and large interactions, and provides a good description for
all interaction strengths. It is a mere coincidence that
second-order Hartree-Fock reproduces the exact value for $B_0$
right at the critical interaction strength, $V_{\rm c}=2$.

Neither the kinetic energy nor the ground-state energy are critical quantities,
i.e., their values in the thermodynamic are readily obtained
from DMRG with a high accuracy, and also second-order Hartree-Fock theory
provides a fair estimate for these quantities.

\subsubsection{Finite-size scaling of the ground-state energy}

For the XXZ model, the scaling of the ground-state energy
density as a function of system size~$L$ is
known,~\cite{WoynarovichEckle,Affleck_1989,Rutkevich2021}
\begin{equation}
  \frac{E_0(L,V)}{L} = e_0(V) +\frac{1}{L^2}
  \left(c(V)+\frac{d(V)}{\ln(L)^3} +\ldots\right)\; .
  \label{eq:e0finiteL}
\end{equation}
It is important to note that the approach to the thermodynamic limit depends on
the choice of the boundary conditions. Open boundary conditions introduce
an additional and sizable first-order term that dominates the terms in $1/L^2$
for small system sizes. Therefore, to make use of eq.~(\ref{eq:e0finiteL}), it
is mandatory to employ periodic boundary conditions.

For periodic boundary conditions,
the ambiguity remains whether $L/2$ is
even or odd. To see this, we address the case
of non-interacting spinless fermions.
For even~$L/2$, the ground state is doubly degenerate (open shell)
while it is unique for odd $L/2$ (closed shell).
The corresponding expressions for the ground-state energy for large~$L$ are
\begin{eqnarray}
  \frac{E_0^{\rm os}(L,V=0)}{L} &=& -\frac{2}{L} \sum_{m=-L/4}^{L/4-1}\cos(2\pi m/L)
  \nonumber \\
&=& 
  -\frac{2}{\pi} +\frac{2\pi}{3}\frac{1}{L^2} +
  {\cal O}(1/L^4) \; , \nonumber \\
  \frac{E_0^{\rm cs}(L,V=0)}{L} &=&
-\frac{2}{L} \sum_{m=-(L-2)/4}^{(L-2)/4}\cos(2\pi m/L) \nonumber \\
&=&  -\frac{2}{\pi} -\frac{\pi}{3}\frac{1}{L^2} +
{\cal O}(1/L^4) \; ,
\label{eq:E0openshellclosedshell}
\end{eqnarray}
where we used the Euler-MacLaurin sum formula to expand the finite sums
in powers of inverse system size.
Apparently, $c^{\rm os}(V= 0)=2\pi/3$ and $c^{\rm cs}(V= 0)=-\pi/3$ disagree.

On the other hand, the leading-order correction for the XXZ model
can be calculated
in the metallic regime
from Bethe Ansatz and conformal field
theory~\cite{WoynarovichEckle,Affleck_1989}
\begin{eqnarray}
  c(V)&=&-c\, \frac{\pi}{6}u(V) \;, \nonumber \\
  u(V) &=& 2\sqrt{1-(V/2)^2} \left[\frac{\pi}{2\arccos(V/2)}\right]
  \label{eq:cofVfieldtheory}
\end{eqnarray}
with the central charge $c=1$ for spinless fermions, and
$u(V)$ as the velocity of the elementary
excitations.~\cite{PhysRevA.8.2526,BABELON198313}
For non-interacting fermions,
\begin{equation}
  u(V=0)=\left.\frac{{\rm d}\epsilon(k)}{{\rm d} k}\right|_{k= \pi/2}=2
  \end{equation}
is the particle velocity at the Fermi point $k=\pi/2$.
Therefore, we find the slope $c(V=0)=-\pi/3\approx -1.047$.
Eqs.~(\ref{eq:E0openshellclosedshell}) and~(\ref{eq:cofVfieldtheory})
thus show that a comparison of field-theory/Bethe-Ansatz
predictions for finite-size corrections is only meaningful for DMRG data
obtained for odd $L/2$ (closed shell).

\begin{figure}[ht]
\begin{center}
  \includegraphics[width=8cm]{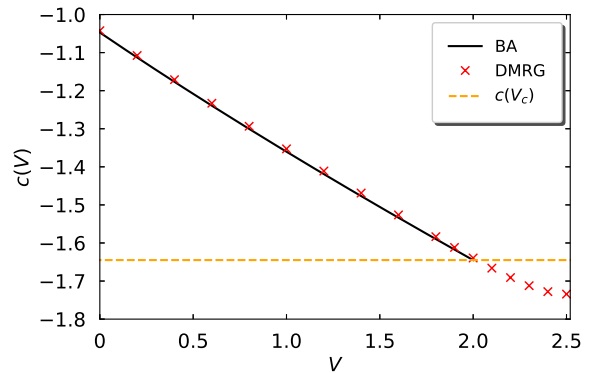}
  %\vspace*{6cm}
\end{center}
\caption{Second-order coefficient $c(V)$ in eq.~(\protect\ref{eq:e0finiteL})
  from the extrapolation of DMRG data ($L/2$ odd, closed shell)
  in comparison with the Bethe-Ansatz
  result~(\protect\ref{eq:cofVfieldtheory}). The horizontal line indicates the
  critical value $c(V_{\rm c})=-\pi^2/6$.\label{fig:centralcharge}}
\end{figure}

In Fig.~\ref{fig:centralcharge} we show the quadratic coefficient $c(V)$
in eq.~(\ref{eq:e0finiteL}) from
the extrapolation of the DMRG data for the ground-state energy $E_0(L,V)/L$
for odd $L/2$, $L=10,30,66,130,258,514$, in comparison with the analytic
result~(\ref{eq:cofVfieldtheory}).
The agreement is very good, and permits to locate
the transition from the criterion
$c(V=V_{\rm c})=-\pi^2/6\approx -1.64493$. A comparison with the
extrapolated numerical data gives $V_{\rm c}^{\rm e}=2.02$,
within about one percent of the exact value.

Note that this very good result is based on several facts.
First, the logarithmic
corrections in eq.~(\ref{eq:e0finiteL}) are known analytically.
This decisively stabilizes the extrapolation of $c(V)$.
Second, the value for the maximal
velocity $u_{\rm c}=\pi$ is used as input. Therefore, a lot of
intelligence from conformal field theory and from Bethe Ansatz
enters the analysis.
Thus, in less fortunate circumstances, the scaling of the ground-state
energy in $1/L$ cannot be used to locate the quantum phase transition.

\subsection{Single-particle gap}
\label{subsec:Delta1}

\subsubsection{Open-shell systems with periodic boundary conditions}

\begin{figure}[t]
\begin{center}
\includegraphics[width=8cm]{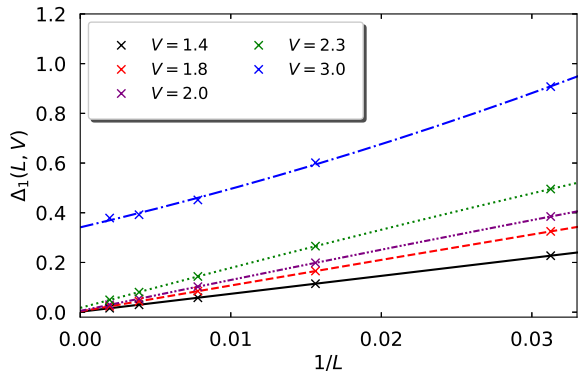}
%  \vspace*{6cm}
\end{center}
\caption{Single-particle gap $\Delta_1(L,V)$
  as a function of $1/L$ for system sizes $L= 32,64,128,256,512$ from DMRG
  for $V=1.4,1.8,2.0,2.3,3$.
Lines are second-order polynomial fits,
see eq.~(\ref{eq:2ndorderpolynomialgapfit}).
\label{fig:gapDMRGfinitesize}}
\end{figure}

In Fig.~\ref{fig:gapDMRGfinitesize}
we show the single-particle gap $\Delta_1(L,V)$
as a function of the nearest-neighbor interaction~$V$
for system sizes $L=32,64,128,256,512$.
Due to finite-size effects, the gap is always finite, of the order $1/L$,
even in the metallic region, $0<V<V_{\rm c}= 2$, and an extrapolation
to the thermodynamic limit is mandatory to
determine the gap in the thermodynamic limit.

\begin{figure}[b]
\begin{center}
\includegraphics[width=8cm]{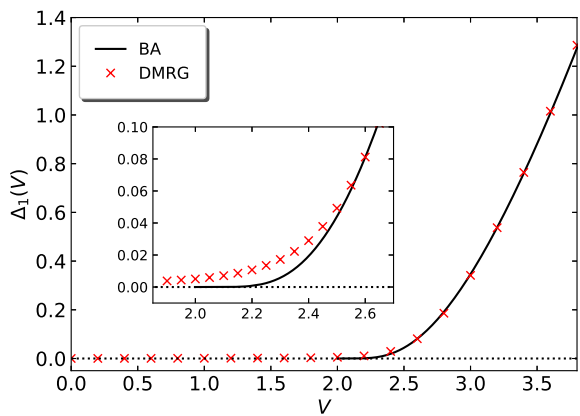}
%  \vspace*{6cm}
\end{center}
\caption{Single-particle gap $\Delta_1(V)$ in the thermodynamic limit
  from the polynomial extrapolation of the
  gap data from DMRG (crosses), in comparison
  with the exact Bethe Ansatz result (line).\label{fig:deltaextrapolated}}
\end{figure}

As standard extrapolation scheme, we apply a polynomial fit,
\begin{equation}
  \Delta_1(L,V) = \Delta_1(V) + \frac{a_1}{L} + \frac{a_2}{L^2}
  \label{eq:2ndorderpolynomialgapfit}
\end{equation}
to the DMRG data for even $L/2$
with $\Delta_1(V)$, $a_1$, and $a_2$ as fit parameters.
This fit appears to be somewhat naive in view of the fact that
the next-to-leading order corrections in the Bethe Ansatz
solution of the XXZ model are not necessarily of order $1/L^2$
but can obey power laws $L^{-\gamma}$ with $\gamma<2$,
or be of the order $1/(L\ln(L))$ at criticality.~\cite{WoynarovichEckle}
However, the simple polynomial fit is the least biased.
We shall discuss other extrapolation schemes 
for DMRG data for open boundary conditions below.

System sizes with an even particle number $N= L/2$ lead to an open-shell
ground state at $V= 0$, i.e., it is doubly degenerate. Therefore,
the gap is exactly zero at $V= 0$ in the absence of a symmetry-breaking term.
In this way, systems with even $L/2$ minimize finite-size effects
for small couplings.

In Fig.~\ref{fig:deltaextrapolated} we show the extrapolated
single-particle gap
as a function of~$V$ from the polynomial fit together with
the exact result from Bethe Ansatz,
see eq.~(\ref{eq:Delta1}).
The polynomial fit leads to a (very small) finite gap for all $V>0$,
and the sharp transition
in the exact solution at $V_{\rm c}= 2$ is smeared out,
as seen from the inset in Fig.~\ref{fig:deltaextrapolated}, so that
it is not possible to determine $V_{\rm c}$ with high accuracy
from the extrapolated gaps.
The standard polynomial extrapolation scheme does not permit
to locate transitions at finite interaction strengths.
This was shown recently for the Mott-Hubbard transition in the
$1/r$-Hubbard model.~\cite{PhysRevB.104.245118}

\subsubsection{Closed-shell systems with periodic boundary conditions}

The Bethe Ansatz solution for the XXZ model
permits to extract the finite-size corrections
to the single-particle gap.~\cite{Hamer_1985,WoynarovichEckle}
As seen in Sect.~\ref{subsc:connnectXXZandtVmodel},
these Bethe Ansatz results can be applied to the model of spinless fermions
only for odd particle numbers. Consequently, we have to study
a closed-shell ground state at half band-filling with odd particle number
$N=L/2$. Now that the excited state must also have an odd particle number,
we must numerically study the ground state with two additional particles,
$N=L/2+2$.

In the XXZ model, the two spin-1 excitations are very far
from each other for large system sizes and we argue that
the two-particle gap
\begin{equation}
  \Delta_2^{\rm XXZ}(L,V) = 2\left(E_0^{\rm XXZ}(S=2,L,V)-E_0(S=0,L,V)\right)
\end{equation}
is twice as large as the single-particle gap in the thermodynamic limit,
\begin{equation}
\Delta_1^{\rm XXZ}(L,V)=\frac{\Delta_2^{\rm XXZ}(L,V)}{2} +{\cal O}(1/L^{\gamma}) \; ,
\end{equation}
where corrections due the interaction of the excitations are of
order $1/L^{\gamma}$ with $\gamma>1$.
If this is the case, we can determine the $1/L$-correction
to the single-particle gap from half of the two-particle gap.
To this end, we extrapolate the DMRG data for spinless fermions
\begin{eqnarray}
  \frac{\Delta_2(L,V)}{2} &=& 
  E_0(N=L/2+2,L,V)\nonumber \\
  && \hphantom{\frac{1}{2}\Bigl(}
  -E_0(N=L/2,L,V)  -2V
  \label{eq:haldtwoparticlegap}
\end{eqnarray}
with a second-order polynomial in $1/L$,
\begin{equation}
  \frac{\Delta_2(L,V)}{2} \approx \frac{\Delta_2(V)}{2} +
  \frac{s_1(V)}{L} +\frac{s_2(V)}{L^2} \; ,
  \label{eq:defslopes1ands2}
\end{equation}
and compare $s_1(V)$ with the Bethe Ansatz
result~\cite{Hamer_1985,WoynarovichEckle}
\begin{eqnarray}
  s_1^{\rm BA}(V)&=&
  4\pi \left(1-\frac{\arccos(V/2)}{\pi}\right)
  \label{eq:slopeBAdelta2}\\
&&\times  \left(2\sqrt{1-(V/2)^2}
  \left[\frac{\pi}{2\arccos(V/2)}\right]\right) \;.\nonumber 
\end{eqnarray}
Note that we work with the gap whereas
the Bethe Ansatz formulae are derived for $\mu_1^{+,{\rm XXZ}}=\Delta_1/2$,
and we adjusted the energy scale.

\begin{figure}[t]
\begin{center}
\includegraphics[width=8cm]{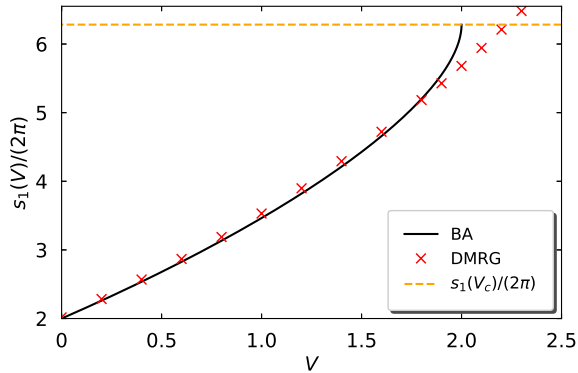}
%  \vspace*{6cm}
\end{center}
\caption{Slope $s_1(V)/(2\pi)$ in $1/L$
  of half the two-particle gap $\Delta_2(V)/2$,
  see eqs.~(\protect\ref{eq:haldtwoparticlegap})
  and~(\protect\ref{eq:defslopes1ands2}) from the extrapolation
  of the DMRG data (dots), in comparison
  with the Bethe Ansatz result (line)
  from eq.~(\ref{eq:slopeBAdelta2}). The horizontal line indicates the
  critical value $s_1(V_{\rm c})=(2\pi)^2$.\label{fig:slopedelta2half}}
\end{figure}

In Fig.~\ref{fig:slopedelta2half}
we compare the results for the slope $s_1(V)$ in eq.~(\ref{eq:defslopes1ands2})
from the polynomial fit of the DMRG data for $\Delta_2(L,V)/2$
and from the Bethe Ansatz expression~(\ref{eq:slopeBAdelta2}).
The agreement is very good for small interactions but it
deteriorates close to the transition. The criterion $s_1(V=V_{\rm c})=(2\pi)^2$,
corresponding to $u(V_{\rm c})=\pi$ in the ground-state energy,
leads to the estimate $V_{\rm c}^{\rm s}\approx 2.3$ from the
extrapolated data for the slope $s_1(V)$.
The result deviates from the
exact result by some 15~percent.
Therefore, the slope estimate is not very accurate, apart from the fact that
additional information from the exact result is necessary to determine
the value $s_1(V_{\rm c})$ at the transition.

\subsubsection{Open boundary conditions}

For open boundary conditions, we must use the par\-ti\-cle-hole symmetric
form of the interaction,
\begin{equation}
\hat{V}_{\rm phs}= \sum_{l= 1}^{L-1} ( \hat{n}_l-1/2)(\hat{n}_{l+ 1}-1/2)\; .
\label{eq:Vphsymm}
\end{equation}
If we used the interaction in eq.~(\ref{eq:definteractionnotphsymmetric})
adopted to a chain,
excited states at the boundaries would interfere so that the bulk gap
cannot be calculated from the ground state energies at half band-filling
and with plus/minus one particle. This is most easily seen
in the atomic limit, and will not be discussed any further.

For the particle-hole symmetric Hamiltonian~(\ref{eq:Hamiltoniandef1})
on a chain,
analytic finite-size corrections
to the single-particle gap are not available.
Therefore, we employ three different extrapolation schemes:
polynomial, see eq.~(\ref{eq:2ndorderpolynomialgapfit}),
logarithmic,
\begin{equation}
  \Delta_1^{\rm ln}(L,V) = \Delta_1(V) + \frac{b_1}{L}\left(1 +
  \frac{b_2}{\ln(L)} \right) \; ,
  \label{eq:logggapfit}
\end{equation}
and Mishra, Carrasquilla, and Rigol~\cite{PhysRevB.84.115135}
\begin{equation}
  \Delta_1^{\rm MCR}(L,V) = \frac{\Delta_1(V) + c_1/L}{1 +
1/(2\ln(L)+c_2)} \; ,
  \label{eq:Mishragapfit}
\end{equation}
where $b_{1,2}$ and $c_{1,2}$ are fit parameters.

\begin{figure}[t]
\begin{center}
\includegraphics[width=8cm]{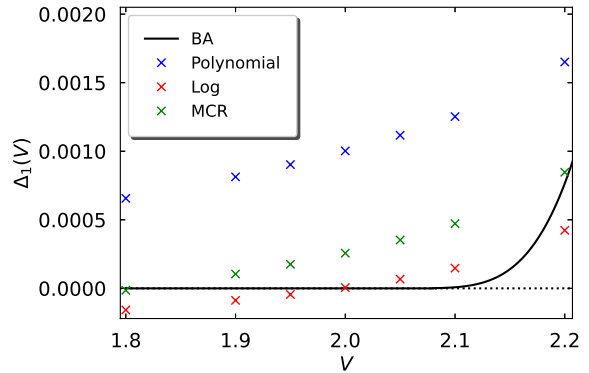}
%  \vspace*{6cm}
\end{center}
\caption{Exact single-particle gap $\Delta_1(V)$ in the thermodynamic limit
  in the range $1.8\leq V\leq 2.2$, in comparison
  with the result of three extrapolations of the DMRG data for
  open boundary conditions for $L=64,128,256,512$:
  polynomial (blue), eq.~(\protect\ref{eq:2ndorderpolynomialgapfit}),
  logarithmic (red), eq.~(\protect\ref{eq:logggapfit}),
  and Mishra et al.\ (green),
  eq.~(\protect\ref{eq:Mishragapfit}).\label{fig:obcgap}}
\end{figure}

In Fig.~\ref{fig:obcgap} we compare the resulting gaps
in the critical region, $1.8 \leq V \leq 2.2$
with the analytic result.
Apparently, neither of the extrapolations can reliably
determine the critical interaction because the extrapolated gaps always
open smoothly. Without the exact result for comparison, we cannot decide
which of the three schemes is superior to the two others.
We examine extrapolation schemes for the single-particle gap
in more detail in the supplemental material~\cite{suppmat}
(see, also, reference~[\onlinecite{
    PhysRevA.87.043606}] therein).

\subsection{Order parameter}

\begin{figure}[t]
\begin{center}
\includegraphics[width=8cm]{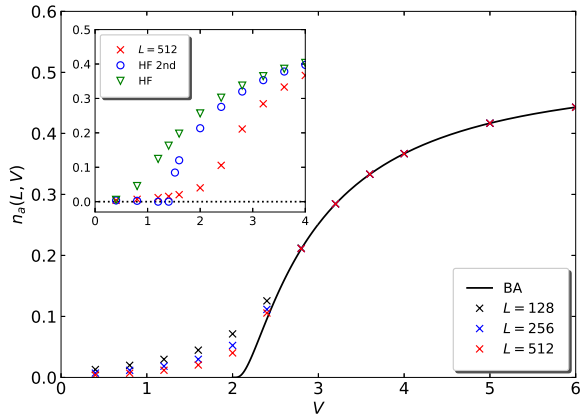}
\end{center}
\caption{Charge-density wave order parameter $n_a(L,V)$ 
  as a function of the nearest-neighbor interaction~$V$ from
Bethe Ansatz (BA) and DMRG for $L=128,256,512$ sites.
  Inset: comparison of (second-or\-der)
  Hartree-Fock and DMRG data for $L=512$ sites.\label{fig:nawithinset}}
\end{figure}

In Fig.~\ref{fig:nawithinset} we show the CDW order parameter
from DMRG as a function of the interaction 
for system sizes $L=128,256,512$.
It is seen that the finite-size corrections are large
for $V\lesssim 2.5$ but marginal for $V\gtrsim 3$.
This indicates that very large system sizes are required to
perform an accurate extrapolation to the thermodynamic limit
in the vicinity of the critical interaction, $V_{\rm c}=2$.

As seen from the inset, Hartree-Fock theory predicts
a continuous increase of the order parameter
for $V>V_{\rm c}^{\rm HF}= 0^+ $.
Second-order Hartree-Fock theory predicts a jump to a substantial CDW order
at $V_{\rm c,2}^{(2)}\approx 1.5$. The curves start to coalesce around $V\gtrsim 4$
where the strong-coupling expansion becomes applicable.
In general, second-order Hartree-Fock theory overestimates the CDW order
parameter but less severely than the standard Hartree-Fock approximation.

Finite-size effects are prominent in the DMRG data for the charge-density wave
order parameter
even for systems with $L= 512$ sites. This does not come as a surprise
because the
CDW order parameter displays the same essential singularity
as the single-particle gap,
see eq.~(\ref{eq:nalimitsexact}). As in the case of the single-particle gap,
the second-order polynomial fit for the finite-size
extrapolation,
\begin{equation}
  n_a(L,V) = n_a(V) + \frac{d_1}{L} + \frac{d_2}{L^2}
  \label{eq:2ndorderpolynomialCDWfit}
\end{equation}
with $n_a(V)$, $d_1$, and $d_2$ as fit parameters, leads to a smooth curve for
$n_a(V)$, in contrast to the exact solution
where the order sets in at $V_{\rm c}= 2$.
Therefore, the critical interaction strength cannot be deduced
from the order parameter.
We face the same difficulties for the single-particle gap
that also displays an essential singularity
at the transition.

\subsection{Correlation energy}
\label{subsec:CEcomparion}

The correlation energy can be calculated exactly from Bethe Ansatz results,
see Sect.~\ref{subsec:correlationenergy}. It goes to zero both for
small and large interactions because the ground state is given by
a single-particle product state in both cases, namely,
a Slater determinant for free fermions at $V= 0$ and a charge-density wave
with a particle on every other lattice site for $V\to \infty$.
Therefore, there is (at least) one extremum for finite $V$ at
$V_{\rm corr}>0$.

\begin{figure}[t]
\begin{center}
  \includegraphics[width=8cm]{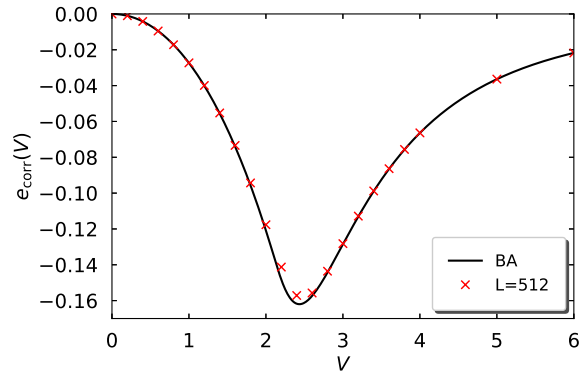}
\end{center}
\caption{Correlation energy 
  at half band-filling
  as a function of the nearest-neighbor interaction~$V$ from 
  Bethe Ansatz (full line) and from DMRG for $L= 512$
  sites (crosses).\label{fig:correnergy}}
\end{figure}

In Fig.~\ref{fig:correnergy} we show the correlation energy as a function
of the interaction strength from Bethe Ansatz and from DMRG for $L= 512$ sites.
The overall agreement is very good. It is seen that the
correlation energy is always negative. The
single-particle contributions generically overestimate the interaction
because they do not take the correlation hole into account that forms
around the particles but only the exchange hole.
The correlation energy has a (single) minimum but it is {\em not\/}
located at the critical interaction but at $V_{\rm corr,min}\approx 2.4$,
larger than $V_{\rm c}= 2$ by some twenty percent.

Eq.~(\ref{eq:ecorrexactBA}) shows that various quantities
contribute to the correlation energy. The ground-state energy and
its derivative do not signal the metal-insulator
transition whereas the order parameter $n_a(V)$ is finite for $V>V_{\rm c}$.
The mixture of regular and critical quantities
shifts the minimum of the correlation energy
away from $V_{\rm c}$.
This example shows that not every extremum in a physical quantity
can be used to locate $V_{\rm c}$ with high precision.

\subsection{Momentum distribution}

\begin{figure}[t]
    \begin{tabular}[b]{@{}ll@{}}
    (a) &\\[12pt]
         &      \includegraphics[width=8cm]{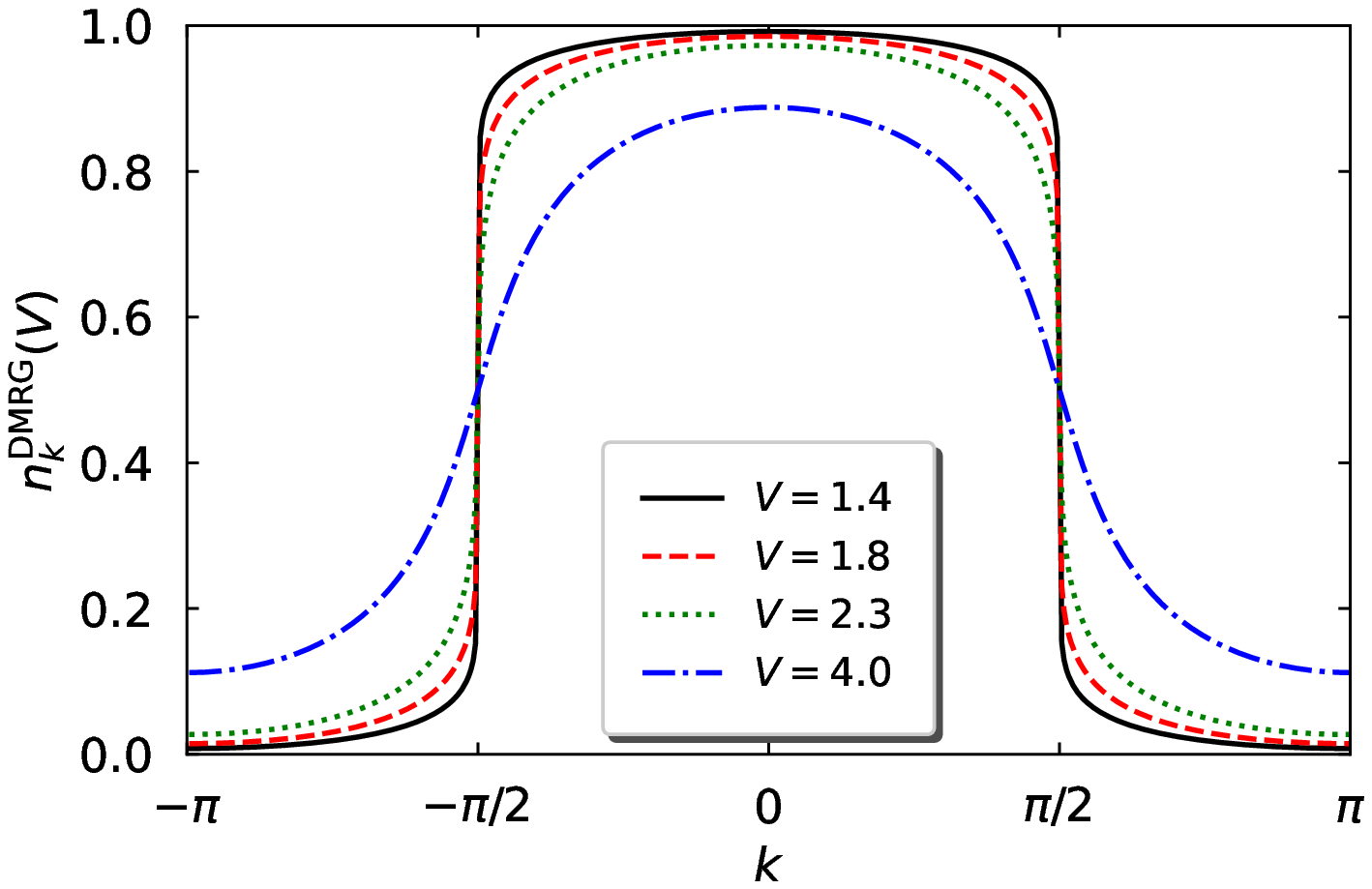}\\
      (b) &\\[12pt]
      & \includegraphics[width=8cm]{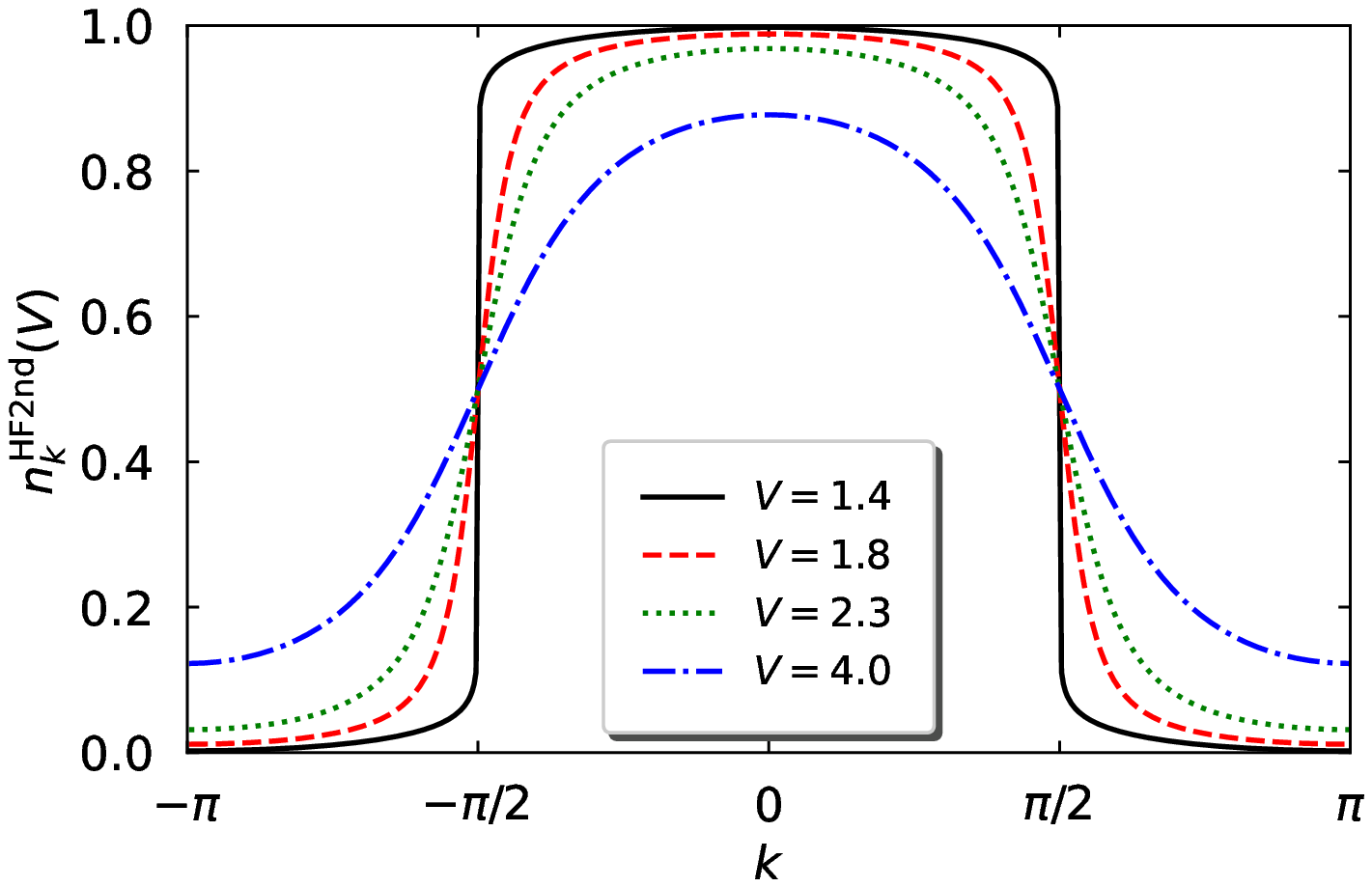}
    %  \vspace*{6cm}
    \end{tabular}
\caption{Momentum distribution $n_k$ for $V= 1.4,1.8,2.3,4$
  (a) from DMRG for $L= 512$ lattice sites and (b)
  from second-order Hartree-Fock approximation.\label{fig:nkDMRG}}
\end{figure}

Next, we discuss the momentum distributions which have not been determined
analytically from Bethe Ansatz for all $k$ and $V$ thus far.
In Fig.~\ref{fig:nkDMRG} we show the momentum distribution
$n_k= \langle \hat{a}_k^+
\hat{a}_k^{\vphantom{+ }}\rangle$. The points are dense enough to warrant
continuous lines.
It is known that the curves for $V>0$ are continuous
in the thermodynamic limit with $n_{k= \pm\pi/2}= 1/2$ due to
particle-hole symmetry, see eq.~(\ref{eqphforkdistribution}) in
Sect.~\ref{subsec:momdisexact}.
The momentum distributions from DMRG in Fig.~\ref{fig:nkDMRG}(a)
and from second-order Hartree-Fock theory
in Fig.~\ref{fig:nkDMRG}(b)
look very similar for $V= 1.4,1.8,2.3,4$
but deviations close to the Fermi wave numbers $\pm \pi/2$ are
clearly visible. Only for weak interactions, $V\lesssim 1$, and
for large interactions, $V\gtrsim 6$, 
the curves for $n_k(V)$ in DMRG and (second-order) Hartree-Fock theory
coalesce.

\begin{figure}[t]
\begin{center}
  \includegraphics[width=8cm]{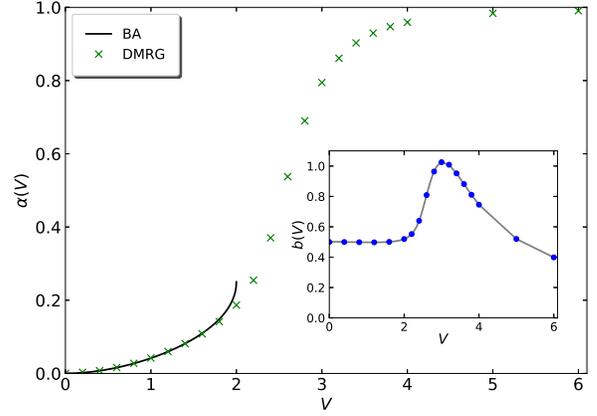}
\end{center}
\caption{Generalized Luttinger liquid exponent $\alpha(V)$
  for spinless fermions
  at half band-filling
  as a function of the nearest-neighbor interaction~$V$ extrapolated from 
  DMRG and Bethe Ansatz
  (full line for $0\leq V\leq V_{\rm c}= 2$).
  Inset: parameter $b(V)$.\label{fig:alphabextrapolation}}
\end{figure}

To identify the quantum phase transition from the DMRG data for
the momentum distribution, we analyze $n_k(V)$
in the vicinity of the Fermi point~$k_{\rm F}= \pi/2$.
We rewrite eq.~(\ref{eq:momdisgeneralexact}) as
%\begin{eqnarray}
%  \ln \left( \frac{1}{2}- n_{\pi/2+ 2\pi/L}(V)\right)
%  &=& \ln[b(V)] + \alpha(V)\ln\left(\frac{2\pi}{L}\right) \;,
%  \nonumber \\
%  \label{eq:lognkvslogL}
%\end{eqnarray}
\begin{equation}
\ln \left( 1/2- n_{\pi/2+ 2\pi/L}(V)\right)
= \ln[b(V)] + \alpha(V)\ln\left(2\pi/L\right) \;,
 \label{eq:lognkvslogL}
\end{equation}
and extrapolate the DMRG data for the left-hand-side
of eq.~(\ref{eq:lognkvslogL})
in $\ln(L)$ to determine the fit parameters $\alpha(V)$ and $b(V)$.
The result is shown if Fig.~\ref{fig:alphabextrapolation}.

The analytic Luttinger exponent $\alpha(V)$
from eqs.~(\ref{eq:momdisLLexact}) and~(\ref{eq:exactLLK})
is reproduced from DMRG for $V\leq 1.9$ but it is underestimated
close to the transition so that
the condition $\alpha^{\rm DMRG}(V_{\rm c}^{\alpha})= 1/4$
leads to $V_{\rm c}^{\alpha}=2.2$.
Likewise, the parameter $b(V)= 1/2$ is observed
with an accuracy of $10^{-3}$ deep in the Luttinger liquid
but deviations of more than one percent occur for $V\gtrsim
V_{\rm c}^{\beta}= 1.8$.
In this way, 
we locate the transition in the region $1.8= V_{\rm c}^{\beta}<V_{\rm c}<
V_{\rm c}^{\alpha}= 2.2$, within ten percent of the critical interaction.

\subsection{Quasi-particle distribution}

More intriguing than the momentum distribution
is the quasi-particle distribution $n_{k,\beta}$.
As we discussed in Sect.~\ref{subsubsec:definenkalphankbeta},
$n_{k,\alpha/\beta}$ describes the occupation numbers for the natural orbitals
that we identify with the lower $(k,\alpha)$ and upper $(k,\beta)$
Hartree-Fock bands.

\begin{figure}[t]
    \begin{tabular}[t]{@{}ll@{}}
      (a) & \\
      & \includegraphics[width=8cm]{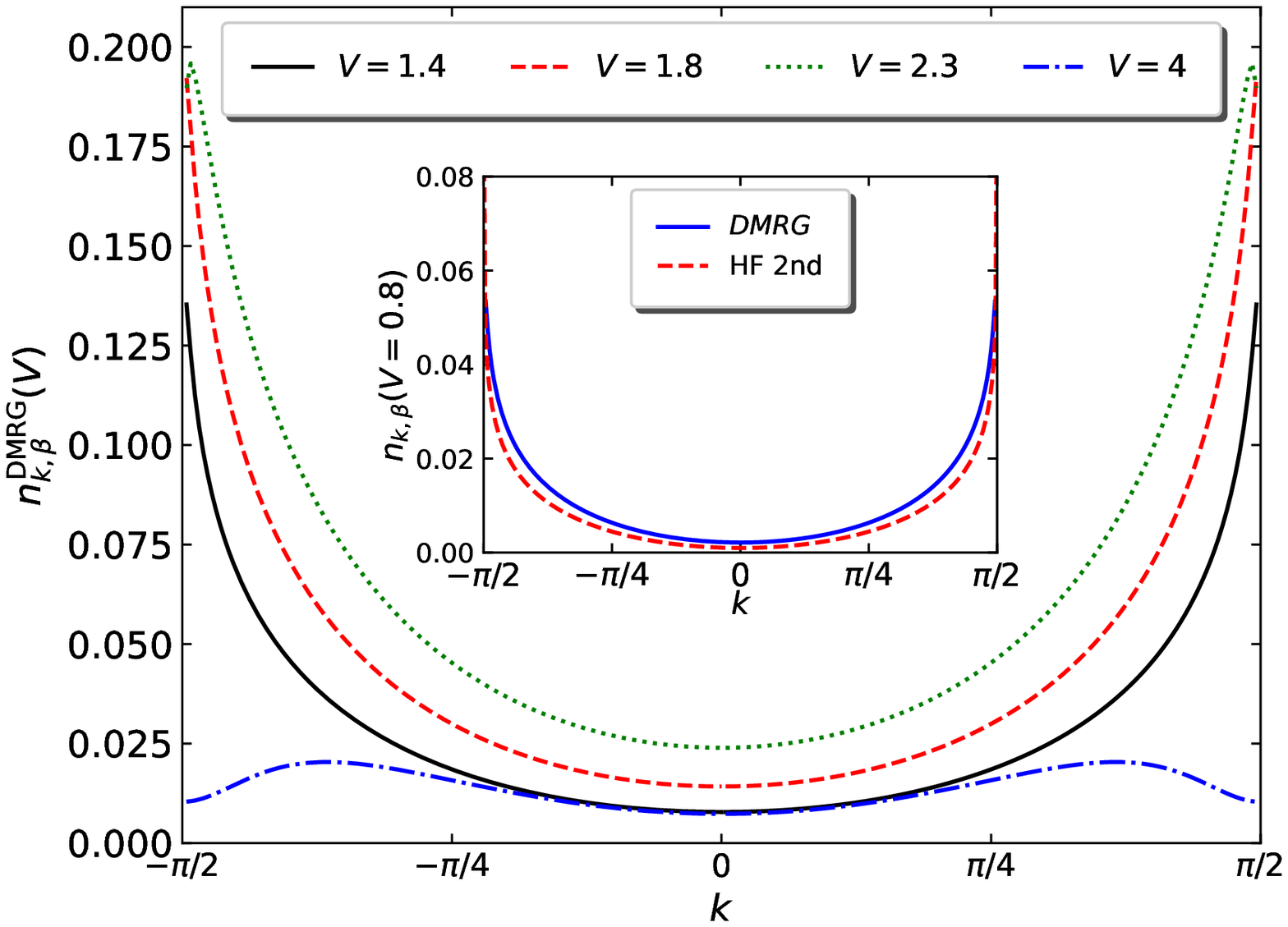}\\
      (b) & \\
      & \includegraphics[width=8cm]{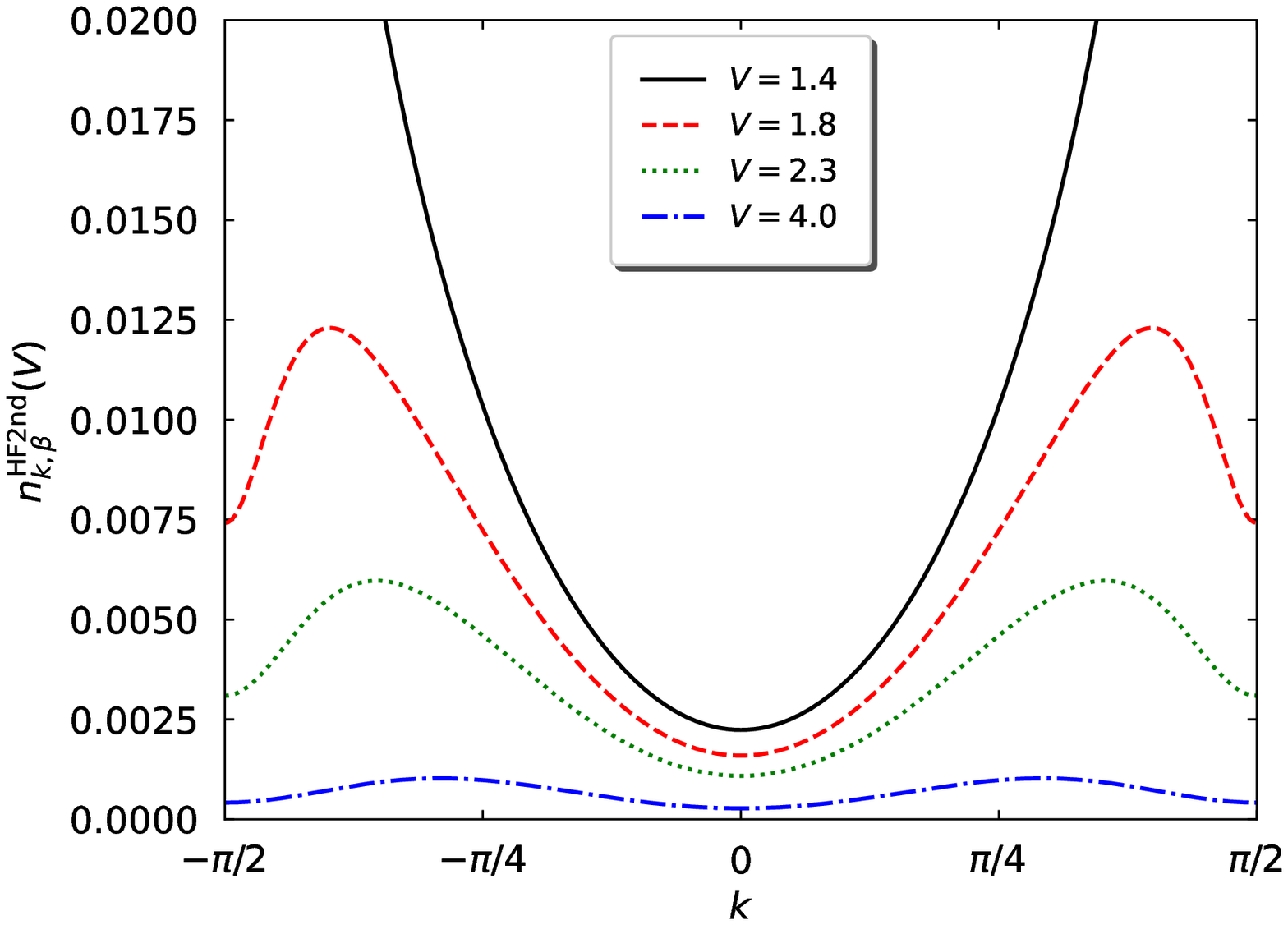}
%      \vspace*{6cm}
\end{tabular}
\caption{Quasi-particle distribution function $n_{k,\beta}$ for $V= 1.4,1.8,2.3,4$
  (a) from DMRG for $L= 512$ lattice sites and
  (b) from second-order Hartree-Fock
  approximation. Note the factor ten difference in the values on the ordinate.
  Inset: DMRG and second-order Hartree-Fock
  for $V=0.8$.\label{fig:nkbetaDMRG}}
\end{figure}

We show the quasi-particle distribution from DMRG
in Fig.~\ref{fig:nkbetaDMRG}(a) 
and from second-order Hartree-Fock theory in Fig.~\ref{fig:nkbetaDMRG}(b).
For $V<V_{\rm c}$ the DMRG data
in Fig.~\ref{fig:nkbetaDMRG}(a) display a maximum at the band edges
whereas in the insulating phase there are two maxima. Therefore,
the onset of two maxima indicates the CDW transition, and a
first estimate for the critical interaction strength can be deduced from
the finite-size data, $V_{\rm c}^{\rm tm}\approx 2.15$.

The inset shows that the second-order results are in quantitative agreement
with those from DMRG at weak coupling, $V=0.8$, which
serves as a significant consistency check for both methods.
As seen from a comparison of the main Figs.~\ref{fig:nkbetaDMRG}(a) and~(b),
the agreement between
DMRG and second-order Hartree-Fock rapidly deteriorates for larger interactions,
$V\gtrsim 1$.
Even in the limit of strong interactions, the second-order Hartree-Fock
approximation
does not reproduce the DMRG data for the quasi-particle distribution.
Although the curves look similar, they substantially differ quantitatively,
by a factor
of ten and more for $V\gtrsim 2$.
In essence, Hartree-Fock theory severely underestimates the total
density of quasi-particle excitations $n_{\beta}(V)$
defined in eq.~(\ref{eq:deftotalqpdensity}).

\begin{figure}[t]
  \begin{center}
    \includegraphics[width=8cm]{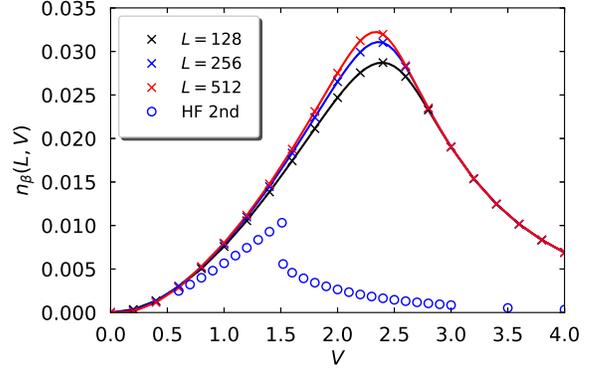}
\end{center}
  \caption{Quasi-particle density $n_{\beta}(V)$
    as a function of $V$ from DMRG for $L= 128,256,512$,
    and from second-order Hartree-Fock theory.\label{fig:nbetaofV}}
\end{figure}

To see this in more detail, we show the
density of quasi-particle excitations $n_{\beta}(V)$ as a function
of the interaction~$V$ in Fig.~\ref{fig:nbetaofV}.
It is seen that the second-order Hartree-Fock
theory is reliable only for $V\lesssim 1$. The quasi-particle density
in Hartree-Fock theory displays a maximum just before
and a jump discontinuity right at $V_{\rm c,2}^{(2)}\approx 1.5$,
in agreement with the results in Sect.~\ref{sec:MITin2ndorder}.
This observation indicates that $n_{\beta}(V)$ is a sensitive quantity to locate
the CDW transition. Moreover, we see that $n_{\beta}^{(2)}(V)<0.011$
so that the condition for a dilute gas of quasi-particles,
$n_{\beta}^{(2)}(V)\ll 1/2$ is always fulfilled. Therefore,
second-order Hartree-Fock theory is applicable for all interaction
strength and is `almost variational', see Sect.~\ref{subsec:almostvariational}.

The DMRG data for the quasi-particle density in Fig.~\ref{fig:nbetaofV} show
that $n_{\beta}(V) < 0.035$, i.e., it is never more than seven percent of its
maximal value of one half. Therefore, the system can be viewed as a
vacuum state with a dilute gas of quasi-particle excitations, even though
second-order Hartree-Fock theory is not sufficient for its description beyond 
weak interactions.

As seen in Fig.~\ref{fig:nbetaofV}, the quasi-particle density is maximal
close to the critical interaction strength, $V_{\rm c}=2$,
so that we could use the maximum of the quasi-particle density
to locate the exact CDW transition from a finite-size extrapolation.
It turns out, however, that
the finite-size scaling is logarithmic
which limits the accuracy to several percent.

\begin{figure}[b]
  \begin{center}
    \includegraphics[width=8cm]{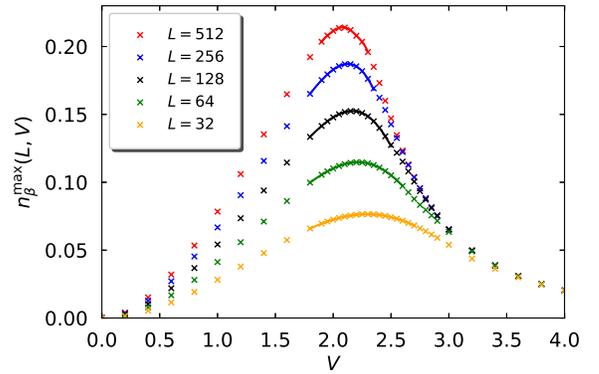}
%    \vspace*{6cm}
\end{center}
  \caption{Values of the maxima in the quasi-particle
    density $n_{\beta}^{\rm max,k}(L,V)$
    as a function of $V$ from DMRG for $L= 32,64,128,256,512$.
    The continuous
    lines are a 6th-order
    polynomial fit for the region  around the maximum.\label{fig:Vmaxpositions}}
\end{figure}

To determine $V_{\rm c}$ more accurately,
we recall that
\begin{equation}
  n_a(V) =\frac{1}{L}\sum_{k\in {\rm RBZ}}\left(\langle
      \hat{a}_k^+\hat{a}_{k+\pi}^{\vphantom{+}}\rangle -
  \langle  \hat{a}_{k+\pi}^+\hat{a}_k^{\vphantom{+}}\rangle \right)\; .
\end{equation}
The exponential behavior of $n_a(V)$ close to the transition
implies that most terms in the sum have a logarithmic dependence on
system size. However, this does not exclude that some
terms have an algebraic scaling in $1/L$ that is more suitable
for finite-size extrapolations. In our analysis, we use
the maximal value of the quasi-particle distribution
\begin{equation}
  n_{\beta}^{\rm max,k}(L,V)=
  \mathop{\rm Max}_k n_{k,\beta}(L,V)
  \label{eq:defnbetamax}
\end{equation}
to locate such special $k$-values for a given interaction strength,
see Fig.~\ref{fig:nkbetaDMRG}. 

As seen in Fig.~\ref{fig:Vmaxpositions},
the maximal value   $n_{\beta}^{\rm max,k}(L,V)$
increases from zero for weak interactions up to a maximal value
near the critical interaction strength
and decreases down to zero for large interactions.
We thus determine the maximum of
$n_{\beta}^{\rm max,k}(L,V)$,
\begin{equation}
  V_{\beta}^{\rm max}(L)=
  \mathop{\rm Max}_V   n_{\beta}^{\rm max,k}(L,V) \; .
  \label{eq:defVmaxfromnbetamax}
\end{equation}
In Fig.~\ref{fig:Vmaxpositions} we show
$n_{\beta}^{\rm max,k}(L,V)$ together with
a 6th-order polynomial fit in the vicinity of $V=2$
to locate the positions $V_{\beta}^{\rm max}(L)$
for system sizes $L=32,64,128,256,512$.
In this way, we locate $V_{\beta}^{\rm max}(L)$ with high accuracy.

\begin{figure}[hb]
  \begin{center}
    \includegraphics[width=8cm]{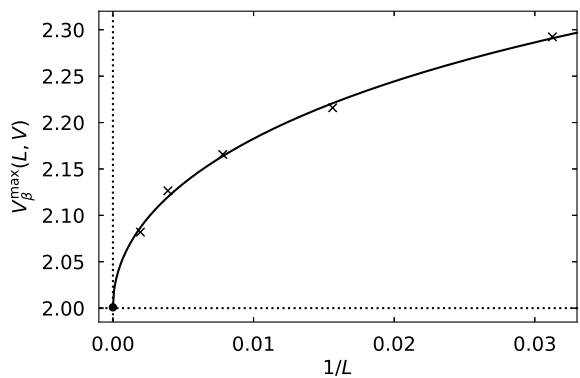}
\end{center}
  \caption{Extrapolation of the maxima position in the quasi-particle density
    as a function of $1/L$ from DMRG for $L= 64,128,256,512$.
    The continuous
    line is a fit to a second-order polynomial
    in $\sqrt{1/L}$.\label{fig:Vmaxextrap}}
\end{figure}

Next, we extrapolate the positions of the maxima to the thermodynamic limit.
In Fig.~\ref{fig:Vmaxextrap} we show $V_{\beta}^{\rm max}(L)$
as a function of inverse system size together with a square-root fit,
\begin{equation}
  V_{\beta}^{\rm max}(L) = V_{\rm c}^{\rm qp}+ \frac{v_1}{\sqrt{L}}
  + \frac{v_2}{L}
  \; ,
  \label{eq:logfitVmax}
\end{equation}
where $V_{\rm c}^{\rm qp}$, $v_1$, and $v_2$ are fit parameters.
The square-root extrapolation is motivated by the fact that
the Luttinger liquid is characterized by algebraic
singularities. Indeed, the Luttinger parameter is $K(V_{\rm c})= 1/2$
at the transition.
The extrapolation results in $V_{\rm c}^{\rm qp}= 2.0008\pm 0.019$,
in agreement with the exact value for the critical interaction
with at most one percent deviation.
The extrapolation of the maxima position in the quasi-particle density
provides a successful route to determine the critical interaction strength
with high accuracy.

A more traditional route to determine the transition traces
the breakdown of the Luttinger liquid, as already utilized
for the finite-size corrections of the ground-state energy and of
the gap. The Luttinger parameter $K(V)$ directly
monitors the Luttinger liquid, as seen from the momentum distribution.
Indeed, an accurate calculation
of Luttinger exponent $K(V)$ from the density-density correlation function
permits to locate the transition with an accuracy of three percent, as
we shall show next.

\subsection{Density-density correlation function}
\label{subsec:CNNdiscussion}

\begin{figure}[t]
  \begin{center}
    \includegraphics[width=8cm]{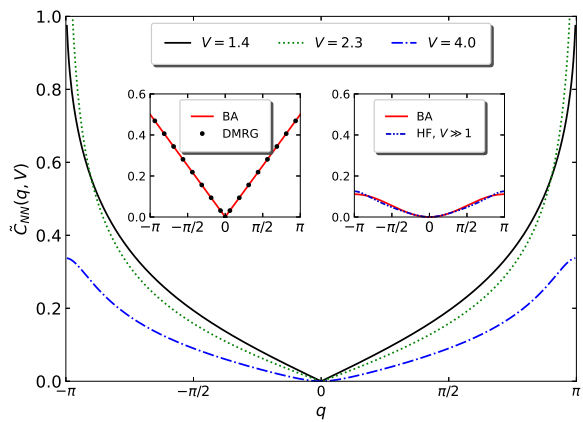}
%    \vspace*{6cm}
\end{center}
  \caption{Structure factor 
    $\tilde{C}^{\rm NN}(q,V)$ from DMRG for $L= 512$ sites
    for $V= 1.4$ (black), $V= 2.3$ (green), and $V= 4$ (blue).\\
    Left inset: Structure factor
    $\tilde{C}_0^{\rm NN}(q)$ from DMRG for $L= 32$ sites
  and the analytic result~(\protect\ref{eq:CNNVequalzero})
  for $V=0$ (red line).\\
  Right inset: Structure factor
    $\tilde{C}^{\rm NN}(q,V)$ from DMRG for $L= 512$ sites
  for $V= 6$ and the analytic result~(\protect\ref{eq:CNNtildelargeq})
  for strong coupling (red line).\label{fig:CNNtilde}}
\end{figure}

Lastly, we address the density-density correlation function,
eq.~(\ref{eq:CNNdef}).
We show its Fourier transform, eq.~(\ref{eq:CNNtildedef}),
from DMRG for $L= 512$ sites and $V= 1.8,2.3,4$ in Fig.~\ref{fig:CNNtilde}.
It is seen that the structure factor $\tilde{C}^{\rm NN}(q,V)$ shows the expected behavior,
see Sect.~\ref{subsec:CNNexact}. It vanishes at $q= 0$ with a finite slope
for all $V$. It diverges for $|q|\to \pi$ in the Luttinger-liquid phase,
and remains finite for all $q$ in the CDW phase.

The insets of Fig.~\ref{fig:CNNtilde}
show $\tilde{C}^{\rm NN}(q,V)$ for $V= 0$ and for
$V= 6$, in comparison with the leading-order results for weak and
strong coupling,
see eqs.~(\ref{eq:CNNVequalzero}) and~(\ref{eq:CNNtildelargeq}).
At $V= 0$, the agreement is  excellent already for $L= 32$ sites.
For strong coupling, the agreement at $V= 6$ is already very good
but it is clearly seen that the corrections to order $1/V^3$ are important.
This not only quantitatively
applies at the Brillouin zone boundaries, $q= \pm \pi$, but
also qualitatively close to $q= 0$. Within Hartree-Fock theory,
$\tilde{C}^{\rm NN}_{\rm BA}(q\to 0)\sim q^2$ whereas
the exact density-density correlation function
displays a kink at $q= 0$,
$\tilde{C}^{\rm NN}_{\rm HF}(q\to 0)\sim |q|$. This reflects the fact that
the domain walls are mobile in the exact solution but rigid
within the Hartree-Fock approximation.
The freely mobile quasi-particles lead to a small-$q$ behavior
resembling that of free fermions.

\begin{figure}[t]
  \begin{center}
    \includegraphics[width=8cm]{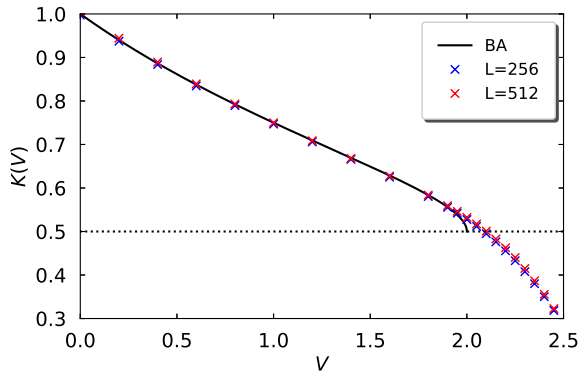}
%    \vspace*{6cm}
\end{center}
  \caption{Luttinger parameter $K(V)$ as a function
    of the interaction from the Fourier-transformed
    density-density correlation function at small~$q$ from DMRG,
    eq.~(\protect\ref{eq:KfromCNN}), for $L= 256$ sites (blue crosses)
    and $L= 512$ sites (red crosses), in comparison with
    the exact result~(\protect\ref{eq:exactLLK}) from Bethe Ansatz
    for $0\leq V\leq V_{\rm c}= 2$ (black line).\label{fig:KrhofromCNNtilde}}
\end{figure}

The main advantage of the density-density correlation function
lies in the fact that it permits to determine the Luttinger parameter
$K(V)$ with high accuracy.
In Fig.~\ref{fig:KrhofromCNNtilde} we show the exact result
for $K(V)$ as a function of~$V$ from Bethe Ansatz,
eq.~(\ref{eq:exactLLK}), in comparison with DMRG data
for $L= 256$ and $L= 512$ sites. It is seen that the finite-size effects
are of the same order of magnitude as the accuracy of the data.
The agreement with the exact result is very good for $V\leq 1.95$,
with deviations close to the transition. The field-theory
criterion,
$K(V_{\rm c})= 1/2$,~\cite{Thierrybook} leads to $V_{\rm c}^{\rm LL}= 2.06$. The
analysis of $K(V)$ permits to locate
the critical interaction with an accuracy of three percent.

\section{Conclusions}
\label{conclusions}

A summary and a short outlook close our presentation
on the charge-density wave transition for spinless fermions in one dimension.

\subsection{Summary}

In this work, we study
spinless fermions in one dimension with nearest-neighbor interaction~$V$
and nearest-neighbor transfer matrix element~$(-t)$ ($t\equiv 1$)
at half band-filling.
We use the Hartree-Fock approximation
to first and second order in the interaction
and the numerical density-matrix renormalization group (DMRG)
for rings with up to 514 sites and compare the data with
exact results from Bethe Ansatz in the thermodynamic limit.
In particular, we investigate
the ground-state energy per lattice site $e_0(V)$,
the nearest-neighbor single-particle density matrix $B_0(V)$,
the single-particle gap $\Delta_1(V)$, and the
charge-density wave order parameter $n_a(V)$.
For the ground-state energy and the gap,
exact analytical formulae are available for the leading finite-size corrections
in the metallic phase.

In addition, DMRG and second-order Hartree-Fock theory permit to
calculate the single-particle density matrix and the density-density
correlation function for all distances and thus
provide the momentum distribution $n_k$,
the quasi-particle distribution $n_{k,\beta}$, and the structure factor
$\tilde{C}^{\rm NN}(q,V)$.

Hartree-Fock theory provides a good upper bound
to the ground-state energy that is
improved for all interaction strengths by including second-order corrections.
Second-order Hartree-Fock theory is applicable for all interaction
strengths because the density of quasi particles is very small so that
second-order Hartree-Fock theory is almost variational.

In contrast to other exactly solvable one-dimensional
models, spinless fermions display a charge-density-wave (CDW) transition
at a finite value, $V_{\rm c}=2$. In standard Hartree-Fock theory,
the CDW transition is predicted to set in at any finite interaction, reflecting 
the perfect nesting situation at half band-filling.
Second-order Hartree-Fock theory predicts 
a discontinuous CDW transition at $V_{\rm c,2}^{\rm (2)}\approx 1.51$;
the ordered phase around $V=0$ is reduced to the region
$0<V<V_{\rm c,1}^{(2)}\approx 0.21$ and is characterized by a tiny order parameter.
Therefore, second-order Hartree-Fock theory improves the description
of spinless fermions considerably, both qualitatively and quantitatively.

Quantitatively reliable information about the CDW transition 
is obtained from DMRG on large systems.
The exact ground-state energy density and 
nearest-neighbor single-particle density matrix
do not display any singularities and are almost perfectly reproduced
for all interaction strengths by DMRG for up to 514~sites.
Likewise, the gap and the CDW order parameter
are obtained with good accuracy from a finite-size
extrapolation of the DMRG data, except for the critical region where
the gap and the CDW order parameter display essential singularities.
Therefore, different strategies have to be designed to locate
the quantum phase transition accurately.

In this work, two strategies are designed that permit to
determine the critical interaction strength. The traditional route
focuses on the breakdown of the Luttinger liquid. Results from
conformal field theory and the Bethe Ansatz
for the finite-size corrections of the ground-state energy and the gap
lead to useful but not very accurate estimates for $V_{\rm c}$.
Moreover, these estimates require a lot of a-priori knowledge from the exact
solution. Instead, the traditional derivation of the Luttinger parameter from
the momentum distribution, $n_k(L,V)$, and, more accurately, from
the structure factor at small momenta, $\tilde{C}^{\rm NN}(q\to 0,V)$,
leads to $V_{\rm c}^{\rm LL}= 2.06$, only three percent off the exact result.
The second strategy to determine the CDW transition point with high
accuracy utilizes the maxima of the quasi-particle distribution
$n_{k,\beta}(L,V)$. For $L\to \infty$, $n_{\beta}^{\rm max,k}(V)$
peaks at $V_{\rm c}$. The finite-size extrapolation
of DMRG data for up to $L=512$ sites leads to an agreement
with one percent accuracy, $V_{\rm c}^{\rm qp}= 2.0008\pm 0.019$.

The density of quasi-particles is small also for the exact solution,
$n_{\beta}^{\rm max,k}(V) \leq 0.035 \ll 0.5$.
This implies that the system may be viewed as a vacuum state
with a dilute gas of excitations. This observation ties in with the fact that
dynamic correlation functions for the XXZ model
can be expressed in terms as a series of $2n$-spinon excitations that
is dominated by the first few terms.

\subsection{Outlook}

The comparison with exact results from the Bethe Ansatz for
spinless fermions in one dimension demonstrates that it is possible to locate
Kosterlitz-Thouless transitions at finite interaction strengths 
from sophisticated extrapolations of DMRG data.
Therefore, the strategies and extrapolation schemes proposed here
can reliably be applied to non-integrable models in one dimension.
The gapless phase of such models is described by a Luttinger
  liquid. As shown in this work,
  the quantum phase transition to a gapped phase
  can be detected by monitoring the Luttinger parameter
  obtained from the static density-density correlation function.
  Moreover, the quasi-particle densities depend on the ground-state phase,
so that the occupation numbers of the natural orbitals
provide a sensitive probe for locating Kosterlitz-Thouless-type
phase transitions in generic one-dimensional many-particle models.
  
Our analysis also shows that second-order Hartree-Fock theory provides
a reasonable description for all interaction strengths even in one spatial
dimension. Therefore, we expect that it is useful to extend and apply
the method to
two and three dimensions where some peculiarities of one dimension
are absent, e.g., freely moving domain walls in the strong-coupling limit.
Work in this direction is in progress.

\begin{acknowledgments}
  We thank Frank G\"ohmann for sharing his expertise on the history
of Bethe Ansatz results for the XXZ model.
F.G.\ thanks him, Andreas Kl\"umper, and Sergei Rutkevich
  for interesting and helpful discussions on the physics of the XXZ model.
  
  \"O.L. has been supported by the Hungarian National Research,
  Development, and Innovation Office (NKFIH) through Grants No.~K120569,
  No.~K134983, and TKP2021-NVA
  by the Hungarian Quantum Technology National Excellence
  Program (Project No.~2017-1.2.1-NKP-2017-00001)
  and by the Quantum Information National Laboratory of Hungary.
  \"O.L. also acknowledges financial support from the Alexander von Humboldt
  foundation and the Hans Fi\-scher Senior Fellowship program
  funded by the Technical University of Munich -- Institute for Advanced Study.
 
  The development of DMRG libraries has been supported by the Center for
  Scalable and Predictive methods for Excitation and Correlated phenomena
  (SPEC),
  which is funded as part of the Computational Chemical Sciences Program
  by the U.S.\
  Department of Energy (DOE), Office of Science, Office of
  Basic Energy Sciences,
  Division of Chemical Sciences, Geosciences, and Biosciences
  at Pacific Northwest National Laboratory.
\end{acknowledgments}

%%%%%%%%%%%%%%%%%%%%%%%%%%%%%%%%%%%%%%%%%%%%%%%%%%%%%%%%%%%%%%%

%\appendix

%\newpage

%\bibliographystyle{unsrt}
%\bibliographystyle{pss}
%\bibliography{tVmodel}% Produces the bibliography via BibTeX.


\clearpage

\begin{widetext}

\begin{center}
{\large\sc Supplemental material}\\[6pt]
{\large\bf Accurate localization of Kosterlitz-Thouless-type quantum phase
  transitions}\\[3pt]
{\large\bf   for one-dimensional spinless fermions}
  \end{center}

{\small The supplemental material consists of six parts. In supplement~I,
  we examine various extrapolation schemes to extract
  the critical interaction strength from finite-size data
  for the charge-density wave
  order parameter and the single-particle gap.
  In supplements~II, III, and~IV, we provide the series expansion
  for the exact ground-state energy, order parameter, and single-particle gap,
  respectively.
  In supplement~V, we calculate the matrix elements
  in second-order Hartree-Fock theory.
  In supplement~VI, we derive observables in second-order Hartree-Fock
  approximation.}
%\vspace*{-24pt}

%\pacs{72.15.Qm,75.20.Hr,75.30.Hx}

%\keywords{Impurity and Kondo physics, DMRG, Gutzwiller variational approach.}
\end{widetext}

\appendix

\renewcommand{\theequation}{\Roman{section}-\arabic{equation}}
\renewcommand{\appendixname}{Supplement}
\renewcommand{\thesection}{\Roman{section}}

\section{Multi-parameter extrapolations}
\label{supp:elephant-games}

In the first section of the supplemental material, we scrutinize
extrapolation schemes that aim to locate accurately
the Kosterlitz-Thouless transition
from finite-size data for the charge-density-wave (CDW) order parameter
or the single-particle gap.

\subsection{Extrapolation of the order parameter}
\label{CDW-parameter-extrapolation}

In the main text, we use a polynomial extrapolation
\begin{equation}
  n_a(L,V) = n_a(V) + \frac{d_1}{L} + \frac{d_2}{L^2}
  \label{eq:2ndorderpolynomialCDWfit-supp}
\end{equation}
with $n_a(V)$, $d_1$, and $d_2$ as fit parameters
to estimate the charge-density-wave order parameter $n_a(V)$
in the thermodynamic limit.
A much better fit for the CDW order parameter
can be obtained from 
a logarithmic extrapolation of the DMRG data,
\begin{equation}
  n_a^{\rm log}(L,V) = n_a^{\rm log}(V) + \frac{c}{(d+ \ln(L))^{\gamma}}
  \label{eq:logCDWfit}          
\end{equation}
with the four fit parameters $n_a^{\rm log}(V)$, $c$, $d$, and $\gamma$.

\begin{figure}[b]
\begin{center}
\includegraphics[width=8cm]{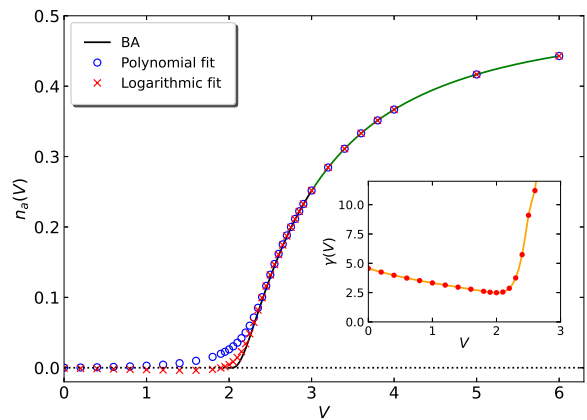}
\end{center}
\caption{Charge-density wave order parameter $n_a(V)$ 
  as a function of the nearest-neighbor interaction~$V$ extrapolated from
  DMRG data (polynomial fit and logarithmic fit), in comparison with the exact
  Bethe-Ansatz solution (BA).
  Inset: exponent $\gamma(V)$: it is minimal at the exact critical
  interaction strength~$V_{\rm c}= 2$.\label{fig:naTDL}}
\end{figure}

We extrapolate the DMRG data for the charge-density-wave order parameter
$n_a(L,V)$ obtained for periodic boundary conditions and $L=64,128,256,512$,
and plot $n_a^{\rm log}(V)$ in Fig.~\ref{fig:naTDL}, together with the result
from the second-order polynomial extrapolation $n_a(V)$
and the analytic result from Bethe Ansatz.
As seen from the figure, the logarithmic fit procedure
gives a much better agreement with the exact result
even though the extrapolated order parameter
is slightly negative below the transition
and becomes positive around $V_{\rm c}$.
This behavior is typical for transitions at finite interaction strengths.
We may use the re-entrance criterion of a zero order parameter,
$n_a(V_{\rm c}^{\rm re})=0$, to estimate the critical interaction strength.
In this way we find $V_{\rm c}^{\rm re}=1.98$.

The critical interaction strength $V_{\rm c}$
can be determined with the same accuracy
from the minimum of the exponent $\gamma(V)$.
For our data set, the minimum of $\gamma(V)$
occurs at the exact value $V_{\rm c}= 2$ so that
the logarithmic extrapolation of the order parameter
leads to $V_{\rm c}= 2.00\pm 0.02$, with an accuracy of
one percent.

However, this observation deserves some critical comments. First, the
extrapolation includes {\em four\/} fit parameters
which always carries the risk of over-fitting the data.
Second, the functional form~(\ref{eq:logCDWfit}) is purely phenomenological.
%i.e., it is not backed up
%by any theory for finite-size corrections in the tV~model.
In fact, the same approximation leads to nonsensical results when applied to
the single-particle gap. The logarithmic scheme~(\ref{eq:logCDWfit})
only coincidentally works for the CDW order parameter.

%\clearpage

\begin{figure*}[ht]
  \begin{flushleft}
\begin{tabular}[b]{@{}ll@{}}
  (a) & (b) \\
  \includegraphics[width=\columnwidth]{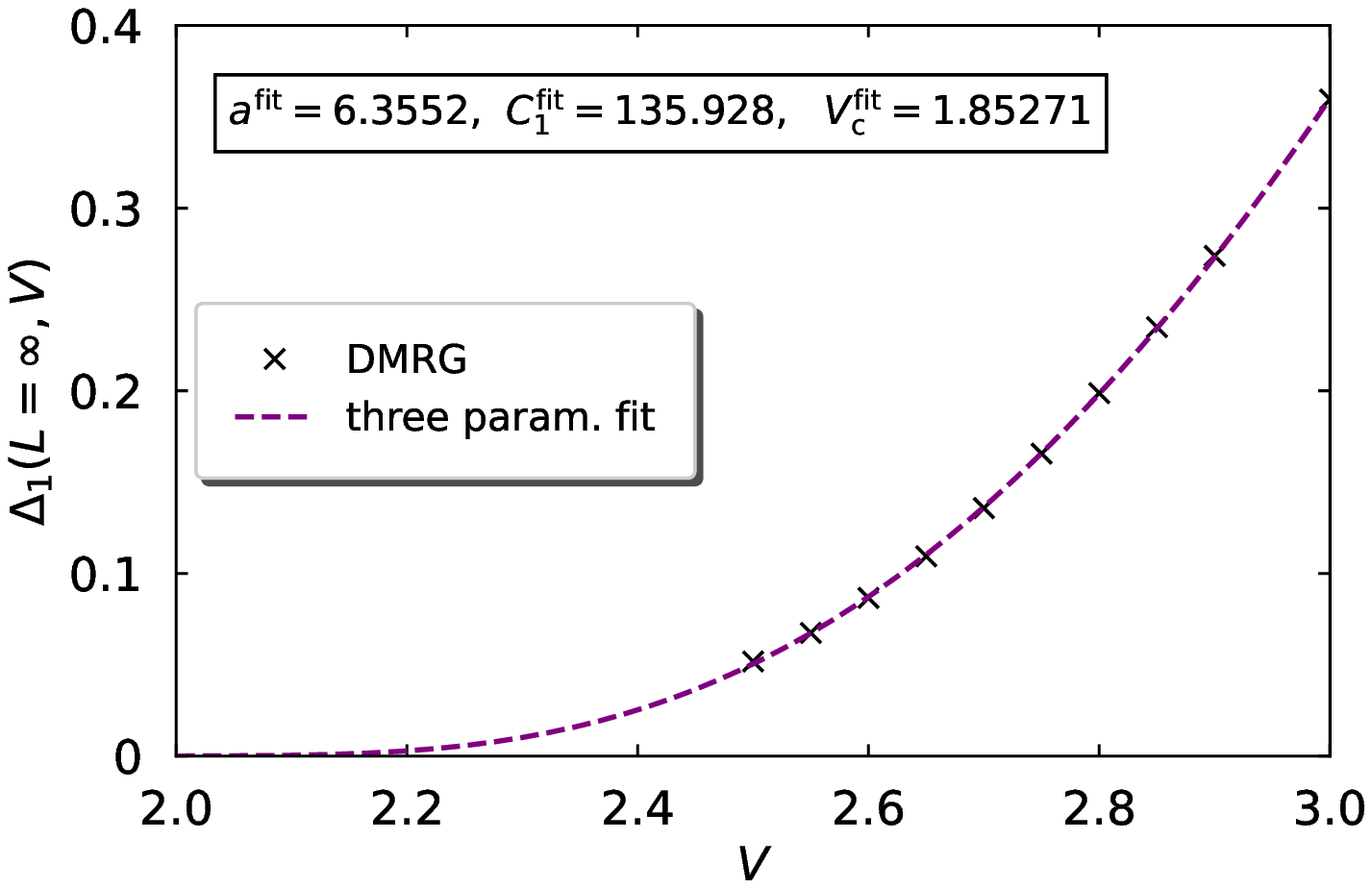} &
  \includegraphics[width=\columnwidth]{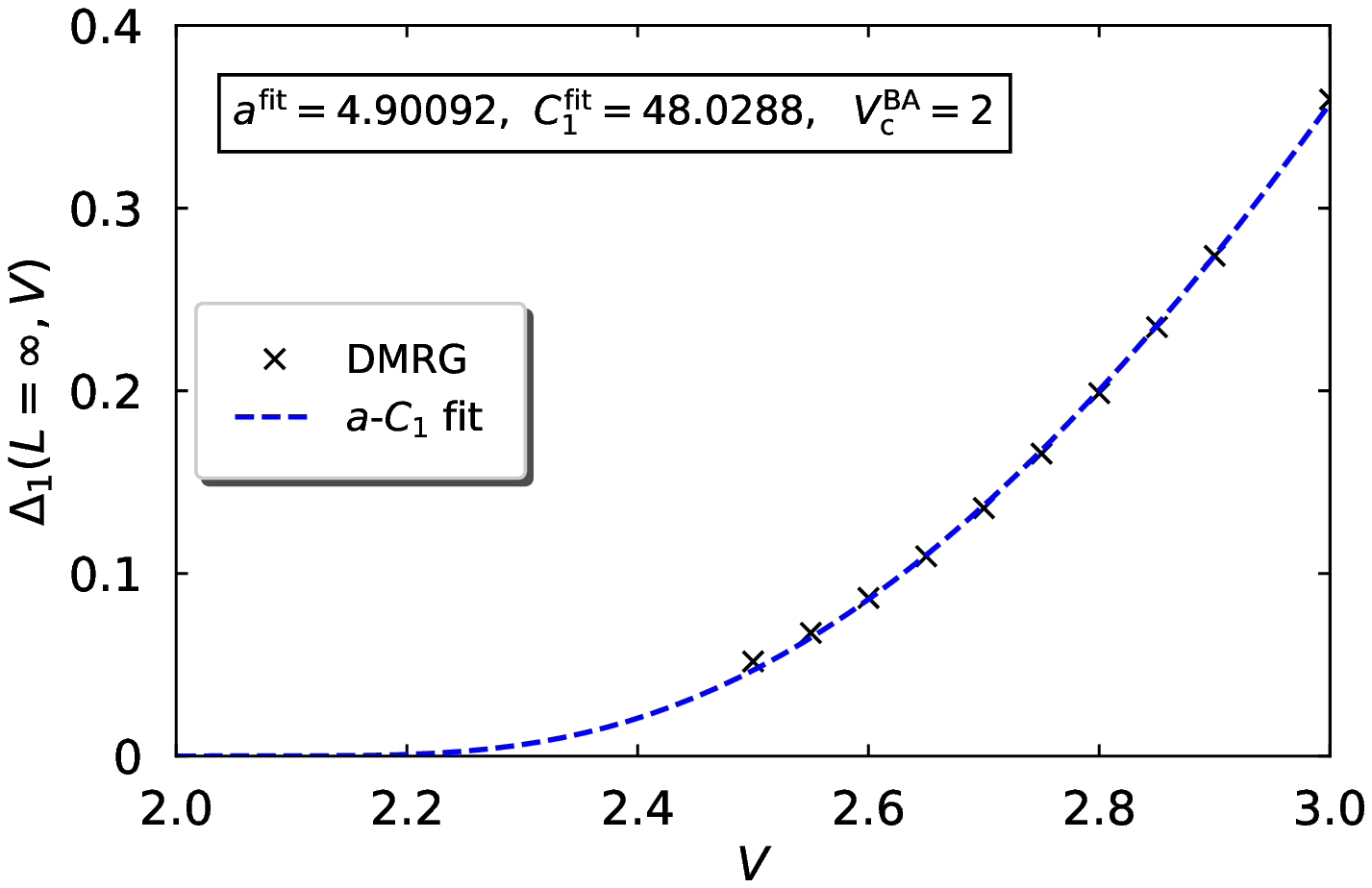}\\[12pt]
  (c) & (d) \\
  \includegraphics[width=\columnwidth]{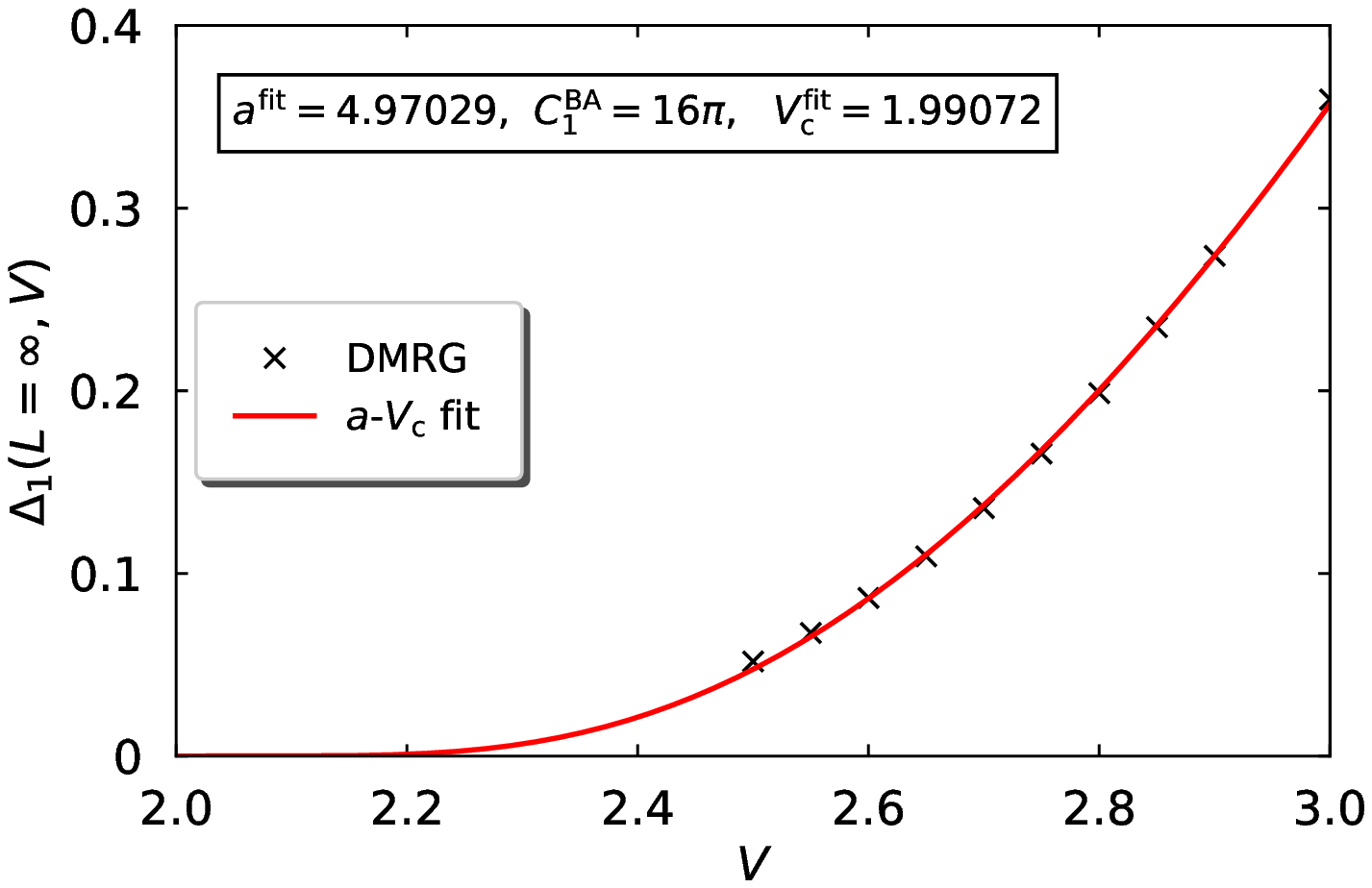} &
   \includegraphics[width=\columnwidth]{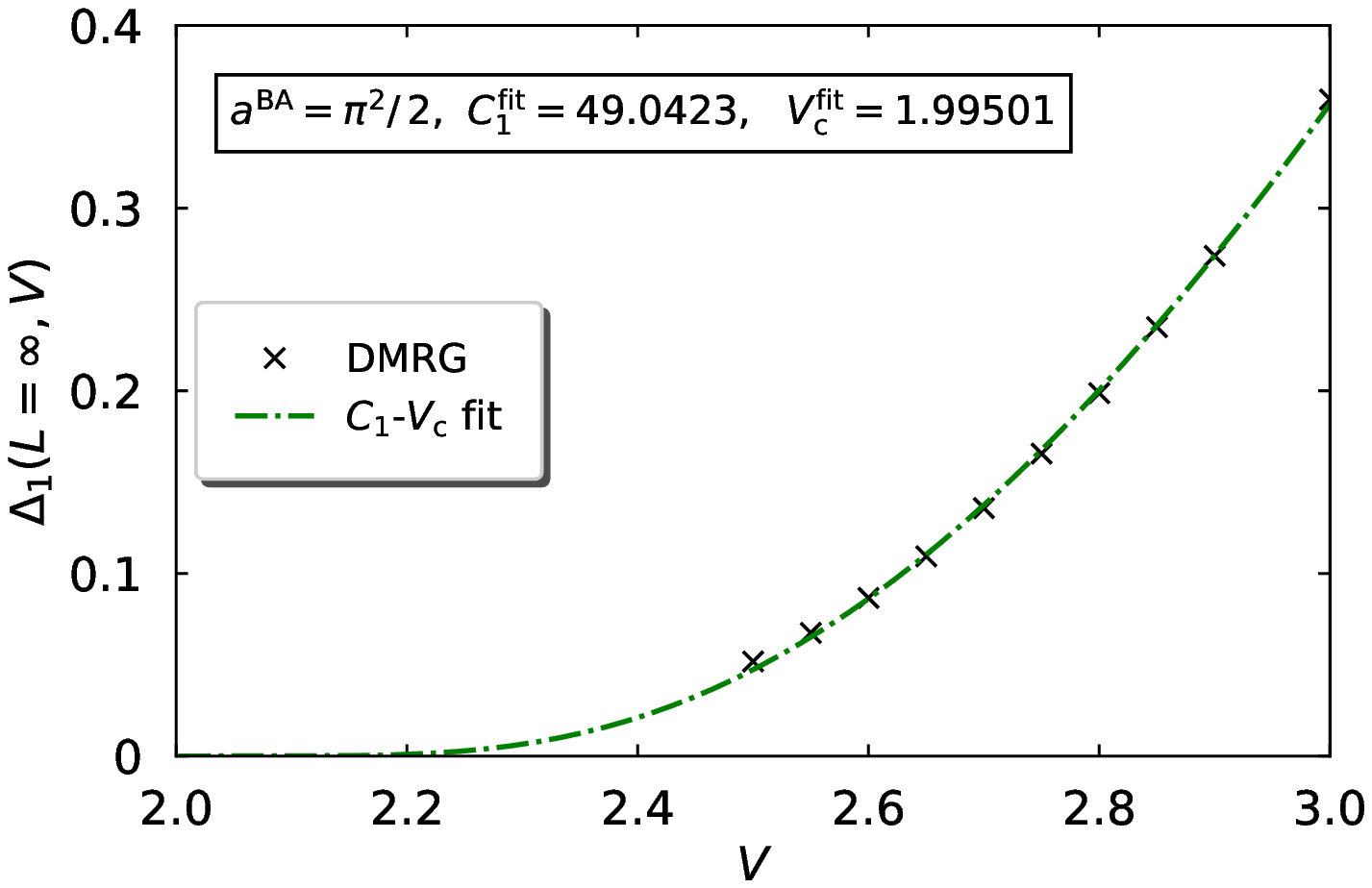}
\end{tabular}%\vspace*{8cm}
\end{flushleft}
\caption{DMRG single-particle gap $\Delta_1(V)$ 
  as a function of the nearest-neighbor interaction~$V$ extrapolated from
  DMRG data (polynomial fit), in comparison with
  (a) a three-parameter fit; (b) a two-parameter fit with fixed $V_{\rm c}=2$;
  (c) a two-parameter fit with fixed $C_1=16\pi$;
  (d) a two-parameter fit with fixed
  $a=\pi^2/2$.\label{fig:Delta1extrapolations}}
\end{figure*}

\subsection{Extrapolation of the single-particle gap}
\label{Gap-extrapolation}

In the exact solution, the gap is exponentially small close to the transition,
\begin{equation}
  \Delta_1(V\gtrsim 2)= 16 \pi \exp\left(-\frac{\pi^2/2}{\sqrt{V-2}}\right)
\end{equation}
with an accuracy of better than seven percent for $V\lesssim 3$.
Therefore, it is natural to fit the 
DMRG gaps, extracted from a second-order polynomial fit for the gap data
for $L=64,128,256,512$, to the expression
\begin{equation}
  \Delta_1(a,C_1,V_{\rm c})= C_1 \exp\left(-\frac{a}{\sqrt{V-V_{\rm c}}}\right)
  \label{supp:threeparametergapfit}
\end{equation}
with $C_1$, $a$, and $V_{\rm c}$ as fit parameters,
where
$a=\pi^2/2\approx 4.93$, $C_1=16\pi \approx 50.3$ and $V_{\rm c}=2$
are the exact values.

The application of the fit formula~(\ref{supp:threeparametergapfit})
generates several problems.
First, for which parameters~$V$ can the fit formula be applied?
If $V$ is too large, we leave the asymptotic range; if $V$ is too small,
the error in the extrapolated values from DMRG are too large.
Below, we choose $2.5 \leq V\leq 3$ to obtain meaningful results.

Second, an unbiased three-parameter fit leads to unsatisfactory results,
as seen in Fig.~\ref{fig:Delta1extrapolations}(a).
The DMRG gaps for $2.5\leq V \leq 3$ are faithfully reproduced
but $C_1=136$, $a=6.36$, and $V_{\rm c}=1.85$ are far off the exact results.
For a stable result, either $C_1$ or $a$ has to be close to its exact value.
Using the guess $V_{\rm c}=2$, we obtain $C_1=48$ and $a=4.9$ as optimal values,
quite close to the exact values, see Fig.~\ref{fig:Delta1extrapolations}(b).
The two-parameter fit with the choice $C_1=16\pi$ leads to
$a=4.97$ and $V_{\rm c}=1.99$, see Fig.~\ref{fig:Delta1extrapolations}(c),
and for fixed $a=\pi^2/2$ we obtain
$C_1=49$ and $V_{\rm c}=1.995$, and the values for all three parameters are
very satisfactory. Note, however, that the least-square fit remains best
for the three-parameter fit, case~(a).

It is seen that the fitting of the gap requires a lot of a-priori
information to obtain
a good estimate for the critical interaction.
Apparently, this is not a viable approach.

\subsection{Scaling approach to the single-particle gap}
\label{Rigol-extrapolation}

To overcome the above limitations,
Mishra, Carrasquilla, and Rigol
designed a scaling approach~\cite{PhysRevB.84.115135APP}
to extract $V_{\rm c}$ from the finite-size data so that
an unspecific polynomial extrapolation is avoided and all raw data
are included in the fit for $V_{\rm c}$.

The finite-size data are supposed to lie on a scaling curve,
\begin{equation}
  L \Delta_1(L,V) \left(1+ \frac{1}{2\ln(L)+ C}\right)
  = F(L/\xi) \; , 
  \label{eq:scalingassumption-supp}
\end{equation}
where $\xi$ is the correlation length, and $F(x)$ is 
the unknown scaling function.
Here, $C$ is a number that is independent of $L$ and $V$
in the critical regime. The functional form on the left-hand-side of
eq.~(\ref{eq:scalingassumption-supp}) is motivated from
the scaling behavior of the conductivity in two-dimensional systems
with a Kosterlitz/Thouless
transition at finite temperatures.~\cite{PhysRevB.84.115135APP}

Close to the transition, 
the correlation length on the insulating side
should diverge proportional to the inverse gap,
\begin{equation}
  \ln \xi = \frac{a}{\sqrt{V-V_{\rm c}}} \sim -\ln(\Delta) \;.
  \label{suppeq:xiandDelta}
\end{equation}
Therefore, it is advisable to employ not $L/\xi$ but $x_L= \ln(L/\xi)=
\ln(L)-a/\sqrt{V-V_{\rm c}}$ as variable in the scaling function,
i.e., $F(L/\xi)\equiv f(x_L)$ in eq.~(\ref{eq:scalingassumption-supp}).
To make progress, it must be assumed that in the region
of accessible values for $L$ and $V$, the scaling function $f(x_L)$
can reliably be represented by a polynomial 
of order~$R$
\begin{equation}
f(x_L) \approx \sum_{r= 0}^R \alpha_r (x_L)^r \; ,
\end{equation}
where the coefficients $\alpha_r$ remain to be determined.
Here, $R$ should be chosen such that the absolute value of the
real coefficient $\alpha_r\sim 10^{-r}$ is small.
Below, we shall use $R= 5$.

The remaining task is to find the optimal fit of the numerical data to
the scaling form, i.e., we minimize the penalty
function~\cite{PhysRevA.87.043606APP}
\begin{eqnarray}
  S(a,C,v_{\rm c},\left\{\alpha_r\right\}) &=& \sum_{i,j}\biggl[
    L_i\Delta_1(L_i,V_j)
    \Bigl[1+ \frac{1}{2\ln(L_i)+ C}\Bigr] \nonumber \\
    && %\hphantom{\sum_{i,j}\biggl[}
    - \sum_{r= 0}^R \alpha_r \Bigl(\ln(L_i)-\frac{a}{\sqrt{V_j-V_{\rm c}}}\Bigr)^r
    \biggr]^2\nonumber \\
  \label{suppeq:penaltyfunction}
\end{eqnarray}
with respect to the Taylor coefficients $\alpha_r$ and the
parameters $a,C,V_{\rm c}$.
Here, we include the gap data $\Delta_1(L_i,V_j)$
for the system sizes $L_i\in \{ 128,256,512\}$
and interaction strengths $V_j \in \{ 2.15, 2.2, 2.25,2.3,2.35,2.4,2.45,2.5\}$.
As seen from eq.~(\ref{suppeq:penaltyfunction}),
we cannot select values $V_i$ that are too close to the transition
because,  due to the factor $1/\sqrt{V_i-V_{\rm c}}$,
the range of values for $x_L$ becomes too large for
the series expansion. 

The minimization task posed by eq.~(\ref{suppeq:penaltyfunction})
is non-trivial. It is done in two steps.
First, for fixed $(C,a,V_{\rm c})$, the minimization implies
a least-square fit of the data to the scaling function.
This quadratic problem is readily solved by using {\sc FindMinimum}
of {\sc Mathematica}.\cite{Mathematica12APP}.

In the second step, $(C,a,V_{\rm c})$ are optimized. This can either be done
by using a dense grid of values for 
$(C,a,V_{\rm c})$ around some educated guess
$(C,a,V_{\rm c})_0$, and picking the lowest values for
the penalty function to determine the optimal
set of parameters.~\cite{PhysRevA.87.043606APP}
Alternatively, we apply a sequence of optimizations by first
optimizing $C$ for given $(a,V_{\rm c})$, then we optimize~$a$ while
re-optimizing $C$ whenever $a$ changes. As soon as $(C,a)$ are optimal
for the given $V_{\rm c}$, we optimize $V_c$ for this set.
The procedure is repeated with the optimization of
$(C,a)$ for the new value for $V_{\rm c}$ and terminates when
the penalty function changes by less than $10^{-5}$.
The final result does not depend on the sequence of optimizations.
The results agree with those obtained from
the dense-grid method.

\begin{figure}[b]
\begin{center}
  \includegraphics[width=8cm]{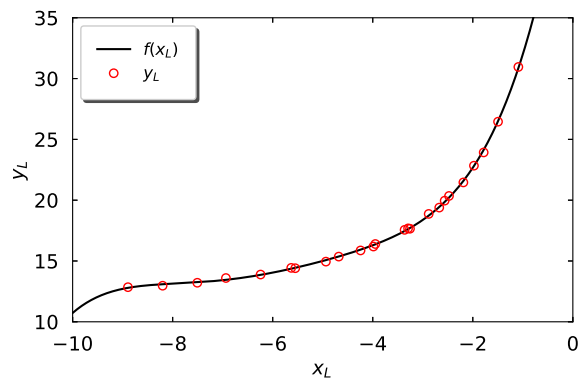}
  %\vspace*{6cm}
\end{center}
\caption{Finite-size data for the single-particle gap, plotted
  as  $y_L(L_i,V_j)=L_i \Delta_1(L_i,V_j) \left[1+ 1/(2\ln(L_i)+ C)\right]$
  versus the scaling parameter $x_L= \ln(L_i)-a/\sqrt{V_j-V_{\rm c}}$,
  together with the corresponding
  scaling function $f(x_L)$. For the chosen parameter sets,
  see text.\label{fig:scalingcurve}}
\end{figure}

The best result is $S= 0.226466$ for the parameter set $C=-15.758$,
$a= 5.12388$, and $V_{\rm c}= 2.01105$. The data and the scaling curve
are shown in Fig.~\ref{fig:scalingcurve}.
Apparently, a nice agreement with the exact data, $a= \pi^2/2\approx 4.93$,
and $V_{\rm c}= 2$ is obtained. The agreement is slightly worse,
$S= 0.413908$ for the parameter set $C=-15.682$,
$a= 5.1743$, and $V_{\rm c}= 2.02$ when the data for $L= 64$ are included.
This signals that the inclusion of larger system sizes systematically lead to
better results for the critical interaction strength.

The degree~$R$ of the polynomial is a more sensitive parameter.
When we choose $R= 4$ instead of $R= 5$, we obtain
$S= 0.623039$ for the parameter set $C=-15.9479$, 
$a= 6.332$, and $V_{\rm c}= 1.93$. At the same time,
$R= 6$ leads to
$S=0.0779472$ for the parameter set $C=-15.588$,
$a= 4.5168$, and $V_{\rm c}= 2.06$.
Apparently, the least-square fit needs to be stabilized by providing
more data points, i.e., DMRG data for more
system sizes and interaction strengths ($L_i,V_j)$.

More problematic than the selection of~$R$ is the sign of the constant~$C$.
Since it is negative, there is a system size $L_C= \exp(-C/2)\approx 2600$
where the assumption of a logarithmic correction in
eq.~(\ref{eq:scalingassumption-supp}) becomes meaningless.
Apparently, the logarithmic term is needed
to stabilize the extrapolation scheme for $L\ll L_C$. Indeed, without
the logarithmic term, the critical interaction would be far off the
exact value, $V_{\rm c}= 1.6$. We also found that the factor $L$
in front of $\Delta_1(L,V)$ in eq.~(\ref{eq:scalingassumption-supp})
is essential to obtain meaningful values for $V_{\rm c}$.
Moreover, the dependence of the correlation length
on $1/\sqrt{V-V_{\rm c}}$ in eq.~(\ref{suppeq:xiandDelta}) is taken for granted. 
We come to the conclusions that
the scaling approach cannot be used as a black box tool
to derive $V_{\rm c}$ with good accuracy.

\section{Series expansions for the exact ground-state energy}
\label{app:gsseries}

In this appendix, we give the Taylor series expansions
for weak and strong interactions, and asymptotic series expansions around
the critical interaction strength for the ground-state energy at
half band-filling.
To this end, we use the original expressions by Yang and Yang.
A more sophisticated
approach employs special functions.~\cite{GoehmannprivatecommAPP}

\subsection{Approximation for weak interactions}

For weak interactions, we use the representation of $g(\mu)$
given by Yang and Yang,~\cite{PhysRev.150.321APP,PhysRev.150.327APP}
\begin{eqnarray}
  g(\mu)&=& \frac{\cos(\mu)}{4}-\frac{2 \sin(\mu)}{\mu}Y(\mu)\nonumber \; , \\
  Y(\mu)&= & \ln(2) -2 \int_0^{\infty} {\rm d} y
  \frac{\tanh(y)}{\exp(2\pi y/\mu)-1} \; .
\end{eqnarray}
We introduce $\gamma= \pi/2-\mu$ so that
\begin{equation}
  \sin(\gamma) = \frac{V}{2} \; ,
  \label{eq:gammaandV}
\end{equation}
and write
\begin{eqnarray}
  Y(\gamma)&= & \ln(2) -  \left(\frac{1}{2}-\frac{\gamma}{\pi}\right)
  Z(\gamma) \;, \nonumber \\[3pt]
  Z(\gamma) &= & \int_0^{\infty} {\rm d}u
  \frac{\tanh[u/4-\gamma u/(2\pi)]}{e^u-1} \; .
\end{eqnarray}
Using {\sc Mathematica},~\cite{Mathematica12APP}
the series expansion for $Z(\gamma)$ around $\gamma= 0$ reads
\begin{eqnarray}
  Z(\gamma)&= & 2\ln(2)-1 -\left( \frac{1-4\ln(2)}{\pi}
  + \frac{\pi}{4} \right)\gamma \nonumber \\[3pt]
&&  -\frac{4 (3 + \pi^2 - 18 \ln(2))}{9\pi^2}\gamma^2 \nonumber \\[3pt]
  &&\frac{-48 - 32 \pi^2 + \pi^4 + 384 \ln(2)}{24 \pi^3} \gamma^3
  \nonumber \\[3pt]
&&  + \frac{16 (-135 - 150 \pi^2 + 7 \pi^4 + 1350 \ln(2))}{675 \pi^4}\gamma^4
  \nonumber \\[3pt]
%  &&+  \frac{-2400 \pi^2 + 224 \pi^4 - 9 \pi^6 + 
%    1440 (-1 + 12\ln(2))}{270 \pi^5}\gamma^5\nonumber \\[3pt]
%  &&-\frac{64 (6615 \pi^2 - 1029 \pi^4 + 62 \pi^6 + 
%    2835 (1 - 14 \ln(2)))}{19845 \pi^6}  \gamma^6\nonumber \\[3pt]
%  &&+ \frac{-725760 - 2257920 \pi^2 + 526848 \pi^4 - 63488 \pi^6 + 
%    2601 \pi^8 + 11612160 \ln(2)}{45360 \pi^7}  \gamma^7 \nonumber \\[3pt]
%&& + \frac{256 (-94500 \pi^2 + 30870 \pi^4 - 6200 \pi^6 + 381 \pi^8 + 
%          23625 (-1 + 18 \ln(2))}{212625 \pi^8}\gamma^8
%  \nonumber \\[3pt]
%  &&+ \left(
  %  -\frac{256}{\pi^7} + \frac{25088}{225 \pi^5} - \frac{31744}{945 \pi^3}
  %+ \frac{32512}{7875 \pi}
%  - \frac{961 \pi}{5670} + \frac{256 (-1 + 20 \ln(2))}{5 \pi^9}\right) \gamma^9
%  \nonumber \\[3pt]
%  &&+ \left(-\frac{149504}{88209} - \frac{1024}{11 \pi^{10}} - \frac{5120}{9 \pi^8}
%  + \frac{14336}{ 45 \pi^6} - \frac{126976}{945 \pi^4} + \frac{130048}{4725 \pi^2} +
%  \frac{2048 \ln(2)}{\pi^{10}}\right)\gamma^{10} \nonumber \\[3pt]
&& +     {\cal O}\left(\gamma^{5}\right) \; .
\end{eqnarray}
Collecting all terms and using eq.~(\ref{eq:gammaandV}),
{\sc Mathematica}~\cite{Mathematica12APP} gives
\begin{eqnarray}
  e_0(V,N= L/ 2)&= & -\frac{2}{\pi} + \left(\frac{1}{4}-\frac{1}{\pi^2}\right)V
  \nonumber \\[3pt]
  &&+ \left(-\frac{2}{3 \pi^3} + \frac{1}{36 \pi}\right)V^2
  \nonumber \\[3pt]
  && + \left(\frac{1}{32} - \frac{1}{2 \pi^4} - \frac{1}{4 \pi^2}\right)  V^3
\nonumber \\[3pt]
&& + \left(-\frac{2}{5 \pi^5} - \frac{5}{12 \pi^3}
+ \frac{73}{1600 \pi}\right)V^4
\nonumber \\[3pt]
%+ \left(-\frac{1}{3 \pi^6} - \frac{5}{9 \pi^4} + \frac{13}{216 \pi^2}\right)V^5
%\nonumber \\[3pt]
%&& + \left(-\frac{2}{7 \pi^7} - \frac{41}{60 \pi^5} + \frac{259}{2880 \pi^3}
%- \frac{10439}{ 5644800 \pi}\right)V^6\nonumber \\[3pt]
%&& + \left(\frac{1}{1024} - \frac{1}{4 \pi^8} - \frac{29}{36 \pi^6} + \frac{593}{4320 \pi^4}
%- \frac{97}{ 6480 \pi^2}\right)V^7
%\nonumber \\[3pt]
%&& + \left(-\frac{2}{9 \pi^9} - \frac{233}{252 \pi^7} + \frac{1787}{8640} \pi^5
%- \frac{194939}{4354560 \pi^3} + \frac{1838239}{541900800 \pi}\right)V^8 \nonumber \\[3pt]
%&& + \left(-\frac{1}{2048} - \frac{1}{5 \pi^{10}} - \frac{25}{24 \pi^8}
%+ \frac{1093}{3600 \pi^6} - \frac{701}{6720 \pi^4} + \frac{10117}{756000 \pi^2}
%\right)V^9
%\nonumber \\[3pt]
%&& + \left(-\frac{2}{11 \pi^{11}} - \frac{125}{108 \pi^9} +
%\frac{26171}{60480 \pi^7} - \frac{61489}{290304 \pi^5} +
%\frac{91021513}{2090188800 \pi^3}\right)V^{10} \nonumber\\[3pt]
%&&- \frac{269070671}{104911994880 \pi} V^{10}
&& + {\cal O}\left(V^{5}\right) \; .
\end{eqnarray}
The accuracy of the Taylor series is better than one percent
for $0\leq V\leq 2$.

\subsection{Approximation for large interactions}

For $V>2$ we write the ground-state energy for spinless fermions
at half band-filling in the form
\begin{equation}
  e_0(V) = e^{-\lambda}-4\sinh(\lambda)\sum_{n= 1}^{\infty}
  \frac{1}{1+ \exp(2\lambda n)}
\end{equation}
with $\lambda= \lambda(V)=\arccosh(V/2)$.
  
For large~$V$, only a limited number of terms in the sum needs to be included.
Using {\sc Mathematica}~\cite{Mathematica12APP}
then gives the large-$V$ asymptotics
\begin{eqnarray}
  e_0(V>2)&=&
  -V^{-1} + V^{-3}-V^{-7}  -2V^{-9} \nonumber \\
  && -2 V^{-11} + 2V^{-13} + 17V^{-15} \nonumber \\
  && + 56 V^{-17} + 134 V^{-19}  \nonumber\\
  &&+ {\cal O}\left(  V^{-21}\right) \; .
\end{eqnarray}
The accuracy of the approximation is $10^{-4}$ for $V\geq 2$.

\subsection{Approximation close to the transition}

Around the critical interaction $V_{\rm c}= 2$, the ground-state energy
can be expressed in terms
of an asymptotic series in the metallic and in the insulating region,
respectively.

\subsubsection{Metallic region}

We start from eq.~(16) in the main text.
Yang and Yang~\cite{PhysRev.150.321APP,PhysRev.150.327APP}
provide an asymptotic series around $\mu= 0$.
We can either follow Yang and Yang who give an analytic expression for the
series coefficients,
or we obtain the lowest-order coefficients from a Taylor series
of the integrand in $g(\mu)$. The resulting
integrals can be done with {\sc Mathematica},~\cite{Mathematica12APP}
apart from the leading order,
\begin{equation}
\int_{-\infty}^{\infty} {\rm d}x\frac{1}{(1+ 4x^2)\cosh(\pi x)} = \ln(2)  \; ,
\end{equation}
that requires a contour integral
for its evaluation.~\cite{PhysRev.150.321APP,PhysRev.150.327APP}
We find
\begin{eqnarray}
  g(\mu) &=&  \frac{1}{2}-2\ln(2)+
  \frac{1}{12}\left(4\ln(2)-1\right)\mu^2\nonumber \\
&&  - \frac{1}{240}\left(3+ 4\ln(2)\right)\mu^4 \nonumber \\
  && + \frac{1}{10080}\left(27 + 4\ln(2)\right)\mu^6
  %\nonumber \\
%  -\frac{1}{725760}\left(503 +  4\ln(2)\right) \mu^8 \nonumber \\
  %  && + \frac{1}{79833600}\left(33255 + 4\ln(2)\right) \mu^{10}
  + {\cal O}\left(\mu^{8}\right)
    \; .
\end{eqnarray}
Thus, for $V\lesssim 2$,
we obtain for the ground-state energy of spinless fermions
up to second order 
\begin{eqnarray}
  e_0(V) &= & 
  1- 2 \ln(2) \nonumber \\
  && + \frac{1}{3}\left(1 - \ln(2)\right) \left(V-2\right) \nonumber \\
  &&
  + \frac{1}{360} \left(-7 + 4 \ln(2)\right) \left(V - 2\right)^2
  \nonumber \\
 && + {\cal O}\left((V-2)^3\right)
  \; .
\end{eqnarray}
The approximation is better than two per mille for $0 \leq V\leq 2$.
The next order reduces the accuracy down to five per mille for $0 \leq V\leq 2$.
Recall that the series is only asymptotic.

For comparison, we give the asymptotic result for the ground-state energy up
to fifth order,
\begin{eqnarray}
  e_0(V) &= & 
  1- 2 \ln(2) \nonumber \\[3pt]
  && + \frac{1 - \ln(2)}{3}\left(V-2\right) \nonumber \\[3pt]
  &&  + \frac{4 \ln(2)-7}{360} \left(V - 2\right)^2\nonumber \\[3pt]
  && \frac{1 - 4 \ln(2)}{3024}  \left(V-2\right)^3\nonumber \\[3pt]
  &&+ \frac{184 \ln(2)-487}{907200} \left(V - 2\right)^4  \nonumber \\[3pt]
  && - \frac{2104 \ln(2)+ 14153}{59875200} \left(V-2\right)^5  \nonumber \\[3pt]
 &&  + {\cal O}\left((V-2)^6\right)
  \; . \label{eq:asymptoticmetallic}
\end{eqnarray}
We shall show below
that it agrees with the corresponding expression in the insulating region.

\subsubsection{Insulating region}
  
We start from eq.~(16) of the main text.
First, we write 
\begin{eqnarray}
  S(\lambda)&\equiv&  \sum_{n= 1}^{\infty} \frac{1}{1+\exp(2\lambda n)}
  \nonumber \\
  &=& \sum_{n= 1}^{\infty} \sum_{r= 0}^{\infty}
  (-1)^r   e^{-2\lambda(r+ 1)n} \nonumber \\
  &=& - \sum_{r,n= 1}^{\infty}(-1)^r e^{-2\lambda rn}\nonumber \\
  &=& \sum_{r= 1}^{\infty}  (-1)^r \frac{1}{1-\exp(2\lambda r)} \; .
  \label{eq:Sdefandfirst}
\end{eqnarray}
The rearrangement of the sum shows directly that, in the limit $\lambda\to 0$,
\begin{equation}
  S(\lambda\to 0) = -\frac{1}{2\lambda}\sum_{r= 1}^{\infty}(-1)^r\frac{1}{r}
  = \frac{\ln(2)} {2\lambda}\; .
  \label{eq:Sisoforderlambdaminusone}
  \end{equation}
Therefore,
\begin{equation}
h(0)= \frac{1}{2}-4\lim_{\lambda\to 0}\lambda S(\lambda)= \frac{1}{2}-2\ln(2)
\end{equation}
in eq.~(18) of the main text, as it must be
because the ground-state energy is continuous at $V= 2$.

To make further progress, we define
\begin{eqnarray}
  f_1(\lambda,m) &=& \frac{1}{\exp[2\lambda(2m)]-1}-\frac{1}{2\lambda(2m)}
  \nonumber \; , \\
  f_2(\lambda,m) &=& \frac{1}{\exp[2\lambda(2m-1)]-1}-\frac{1}{2\lambda(2m-1)}
  \nonumber \; , \\
\end{eqnarray}
so that
\begin{eqnarray}
  S(\lambda)&=& \sum_{m=1}^{\infty} \left(f_2(\lambda,m)-f_1(\lambda, m)\right)
  \nonumber \\
  && +\sum_{m=1}^{\infty}  \left( \frac{1}{2\lambda(2m-1)}-\frac{1}{2\lambda(2m)}
  \right) \nonumber \\
  &=& \sum_{m=1}^{\infty} \left(f_2(\lambda,m)-f_1(\lambda, m)\right)
  + \frac{\ln(2)}{2\lambda}\; .
\end{eqnarray}
For the sum we invoke the Euler-Maclaurin sum formula to construct
an asymptotic series expansion
for small~$\lambda$,
\begin{eqnarray}
  \sum_{m=1}^{M} F(m) &\sim& \int_{1}^{\infty}{\rm d} x F(x)
  +\frac{1}{2}\left(F(M)+F(1)\right) \nonumber \\
  &&  +\sum_k \frac{B_{2k}}{(2k)!} \left(F^{(2k-1)}(M)-F^{(2k-1)}(1)\right) \; .
  \nonumber \\
\end{eqnarray}
We must calculate the following three terms. The integral terms reads
\begin{eqnarray}
  T_1(\lambda)&=& \int_1^{\infty}{\rm d}x  f_2(\lambda,x)-\int_1^{\infty}{\rm d}x
  f_1(\lambda,x)
  \nonumber \\
  &=& \int_{1/2}^{\infty}{\rm d}x f_1(\lambda,x) -\int_1^{\infty}{\rm d}x
  f_1(\lambda,x) \nonumber \\
  &=& \frac{1}{4\lambda} \int_{2\lambda}^{4\lambda} {\rm d} u
  \left( \frac{1}{e^u-1}-\frac{1}{u}\right)
  \nonumber \\
  &=& \frac{\ln[\cosh(\lambda)]}{4\lambda}-\frac{1}{4} \; .
\end{eqnarray}
The second term is
\begin{equation}
  T_2(\lambda)= \frac{1}{2}\left( f_2(\lambda,1)-f_1(\lambda,1)\right)
  = \frac{1}{4\sinh(2\lambda)}-\frac{1}{8\lambda} \; .
\end{equation}
The third term involves the derivatives of $f_2(\lambda,x)-f_1(\lambda,x)$
at $x=1$ that we expand
in a Taylor series around $\lambda=0$. We note that 
\begin{equation}
  \left.  \left(\frac{\partial}{\partial x}\right)^{2k-1}
  \left(f_2(\lambda,x)-f_1(\lambda,x)\right)\right|_{x=1}
\sim \lambda^{2k+1}  \; ,
\end{equation}
so that in the remaining series over the derivatives we have
\begin{equation}
  T_3(\lambda) + T_3^{(K)}(\lambda) = {\cal O}\left(\lambda^{2K+3}\right)
\end{equation}
with
%\begin{eqnarray}
\begin{equation}
%  T_3(\lambda) + T_3^{(K)}(\lambda) &=& {\cal O}\left(\lambda^{2K+3}\right) \,
%  \nonumber \\
  T_3^{(K)}(\lambda)  =    \sum_{k=1}^K \frac{B_{2k}}{(2k)!}\left(
  f_2^{(2k-1)}(\lambda,1)-f_1^{(2k-1)}(\lambda,1)   \right) \; .
%  \nonumber \\
  %\end{eqnarray}
  \end{equation}
In fact, {\sc Mathematica} gives
\begin{equation}
  T_3^{(4)}(\lambda)  =    -\frac{\lambda^3}{60} +\frac{\lambda^5}{90}
  -\frac{\lambda^7}{6300}
  +\frac{101\lambda^9}{44550}
\end{equation}
for $\lambda \ll 1$.
The series expansion for $S(\lambda)$ thus becomes
\begin{eqnarray}
  S(\lambda) &=& \frac{\ln(2)}{2\lambda}
  + T_1(\lambda)+T_2(\lambda)+T_3(\lambda) \nonumber \\
  &\sim & \frac{\ln(2)}{2\lambda} -\frac{1}{4} +\frac{\lambda}{24}
  +\frac{\lambda^3}{720}
  +\frac{\lambda^5}{3780}  +\frac{17 \lambda^7}{151200}\nonumber \\
  &&+\frac{31 \lambda^9}{374220}
  +{\cal O}\left(\lambda^{11}\right)\; .
\end{eqnarray}
The approximation is better than $10^{-4}$ for $0\leq \lambda \leq 0.9$.

Thus, the expansion of the ground-state energy
close to the transition in the insulating phase becomes for $V\gtrsim 2$
\begin{eqnarray}
  e_0(V) &=&  
  1- 2 \ln(2)\nonumber \\[3pt]
  && + \frac{1-\ln(2)}{3}  \left(V-2\right)\nonumber \\[3pt]
  && + \frac{4\ln(2)-7}{360}  \left(V-2\right)^2 \nonumber \\[3pt]
 &&  +\frac{1-4\ln(2)}{3024} \left(V-2\right)^3 \nonumber \\[3pt]
 && +\frac{184\ln(2)-487}{907200} \left(V-2\right)^4 \nonumber \\[3pt]
  && -\frac{2104\ln(2)+14153}{59875200} \left(V-2\right)^5  \nonumber \\[3pt]
  && + {\cal O}\left((V-2)^6\right)
  \; .
  \label{eq:asymptoticinsulating}
\end{eqnarray}
The approximation is better than $10^{-3}$ only for $2\leq V\leq 3$, i.e.,
the approximation is not particularly useful.

Note that the asymptotic series in the insulating regime,
eq.~(\ref{eq:asymptoticinsulating})
for $V\geq 2$, agrees with the asymptotic expansion
in the metallic regime, eq.~(\ref{eq:asymptoticmetallic}) for $0<V\leq 2$.
This does not come as a surprise because the ground-state energy
is continuously
differentiable at $V= 2$ to all orders.

\section{Series expansion for the exact single-particle gap}
\label{app:gapexpansion}

In this appendix we provide formulae
for the single-particle gap in the strong-coupling limit
and close to the transition
where it opens exponentially.
A more sophisticated
approach employs Jacobi $\theta$-functions.~\cite{GoehmannprivatecommAPP}

\subsection{Strong coupling}
We rewrite eq.~(32) of the main text as
\begin{equation}
  \frac{V}{2} = \frac{1}{2} \left(e^{r(m)}+e^{-r(m)}\right)
\end{equation}
with
\begin{equation}
r(m)=\frac{\pi K(1-m)}{K(m)} \; .
\end{equation}
For large~$V$, we have $m\to 0$ so that the Taylor series gives
\begin{eqnarray}
  \frac{V}{2} &=& -4 + \frac{8}{m} - \frac{19 m}{32} - \frac{19 m^2}{64}
  - \frac{767 m^3}{4096} -
  \frac{ 1085 m^4}{8192} \nonumber \\
  && 
  - \frac{105227 m^5}{1048576} - \frac{166967 m^6}{2097152}
  - \frac{35047971 m^7}{536870912}\nonumber \\[3pt]
  &&    + {\cal O}\left(m^8\right) \; .
  \label{eq:solveform}
\end{eqnarray}
We set $x= mV$ and use the expansion
\begin{equation}
x(V)=\sum_{r=0}^7 a_r \frac{1}{V^r}
\end{equation}
to solve eq.~(\ref{eq:solveform}) iteratively.
The result is
\begin{eqnarray}
  m(V) &=&
  \frac{16}{V} - \frac{128}{V^2} + \frac{720}{V^3}
  - \frac{3328}{V^4} + \frac{13632}{V^5}
  - \frac{51328}{V^6} \nonumber \\
  && + \frac{181488}{V^7}- \frac{611072}{V^8} +   \frac{1977904}{V^9}
  \nonumber \\[3pt]
&& +{\cal O}\left( V^{-10}\right)\; .
\end{eqnarray}
We insert this solution in eq.~(34) of the main text and
let {\sc Mathematica}~\cite{Mathematica12APP}
do the expansion,
\begin{eqnarray}
  \Delta_1(V\gg  2) &=&  2 V - 8 + \frac{4}{V} + \frac{8}{V^2}
  + \frac{4}{V^3} - \frac{8}{V^6} - \frac{12}{V^7} \nonumber \\
&&  +{\cal O}\left(V^{-8}\right)\; .
\end{eqnarray}
The accuracy of the approximation is better than
$4 \cdot 10^{-4}$ for $V\geq 4$.
The accuracy rapidly deteriorates below $V=3$
where the deviation is about two percent.

\subsection{Transition region}

Close to the transition, we have $m\to 1$ and thus we must solve 
 \begin{equation}
\gamma \approx -\frac{\pi^2}{\ln\left((1-m)/16\right)} 
 \end{equation}
 in eq.~(32) of the main text.
 Therefore, $m$ is exponentially close to unity for $V\to 2^+$,
 \begin{equation}
   m(V) \approx  1-16 \exp\left(-\frac{\pi^2}{\gamma}\right)
   \approx 1-16 \exp\left(-\frac{\pi^2}{\sqrt{V-2}}\right)\; .
 \end{equation}
 Using the series expansion of the elliptic integral we finally
 find for the single-particle gap
 \begin{equation}
   \Delta_1(V\to 2^+) \to 16 \pi \exp\left(-\frac{\pi^2}{2\sqrt{V-2}}\right)
   \; . \label{eq:gapsmallnotbrilliant}
 \end{equation}
 Eq.~(\ref{eq:gapsmallnotbrilliant}) shows that the gap becomes exponential
 small close to the critical interaction.

 The accuracy of the approximate result~(\ref{eq:gapsmallnotbrilliant})
 is not impressive because it is only better than seven percent
 for $2\leq V\leq 3$.
A better approximation is given by
\begin{eqnarray}   
  m(V)&\approx& 1-16 \exp\left(-\frac{\pi^2}{\arccosh(V/2)}\right)
  \nonumber \\
  && +128 \exp\left(-\frac{2\pi^2}{\arccosh(V/2)}\right)
  \; , \nonumber\\
  \Delta_1(V\gtrsim 2) &\approx & \frac{8}{\pi}
  \sqrt{\left[\left(V/2\right)^2-1\right]\left[1-m(V)\right]}
  \nonumber \\
  && \times   \biggl[
    -\frac{1}{2} \ln\left[\frac{1-m(V)}{16}\right]
    \nonumber \\
    && \hphantom{\times   \biggl[}
    +\frac{m(V)-1}{8}\left(2+\ln\left[\frac{1-m(V)}{16}\right] \right)
    \biggr]\; .\nonumber \\
  \label{eq:usethisforsmallV}
\end{eqnarray}
The accuracy of the approximation~(\ref{eq:usethisforsmallV}) is better
than $10^{-5}$ for $2\leq V\leq 4$.
It remains better than one per mille for $2\leq V\leq 6$.

\section{Series expansion for the exact order parameter}
\label{app:orderparameterexpansion}

In this appendix we provide formulae
for the order parameter gap in the strong-coupling limit,
and close to the transition
where it opens exponentially.
A more sophisticated
approach employs Jacobi $\theta$-functions
and their derivatives.~\cite{GoehmannprivatecommAPP}

\subsection{Strong coupling}

For strong coupling we have $q= 2|\Delta|= V$ to leading order,
and only the first term
in the product~(38) of the main text needs to be included,
\begin{eqnarray}
  n_a(V\gg 1) &\approx& \frac{1}{2} \left(1-\frac{2}{q^2}\right)^2\nonumber \\
  &  \approx & \frac{1}{2}\left(1-\frac{4}{V^2}\right)
  = \frac{1}{2}-2 \left(\frac{1}{V}\right)^2
  \;. 
\end{eqnarray}
Using {\sc Mathematica}~\cite{Mathematica12APP}
we can easily obtain the series up to order $V^{-12}$,
\begin{eqnarray}
  n_a(V\gg 1) &\approx&
  \frac{1}{2} -2 \left(\frac{1}{V}\right)^2 -2 \left(\frac{1}{V}\right)^4
  -2 \left(\frac{1}{V}\right)^6 \nonumber \\
  && +2 \left(\frac{1}{V}\right)^8 
  + 24 \left(\frac{1}{V}\right)^{10}+ 114 \left(\frac{1}{V}\right)^{12}
  \nonumber \\
  && + {\cal O}\left(V^{-14}\right) \; .
\end{eqnarray}
The series expansion agrees with the exact expression down to $V= 3$
with an accuracy of better than one half per mille.

\subsection{Transition region}

A rigorous expansion of the product formula in eq.~(38) of the main text
is not trivial. We write
\begin{eqnarray}
s_0(q) &=& \left( \frac{f_1(q)}{f_2(q)}\right)^2 \; ,\nonumber \\
f_1(q) &= & \prod_{m= 1}^{\infty}\left(1-q^{-2m}\right) \; , \nonumber \\
f_2(q) &= & \prod_{m= 1}^{\infty}\left(1+ q^{-2m}\right) \; ,
\end{eqnarray}
and treat the two functions $f_{1,2}(q)$ separately.

\subsubsection{First function}

By definition, $f_1(q)$ can be written as
\begin{equation}
f_1(q) = \left( \frac{1}{q^2};\frac{1}{q^2}\right)_{\infty} \;,
\end{equation}
where $(a;Q)_{\infty}$ is the Q-Pochhammer function.
For $q\to 1^+ $,
Banerjee and Wilkerson provide the result~\cite{BanerjeeWilkersonAPP}
\begin{equation}
  f_1(q) = \sqrt{\frac{\pi}{\ln(q)}}
  \exp\left( -\frac{\pi^2}{12\ln(q)}\right) q^{1/12} \; ,
  \label{eq:f1approx}
\end{equation}
which agrees with the exact numerical value
with an accuracy of $10^{-4}$ up to $q= 8$.

\subsubsection{Second function}

$f_2(q)$ can also be expressed in terms of the Q-Poch\-hammer function,
\begin{equation}
f_2(q) = \frac{1}{2} \left( -1;\frac{1}{q^2}\right)_{\infty} \;.
\end{equation}
Garoufalidis and Zagier provide an approximation for
$(w Q^x;Q)_{\infty}$ for general $Q\to 1^{-}$.\cite{GaroufalidisZagier2021APP}
It is permitted to use their formula for $w= -1$ and $x= 0$ so that
we find
\begin{eqnarray}
  \ln\left[2f_2(q)\right] &=& B_0(0) {\rm li}_2(-1) \frac{1}{\ln(1/q^2)} +
  B_1(0) {\rm li}_1(-1) \nonumber \\
  && +
  \frac{1}{2} B_2(0) {\rm li}_0(-1) \ln(1/q^2) + \ldots \nonumber \\
  &\approx & -\frac{\pi^2}{12\ln(1/q^2)}+\frac{\ln(2)}{2}
  +\frac{1}{12}\Bigl(-\frac{1}{2}\Bigr) \ln\biggl[\frac{1}{q^2}\biggr] ,
  \nonumber \\
\end{eqnarray}
where $B_k(x)$ is the $k$th Bernoulli polynomial and ${\rm li}_k(z)$
is the poly-logarithm of order~$k$.
Therefore, the approximation for $q\to 1^+$ becomes
\begin{eqnarray}
  f_2(q) &\approx& \frac{1}{2}
  \exp\left( \frac{\pi^2}{24\ln(q)}+\frac{\ln(2)}{2} +\frac{\ln(q)}{12}\right)
\nonumber \\
  &=& \frac{\sqrt{2}}{2} q^{1/12}\exp\left(\frac{\pi^2}{24\ln(q)}\right)
  \; .
  \label{eq:f2approx}
\end{eqnarray}
The approximation reproduces the exact result with an accuracy of
$10^{-4}$ up to $q=4$.
  
\subsubsection{Final result close to the transition}

Using eqs.~(\ref{eq:f1approx}) and~(\ref{eq:f2approx}) we find
\begin{equation}
  s_0(q) \approx \frac{2\pi}{\ln(q)} \exp\left(-\frac{\pi^2}{4\ln(q)}\right) \;.
  \label{eq:s0approxsmallfinal}
\end{equation}
The approximation agrees with the exact result with an accuracy of better
than $10^{-3}$ up to $q=3$.

With
\begin{equation}
q(V)=\frac{V}{2} +\sqrt{\left(\frac{V}{2}\right)^2-1}
\end{equation}
we thus have for the order parameter
\begin{equation}
n_a(V) \approx \frac{\pi}{\ln(q(V))} \exp\left(-\frac{\pi^2}{4\ln(q(V))}\right)
\end{equation}
with an accuracy of better than $10^{-4}$ for $2\leq V\leq 3$.
  
\subsubsection{Alternative derivation}

A fast but less rigorous way to derive eq.~(\ref{eq:s0approxsmallfinal})
starts from 
\begin{equation}
  \ln\left(\frac{f_1(q)}{f_2(q)}\right) =
  \sum_{m=1}^{\infty} \ln\left[\tanh(\alpha m)\right] \; ,
\end{equation}
where we set $\alpha=\ln(q)$.
With the help of the Euler-Maclaurin sum formula we find up to order $\alpha$
\begin{eqnarray}
  \ln\left(\frac{f_1(q)}{f_2(q)}\right) &=&
  \int_1^{\infty} {\rm d}x \ln\left[\tanh(\alpha x)\right]
  +\frac{1}{2} \ln\left[\tanh(\alpha)\right] \nonumber \\
  && - \sum_{k=1}^{p} \frac{B_{2k}}{(2k)!}  g^{(2k-1)}(1) +R_{p,k}
  \nonumber \\
  && + {\cal O}\left(\alpha\right)
\end{eqnarray}
with $g(x)=\ln\tanh(\alpha x)$.
For small $\alpha$ we see that $g(x)\approx \ln(\alpha) +\ln(x)$
so that the terms in the second line reduce to the same expression
as obtained for the Stirling formula.
Therefore,
\begin{eqnarray}
  \ln\left(\frac{f_1(q)}{f_2(q)}\right) &\approx&
  \int_1^{\infty} {\rm d}x \ln\left[\tanh(\alpha x)\right]
  +\frac{1}{2} \ln\left[\tanh(\alpha)\right]
  \nonumber \\
  && +\frac{1}{2}\ln(2\pi)-1
  + {\cal O}\left(\alpha\right)\nonumber \\
  &\approx & -\frac{\pi^2}{8\alpha}
  +\frac{1}{2}\ln\left(\frac{2\pi}{\alpha}\right) + {\cal O}\left(\alpha\right)
  \; ,
\end{eqnarray}
in agreement with eq.~(\ref{eq:s0approxsmallfinal}) as $\alpha=\ln(q)$
and $s_0(q)=[f_1(q)/f_2(q)]^2$.
  
\section{Calculation of matrix elements in second-order
  Hartree-Fock approximation}
\label{app:matrixelements}

In this appendix, we calculate the matrix elements required
for second-order perturbation theory.

\subsection{Single-particle excitations}

We have to calculate ($|n_1\rangle\equiv |k;p\rangle$)
\begin{eqnarray}
  \langle 0|  \hat{V}_{\perp} | n_1\rangle &=&
  V \sum_{l=0}^{L-1} \Bigl[ \langle 0 |
    \hat{c}_{l+1}^+\hat{c}_{l+1}^{\vphantom{+}}
    \hat{c}_{l}^+\hat{c}_{l}^{\vphantom{+}}
    \nonumber \\
 &&  \hphantom{\sum_{l=0}^{L-1} \Bigl[}
    - \langle     \hat{c}_{l+1}^+\hat{c}_{l+1}^{\vphantom{+}} \rangle_0
    \hat{c}_{l}^+\hat{c}_{l}^{\vphantom{+}}
    -     \hat{c}_{l+1}^+\hat{c}_{l+1}^{\vphantom{+}} 
    \langle\hat{c}_{l}^+\hat{c}_{l}^{\vphantom{+}}\rangle_0
    \nonumber \\
    && \hphantom{\sum_{l=0}^{L-1} \Bigl[}
      - \hat{c}_{l}^+\hat{c}_{l+1}^{\vphantom{+}}
      \langle        \hat{c}_{l}^{\vphantom{+}}\hat{c}_{l+1}^+\rangle_0
      - \langle \hat{c}_{l}^+\hat{c}_{l+1}^{\vphantom{+}}\rangle_0
      \hat{c}_{l}^{\vphantom{+}}\hat{c}_{l+1}^+\Bigr]\nonumber \\
     &&  \hphantom{V\sum_{l=0}^{L-1} \Bigl[}
    \hat{\beta}_k^+\hat{\alpha}_p^{\vphantom{+}} |0\rangle \; ,
\end{eqnarray}
where we used $\langle 0 | n\rangle=0$. Using Wick's theorem we find
\begin{equation}
  \langle 0|  \hat{V}_{\perp} | n_1\rangle=
  V \sum_{l=0}^{L-1}
  \left|
  \begin{array}{ccc}
    0 & 0 & \langle \hat{c}_{l+1}^+\hat{\alpha}_p^{\vphantom{+}}\rangle_0 \\
    0 & 0 & \langle \hat{c}_{l}^+\hat{\alpha}_p^{\vphantom{+}}\rangle_0 \\
    -\langle \hat{c}_{l+1}^{\vphantom{+}} \hat{\beta}_k^+\rangle_0
    & -\langle \hat{c}_{l}^{\vphantom{+}} \hat{\beta}_k^+\rangle_0 &0 
    \end{array}
  \right|
\end{equation}
because the Hartree-Fock decomposition takes care of all contractions
of the $\hat{c}$-operators with themselves, and $\langle 
\hat{\beta}_k^+\hat{\alpha}_p^{\vphantom{+}}\rangle_0=0$.
The determinant vanishes so that there is no contribution
from single-particle excitations to second order.

\subsection{Two-particle excitations}

We have to calculate ($|n_2\rangle\equiv |k_1,k_2;p_1,p_2\rangle$)
\begin{eqnarray}
  \langle 0|  \hat{V}_{\perp} | n_2\rangle &=&
  V \sum_{l=0}^{L-1} \Bigl[ \langle 0 |
    \hat{c}_{l+1}^+\hat{c}_{l+1}^{\vphantom{+}}
    \hat{c}_{l}^+\hat{c}_{l}^{\vphantom{+}}
    \nonumber \\
    && \hphantom{\sum_{l=0}^{L-1} \Bigl[}
      - \langle     \hat{c}_{l+1}^+\hat{c}_{l+1}^{\vphantom{+}} \rangle_0
    \hat{c}_{l}^+\hat{c}_{l}^{\vphantom{+}}
    -     \hat{c}_{l+1}^+\hat{c}_{l+1}^{\vphantom{+}} 
    \langle\hat{c}_{l}^+\hat{c}_{l}^{\vphantom{+}}\rangle_0\nonumber \\
    && \hphantom{\sum_{l=0}^{L-1} \Bigl[}
      - \hat{c}_{l}^+\hat{c}_{l+1}^{\vphantom{+}}
      \langle        \hat{c}_{l}^{\vphantom{+}}\hat{c}_{l+1}^+\rangle_0
      - \langle \hat{c}_{l}^+\hat{c}_{l+1}^{\vphantom{+}}\rangle_0
      \hat{c}_{l}^{\vphantom{+}}\hat{c}_{l+1}^+\Bigr]\nonumber\\
  && \hphantom{\sum_{l=0}^{L-1} \Bigl[}
    \hat{\beta}_{k_1}^+\hat{\alpha}_{p_1}^{\vphantom{+}}
    \hat{\beta}_{k_2}^+\hat{\alpha}_{p_2}^{\vphantom{+}}
    |0\rangle
\end{eqnarray}
with $k_1<k_2$ and $p_1<p_2$,
where we again used $\langle 0 | n\rangle=0$. Using Wick's theorem we find
\begin{eqnarray}
  \langle 0|  \hat{V}_{\perp} | n_2\rangle
% V \sum_{l=0}^{L-1}
%  \left|
%  \begin{array}{cccc}
%    0 & 0 & \langle \hat{c}_{l+1}^+\hat{\alpha}_{p_1}^{\vphantom{+}}\rangle_0
%& \langle \hat{c}_{l+1}^+\hat{\alpha}_{p_2}^{\vphantom{+}}\rangle_0    \\
%    0 & 0 & \langle \hat{c}_{l}^+\hat{\alpha}_{p_1}^{\vphantom{+}}\rangle_0
%    & \langle \hat{c}_{l}^+\hat{\alpha}_{p_2}^{\vphantom{+}}\rangle_0    \\
%    -\langle \hat{c}_{l+1}^{\vphantom{+}} \hat{\beta}_{k_1}^+\rangle_0
%    & -\langle \hat{c}_{l}^{\vphantom{+}} \hat{\beta}_{k_1}^+\rangle_0 &0 &0 \\% 
%    -\langle \hat{c}_{l+1}^{\vphantom{+}} \hat{\beta}_{k_2}^+\rangle_0
%    & -\langle \hat{c}_{l}^{\vphantom{+}} \hat{\beta}_{k_2}^+\rangle_0 &0 &0 \\% 
%     \end{array}
%  \right| \nonumber \\[6pt]
  &=&
  V \sum_{l=0}^{L-1} M_1^l(p_1,p_2)M_2^l(k_1,k_2)
  \nonumber  \; ,\\
  M_1^l(p_1,p_2) &= &
  \langle \hat{c}_{l+1}^+\hat{\alpha}_{p_1}^{\vphantom{+}}\rangle_0
  \langle \hat{c}_{l}^+\hat{\alpha}_{p_2}^{\vphantom{+}}\rangle_0
  - 
\langle \hat{c}_{l+1}^+\hat{\alpha}_{p_2}^{\vphantom{+}}\rangle_0
  \langle \hat{c}_{l}^+\hat{\alpha}_{p_1}^{\vphantom{+}}\rangle_0 \; ,
  \nonumber \\
M_2^l(k_1,k_2)&= &   
  \langle \hat{c}_{l+1}^{\vphantom{+}}\hat{\beta}_{k_1}^+  \rangle_0
  \langle \hat{c}_{l}^{\vphantom{+}}\hat{\beta}_{k_2}^+\rangle_0
  - 
\langle \hat{c}_{l+1}^{\vphantom{+}}\hat{\beta}_{k_2}^+\rangle_0
\langle \hat{c}_{l}^{\vphantom{+}}\hat{\beta}_{k_1}^+\rangle_0\; .
\nonumber \\
\end{eqnarray}
We have
\begin{eqnarray}
  \langle \hat{c}_n^+\hat{\alpha}_p^{\vphantom{+}} \rangle _0
  &=&\sqrt{\frac{1}{L}} e^{-{\rm i}p n}\left(u_p+(-1)^n v_p\right) \; ,
  \nonumber \\
  \langle \hat{c}_n^{\vphantom{+}}\hat{\beta}_k^+ \rangle _0
  &=&\sqrt{\frac{1}{L}} e^{{\rm i}k n}\left(-v_k+(-1)^n u_k\right) \; ,
\end{eqnarray}
with $u_p=\cos(\varphi_p)$ and $v_p=\sin(\varphi_p)$.
Thus,
\begin{eqnarray}
  \langle 0|  \hat{V}_{\perp} | n_2\rangle &=& \frac{V}{L^2}
\sum_{l=0}^{L-1} e^{-{\rm i}(p_1+p_2)l}
  \hbox{1st}_l \times e^{{\rm i}(k_1+k_2)l} \hbox{2nd}_l \nonumber \; , \\
  \hbox{1st}_l &=& 
    \left(e^{-{\rm i}p_1}-e^{-{\rm i}p_2}\right)(1)_{p_1,p_2}\nonumber \\
&&    +(-1)^l \left(e^{-{\rm i}p_1}+e^{-{\rm i}p_2}\right) (2)_{p_1,p_2}\; ,
  \nonumber \\
  \hbox{2nd}_l &=& 
      -\left(e^{{\rm i}k_1}-e^{{\rm i}k_2}\right)(1)_{k_1,k_2}\nonumber \\
      &&    +(-1)^l \left(e^{{\rm i}k_1}+e^{{\rm i}k_2}\right) (2)_{k_1,k_2} \; ,
      \label{eq:oVperpn2}
\end{eqnarray}
where we introduced the abbreviations
\begin{eqnarray}
  (1)_{k_1,k_2}&=&u_{k_1}u_{k_2}-v_{k_1}v_{k_2} \; , \nonumber \\
  (2)_{k_1,k_2}&=&u_{k_1}v_{k_2}-v_{k_1}u_{k_2} \; .
\end{eqnarray}
We multiply the two terms in eq.~(\ref{eq:oVperpn2})
and regroup them
so that we can perform the sum over the
lattice index~$l$,
\begin{equation}
  \frac{1}{L} \sum_{l=0}^{L-1} e^{{\rm i}(k_1+k_2-p_1-p_2)l}= \delta_{k_1+k_2-p_1-p_2,0}
\end{equation} and
\begin{eqnarray}
  \frac{1}{L} \sum_{l=0}^{L-1} (-1)^le^{{\rm i}(k_1+k_2-p_1-p_2)l}&=&
  \delta_{k_1+k_2-p_1-p_2,-\pi} \nonumber \\
  && + \delta_{k_1+k_2-p_1-p_2,\pi}
    \; .\;\;\label{eq:Kroneckdeltsum}
\end{eqnarray}
This gives
\begin{eqnarray}
  \langle 0|  \hat{V}_{\perp} |n_2\rangle &=&
  \frac{V}{L}A(k_1,k_2;p_1,p_2)  \delta_{k_1+k_2-p_1-p_2,0} 
  \nonumber \\
  && + \frac{V}{L} B(k_1,k_2;p_1,p_2) \\
  &&\times \left(       \delta_{k_1+k_2-p_1-p_2,-\pi} + \delta_{k_1+k_2-p_1-p_2,\pi}\right)
  \nonumber 
\end{eqnarray}
with 
\begin{eqnarray}
  A(n_2)
  &=&
- \left(e^{{\rm i}k_1}-e^{{\rm i}k_2}\right)\left(e^{-{\rm i}p_1}-e^{-{\rm i}p_2}\right)
(1)_{k_1,k_2}(1)_{p_1,p_2} \nonumber \\
&&
+\left(e^{{\rm i}k_1}+e^{{\rm i}k_2}\right)\left(e^{-{\rm i}p_1}+e^{-{\rm i}p_2}\right)
(2)_{k_1,k_2}(2)_{p_1,p_2} \; ,\nonumber \\
  B(n_2)  &=&
\left(e^{{\rm i}k_1}+e^{{\rm i}k_2}\right)\left(e^{-{\rm i}p_1}-e^{-{\rm i}p_2}\right)
(2)_{k_1,k_2}(1)_{p_1,p_2} \nonumber \\
&&
-\left(e^{{\rm i}k_1}-e^{{\rm i}k_2}\right)\left(e^{-{\rm i}p_1}+e^{-{\rm i}p_2}\right)
(1)_{k_1,k_2}(2)_{p_1,p_2} \; ,\nonumber \\
\end{eqnarray}
where $A/B(n_2)\equiv A/B(k_1,k_2;p_1,p_2)$.
Next, we square the matrix elements. The conditions on the momenta allow us
to treat the two terms separately,
\begin{eqnarray}
\left|  \langle 0|  \hat{V}_{\perp} | n_2\rangle\right|^2 &=&
  \frac{V^2}{L^2} \Big(
  \delta_{k_1+k_2-p_1-p_2,0}
  \left|A(n_2)\right|^2 \nonumber \\
  && \hphantom{\frac{V}{L} \Big(}
  +\delta_{k_1+k_2-p_1-p_2,-\pi}\left|B(n_2)  \right|^2\nonumber \\
  && \hphantom{\frac{V}{L} \Big(}
  + \delta_{k_1+k_2-p_1-p_2,\pi} \left|B(n_2)  \right|^2\Bigr)
  \;.\nonumber \\
\end{eqnarray}
\paragraph{Simplification of the first term}
We have
\begin{eqnarray}
  \left|A(n_2)\right|^2 &=&
  16 \sin^2[(k_2-k_1)/2]\sin^2[(p_2-p_1)/2]\nonumber \\
  && \hphantom{\times}\times  (1)_{k_1,k_2}^2
  (1)_{p_1,p_2}^2 \nonumber \\
  &&  +   16 \cos^2[(k_2-k_1)/2]\cos^2[(p_2-p_1)/2] \nonumber \\
  && \hphantom{\times}\times (2)_{k_1,k_2}^2(2)_{p_1,p_2}^2
  \nonumber \\ 
  &&  -8\sin(k_2-k_1)\sin(p_2-p_1)\nonumber \\
  && \hphantom{\times}\times (1)_{k_1,k_2}(1)_{p_1,p_2}(2)_{k_1,k_2}(2)_{p_1,p_2}\; .
\end{eqnarray}
Now, we carry out the products,
\begin{eqnarray}
  (1)_{k_1,k_2}^2 &=& u_{k_1}^2u_{k_2}^2-2u_{k_1}u_{k_2}v_{k_1}v_{k_2}+v_{k_1}^2v_{k_2}^2
  \nonumber\\
  &=&
  \frac{1}{2}\left( 1 +
  \frac{\widetilde{\epsilon}(k_1)\widetilde{\epsilon}(k_2)}{E(k_1)E(k_2)}
  -\frac{(2Vn_a)^2}{E(k_1)E(k_2)}\right)
  \; , \nonumber \\
\end{eqnarray}
\begin{eqnarray}
   (2)_{k_1,k_2}^2 &=& u_{k_1}^2v_{k_2}^2-2u_{k_1}u_{k_2}v_{k_1}v_{k_2}+v_{k_1}^2u_{k_2}^2
  \nonumber\\
  &=& \frac{1}{2}\left( 1 -
  \frac{\widetilde{\epsilon}(k_1)\widetilde{\epsilon}(k_2)}{E(k_1)E(k_2)}
  -\frac{(2Vn_a)^2}{E(k_1)E(k_2)}\right) \; , \nonumber \\
\end{eqnarray}
and
\begin{eqnarray}
  (1)_{k_1,k_2}\ldots(2)_{p_1,p_2} &=&
  (u_{k_1}u_{k_2}-v_{k_1}v_{k_2})(u_{k_1}v_{k_2}-v_{k_1}u_{k_2})
  \nonumber \\
  && \hphantom{\times}
  \times(p_i\leftrightarrow k_i)
\nonumber \\[3pt]
&=&   \frac{1}{2} \left(\frac{2Vn_a}{E(k_2)}-\frac{2Vn_a}{E(k_1)}\right)
\nonumber \\
&&\hphantom{\times}\times
\frac{1}{2} \left(\frac{2Vn_a}{E(p_2)}-\frac{2Vn_a}{E(p_1)}\right)
\; ,
\end{eqnarray}
where we used eq.~(71) of the main text.
The final result for the first term is
\begin{eqnarray}
  \left|A(k_1,k_2;p_1,p_2)\right|^2 &=&
  Q_1(k_1,k_2)Q_1(p_1,p_2)\nonumber \\
  && +   Q_2(k_1,k_2)Q_2(p_1,p_2)\nonumber \\
 && -2Q_3(k_1,k_2)Q_3(p_1,p_2)
  \; ,\nonumber \\
    \label{eq:finalAsq}
\end{eqnarray}
where we introduced the symmetric functions
\begin{eqnarray}
Q_1(k_1,k_2) &= & 
  2\sin^2[(k_2-k_1)/2]\nonumber \\
&& \times \left( 1 +
  \frac{\widetilde{\epsilon}(k_1)\widetilde{\epsilon}(k_2)}{E(k_1)E(k_2)}
  -\frac{(2Vn_a)^2}{E(k_1)E(k_2)}\right)\; , \nonumber \\
Q_2(k_1,k_2) &= &   2\cos^2[(k_2-k_1)/2]\nonumber \\
&& \times \left( 1 -
  \frac{\widetilde{\epsilon}(k_1)\widetilde{\epsilon}(k_2)}{E(k_1)E(k_2)}
  -\frac{(2Vn_a)^2}{E(k_1)E(k_2)}\right)\; ,\nonumber \\
  Q_3(k_1,k_2)  &= &  \sin(k_2-k_1)\nonumber \\
&&  \times \left(\frac{2Vn_a}{E(k_2)}-\frac{2Vn_a}{E(k_1)}\right)\; .
\end{eqnarray}
Therefore,
$  \left|A(k_1,k_2;p_1,p_2)\right|^2$ is symmetric in the interchange
of $k_1\leftrightarrow k_2$ and $p_1\leftrightarrow p_2$,
respectively. Note that
$  \left|A(k,k;p_1,p_2)\right|^2=0= \left|A(k_1,k_2;p,p)\right|^2$
due to the Pauli principle. Formally, $Q_1(k,k)= Q_3(k,k)= 0$ due to the sine
functions. To see that $Q_2(k,k)= 0$,
the definition of $E(k)$, eq.~(71) of the main text,
must be used.

\paragraph{Simplification of the second term}
We have
\begin{eqnarray}
  \left|B(n_2)\right|^2 &=&
  16 \cos^2[(k_2-k_1)/2]\sin^2[(p_2-p_1)/2] \nonumber \\
  && \hphantom{\times}\times
  (2)_{k_1,k_2}^2
  (1)_{p_1,p_2}^2 \nonumber \\
  &&  +   16 \sin^2[(k_2-k_1)/2]\cos^2[(p_2-p_1)/2] \nonumber \\
  && \hphantom{\times}\times
  (1)_{k_1,k_2}^2(2)_{p_1,p_2}^2
  \nonumber \\
  &&  +8\sin(k_2-k_1)\sin(p_2-p_1)
  \nonumber \\
  && \hphantom{\times}\times
  (1)_{k_1,k_2}(1)_{p_1,p_2}(2)_{k_1,k_2}(2)_{p_1,p_2}\; .
\end{eqnarray}
The final result for the second term is
\begin{eqnarray}
  \left|B(k_1,k_2;p_1,p_2)\right|^2 &=&
  Q_1(k_1,k_2)Q_2(p_1,p_2)\nonumber \\
  &&+   Q_2(k_1,k_2)Q_1(p_1,p_2)\nonumber \\
  && + 2Q_3(k_1,k_2)Q_3(p_1,p_2)\; .\; \;
  \label{eq:finalBsq}
\end{eqnarray}
Note that $  \left|B(k_1,k_2;p_1,p_2)\right|^2$ is symmetric in the interchange
of $k_1\leftrightarrow k_2$ and $p_1\leftrightarrow p_2$,
respectively, and
$  \left|B(k,k;p_1,p_2)\right|^2=0= \left|B(k_1,k_2;p,p)\right|^2$
due to the Pauli principle.

\section{Observables in second-order Hartree-Fock approximation}
\label{sec:gsenergy}

In this appendix we derive explicit expressions for the ground-state energy
and the occupation numbers in second-order Hartree-Fock approximation.

\subsection{Calculation of the ground-state energy}

The second-order energy correction
can be represented as a triple sum/integral,
or as double sums/integrals plus an additional parameter integral.

\subsubsection{Triple sums or integrals}

The ground-state energy up to and including the second order
in eq.~(101) of the main text can be written as ($n=1/2$)
\begin{eqnarray}
  e_0^{(2)}(B_0,n_a,V)
  &=&
  V\left(n^2+n_a^2+B_0^2\right)
  -\frac{1}{L} \sum_{k\in{\rm RBZ}} E(k) \nonumber \\
  &&   + \delta e_0^{(2)}(B_0,n_a,V)
  \; ,  \nonumber \\
  \delta e_0^{(2)}(B_0,n_a,V)  &= &
  -\frac{V^2}{4 L^3}\left(\delta e_0^{(2,A)}+ \delta e_0^{(2,B)}\right)\; ,
  \label{eq:deltae0def}
\end{eqnarray}
where
\begin{eqnarray}
  \delta e_0^{(2,A)}
  &=&   \sum_{k_1,k_2,p_1,p_2}
  \frac{\left|A(k_1,k_2;p_1,p_2)\right|^2}{
    E(k_1)+E(k_2)+E(p_1)+E(p_2)}\nonumber \\
  &&     \hphantom{\sum_{k_1,k_2,p_1,p_2}}
  \delta_{k_1+k_2-p_1-p_2,0}
    \; , \nonumber \\
\delta e_0^{(2,B)}
&=&   \sum_{k_1,k_2,p_1,p_2}     \frac{      \left|B(k_1,k_2;p_1,p_2)\right|^2}{
  E(k_1)+E(k_2)+E(p_1)+E(p_2)}\nonumber \\
&& \hphantom{\sum_{k_1,k_2,p_1,p_2}}
\left(       \delta_{k_1+k_2-p_1-p_2,-\pi} + \delta_{k_1+k_2-p_1-p_2,\pi}\right)\; .
\nonumber \\
  \label{eq:main21ndordergsnerergy}
\end{eqnarray}
Here,
the summation restrictions $k_1<k_2$ and $p_1<p_2$ were lifted. This results
in the factor $1/4$ in eq.~(\ref{eq:deltae0def}).
Moreover,
$\left|A(k_1,k_2;p_1,p_2)\right|^2$ and $\left|B(k_1,k_2;p_1,p_2)\right|^2$
can be found in
eqs.~(\ref{eq:finalAsq}) and~(\ref{eq:finalBsq}), respectively.
Of the four sums over the momenta in the reduced Brillouin zone,
one can be eliminated using the Kronecker conditions. 

In the first term, we have the condition $-\pi/2 \leq k_2=p_1+p_2-k_1 < \pi/2$
which imposes $-\pi/2+k_1 \leq p_1+p_2 < \pi/2+k_1$ for $-\pi/2\leq k_1 <\pi/2$
in addition to $-\pi \leq p_1+p_2 <\pi$. Therefore,
$-\pi/2+k_1\leq p_1+p_2 < \pi/2+k_1$ must hold, i.e, the condition on $p_2$ is
$-\pi/2+k_1-p_1\leq p_2 < \pi/2+k_1-p_1$ in addition to
$-\pi/2\leq p_2 <\pi/2$.
Thus, (i), for $k_1\leq p_1<\pi/2$ we have
$-\pi/2 \leq p_2 <\pi/2+k_1-p_1$ and, (ii),
for $-\pi/2 \leq p_1 <k_1$ we have $-\pi/2+k_1-p_1 \leq p_2 <\pi/2$
as the two summation regions with $k_2=p_1+p_2-k_1$.
This gives the summation regions
\begin{eqnarray}
  A_1 &=& 
  \{-\pi/2\leq k_1<\pi/2\} \cup \{ k_1\leq p_1<\pi/2\}\nonumber \\
  && \cup
\{-\pi/2\leq p_2<\pi/2+ k_1-p_1\}  \; ,\nonumber\\[3pt]
A_2 &= &
\{-\pi/2\leq k_1<\pi/2\}\cup \{-\pi/2\leq p_1<k_1\}\nonumber \\
&& \cup
\{-\pi/2+ k_1-p_1\leq p_2<\pi/2\}
\end{eqnarray}
for $k_2= p_1+ p_2-k_1$.
Therefore,
\begin{eqnarray}
  \delta e_0^{(2,A)} \!&= &\! 
%\sum_{-\pi/2\leq k_1<\pi/2} \;\; \sum_{k_1\leq p_1<\pi/2}\;\;
%\sum_{-\pi/2\leq p_2<\pi/2+ k_1-p_1}
%  \sum_{\scriptstyle
%    \begin{array}{ccc}
%\scriptstyle      -\pi/2\leq k_1<\pi/2 \\[-3pt]
%\scriptstyle  k_1\leq p_1<\pi/2 \\[-3pt]
%\scriptstyle  -\pi/2\leq p_2<\pi/2+ k_1-p_1
%  \end{array}}
\sum_{A_1, A_2}
  \frac{
  \left|A(k_1,p_1+ p_2-k_1;p_1,p_2)\right|^2}{E(k_1)+E(p_1+ p_2-k_1)+E(p_1)+E(p_2)}
%\nonumber \\
%  &&  + 
%  \sum_{-\pi/2\leq k_1<\pi/2}\;\; \sum_{-\pi/2\leq p_1<k_1}\;\;
%  \sum_{-\pi/2+ k_1-p_1\leq p_2<\pi/2}
%\sum_{A_2}     \frac{
%  \left|A(k_1,p_1+ p_2-k_1;p_1,p_2)\right|^2}{E(k_1)+E(p_1+ p_2-k_1)+E(p_1)+E(p_2)}
\nonumber \\
\label{eq:e02A}
\end{eqnarray}
for the first term.

In the second term, we have either
the condition $-\pi/2 \leq k_2=p_1+p_2-k_1-\pi < \pi/2$
which imposes $\pi/2+k_1 \leq p_1+p_2 < 3\pi/2+k_1$ for $-\pi/2\leq k_1 <\pi/2$
in addition to $-\pi \leq p_1+p_2 <\pi$. Therefore,
$\pi/2+k_1\leq p_1+p_2 < \pi$ must hold, i.e, the condition on $p_2$ is
$\pi/2+k_1-p_1\leq p_2 < \pi-p_1$ in addition to
$-\pi/2\leq p_2 <\pi/2$.
Thus, for $k_1< p_1<\pi/2$ we have
$\pi/2+k_1-p_1 \leq p_2 <\pi/2$ for
$k_2=p_1+p_2-k_1-\pi$.

Alternatively, in the second term we have 
the condition $-\pi/2 \leq k_2=p_1+p_2-k_1+\pi < \pi/2$
which imposes $-3\pi/2+k_1 \leq p_1+p_2 < -\pi/2+k_1$
for $-\pi/2\leq k_1 <\pi/2$
in addition to $-\pi \leq p_1+p_2 <\pi$. Therefore,
$-\pi\leq p_1+p_2 < -\pi/2+k_1$ must hold, i.e, the condition on $p_2$ is
$-\pi-p_1\leq p_2 < -\pi/2-p_1+k_1$ in addition to
$-\pi/2\leq p_2 <\pi/2$.
Thus, for $-\pi/2\leq p_1<k_1$ we have
$-\pi/2 \leq p_2 <-\pi/2+k_1-p_1$ for
$k=p_1+p_2-k_1+\pi$.
This gives the summation regions
\begin{eqnarray}
  B_1 &=&
  \{-\pi/2\leq k_1<\pi/2\} \cup \{k_1< p_1<\pi/2\} \nonumber \\
  && \cup
\{\pi/2+ k_1-p_1\leq p_2<\pi/2\} \; ,\nonumber\\[3pt]
B_2 &= &
\{-\pi/2\leq k_1<\pi/2\} \cup \{-\pi/2\leq p_1<k_1\}\nonumber \\
&& \cup
\{-\pi/2\leq p_2<-\pi/2+ k_1-p_1\}
\end{eqnarray}
for $k_2^{\pm}= p_1+ p_2-k_1\pm \pi$.

Therefore, %($k_2^{\pm}=p_1+p_2-k_1 \pm \pi$)
\begin{eqnarray}
  \delta e_0^{(2,B)}
  &= & 
  \sum_{B_1} \frac{\left|B(k_1,k_2^-;p_1,p_2)\right|^2}{
    E(k_1)+E(k_2^-)+E(p_1)+E(p_2)}\nonumber \\
  && +
\sum_{B_2} 
\frac{\left|B(k_1,k_2^+;p_1,p_2)\right|^2}{
  E(k_1)+E(k_2^+)+E(p_1)+E(p_2)}\; .\nonumber \\
\label{eq:e02B}
\end{eqnarray}
The calculation of $\delta e_0^{(2,A,B)}$
involves the sum over approximately $(L/2)^3$ terms.
This poses no numerical problems for $L\lesssim 100$ but becomes prohibitive
for $L= 1000$, especially because the function is invoked several times during
the minimization procedure.

In the thermodynamic limit we have
\begin{equation}
  \frac{1}{L}\sum_{-\pi/2\leq k<\pi/2} \to \int_{-\pi/2}^{\pi/2} \frac{{\rm d}k}{2\pi}
  \; , 
\end{equation}
and we find numerically that the two terms in eq.~(\ref{eq:e02A}) and
in eq.~(\ref{eq:e02B}), respectively, are equal.
Therefore, we find in eq.~(\ref{eq:main21ndordergsnerergy})
in the thermodynamic limit ($\delta e_0^{(2)}\equiv \delta e_0^{(2)}(B_0,n_a,V)$)
\begin{eqnarray}
  \delta e_0^{(2)}&= &
  -\frac{V^2}{2} \int_{-\pi/2}^{\pi/2}
  \frac{{\rm d} k_1}{2\pi}
  \int_{-\pi/2}^{k_1}   \frac{{\rm d} p_1}{2\pi}
  \int_{-\pi/2+ k_1-p_1}^{\pi/2}   \frac{{\rm d} p_2}{2\pi}\nonumber \\
  && %\hphantom{-\frac{1}{2} }
\frac{
  \left|A(k_1,p_1+ p_2-k_1;p_1,p_2)\right|^2}{E(k_1)+E(p_1+ p_2-k_1)+E(p_1)+E(p_2)}
\nonumber \\[6pt]
&& -\frac{V^2}{2} \int_{-\pi/2}^{\pi/2}
  \frac{{\rm d} k_1}{2\pi}
  \int_{k_1}^{\pi/2}   \frac{{\rm d} p_1}{2\pi}
  \int_{\pi/2+ k_1-p_1}^{\pi/2}   \frac{{\rm d} p_2}{2\pi}\nonumber \\
  && %\hphantom{-\frac{1}{2}}
\frac{
  \left|B(k_1,p_1+ p_2-k_1-\pi;p_1,p_2)\right|^2}{
  E(k_1)+E(p_1+ p_2-k_1-\pi)+E(p_1)+E(p_2)} \; .\nonumber \\
\label{eq:tripleintegral}
  \end{eqnarray}
For a given set of values for $(B_0,n_a,V)$, the integration takes seconds on
a notebook with an i7 processor.
  
\subsubsection{Two sums plus integration}

To reduce the number of lattice sums, we invert eq.~(\ref{eq:Kroneckdeltsum})
and use
\begin{equation}
\frac{1}{x} = \int_0^{\infty}{\rm d}\lambda e^{-\lambda x}
\end{equation}
for $x>0$. Thus, we can write
\begin{eqnarray}
  \delta e_0^{(2)} &= &
%  -\frac{V^2}{4 L^4}  \int_0^{\infty} {\rm d }\lambda
%  \sum_{l= 0}^{L-1} \biggl[
%    |\tilde{Q}_1(l,\lambda)|^2 +     |\tilde{Q}_2(l,\lambda)|^2 
%    -2     |\tilde{Q}_3(l,\lambda)|^2 \nonumber \\
%    && \hphantom{-\frac{V^2}{4 L^3}}
 %     + (-1)^l \left(
%      \tilde{Q}_1^*(l,\lambda)\tilde{Q}_2(l,\lambda) +
%          \tilde{Q}_2^*(l,\lambda)\tilde{Q}_1(l,\lambda)
%    -2     |\tilde{Q}_3(l,\lambda)|^2
%      \right)
%      \biggr]\nonumber \\
%  &= &
  -\frac{V^2}{4 L^4}  \int_0^{\infty} {\rm d }\lambda
  \biggl[
    \sum_{m= 0}^{L/2-1}
    |\tilde{Q}_1(2m,\lambda) +     \tilde{Q}_2(2m,\lambda)|^2
    \nonumber \\
    && \hphantom{-\frac{V^2}{4 L^3}}
    +\sum_{m= 0}^{L/2-1}
    |\tilde{Q}_1(2m+ 1,\lambda) -     \tilde{Q}_2(2m+ 1,\lambda)|^2\nonumber \\
    && \hphantom{-\frac{V^2}{4 L^3}}
    - 4   \sum_{m= 0}^{L/2-1} |\tilde{Q}_3(2m+ 1,\lambda)|^2
    \biggr],
\end{eqnarray}
where for $i= 1,2,3$ 
\begin{equation}
  \tilde{Q}_i(l,\lambda) = \sum_{k_1,k_2} e^{-\lambda(E(k_1)+E(k_2))}
  e^{{\rm i}(k_1+ k_2)l}Q_i(k_1,k_2) 
\end{equation}
are the weighted Fourier transformed of $Q_i(k_1,k_2)$.

The contributions for even lattice sites, $l= 2m$,
can be cast into the form
\begin{eqnarray}
\frac{  \tilde{Q}_1(2m,\lambda) +     \tilde{Q}_2(2m,\lambda)} {2L^2} &= &
%  2 \sum_{k_1,k_2}e^{-\lambda(E(k_1)+E(k_2))}
 % e^{{\rm i}(k_1+ k_2)2m}   \left(1- \frac{(2Vn_a)^2}{E(k_1)E(k_2)} \right)
 % \nonumber \\
%  &&-  2 \sum_{k_1,k_2}e^{-\lambda(E(k_1)+E(k_2))}
%  e^{{\rm i}(k_1+ k_2)2m}
%  \frac{\widetilde{\epsilon}(k_1)\widetilde{\epsilon}(k_2)}{E(k_1)E(k_2)}
 % \nonumber \\
%  && \hphantom{-  2 \sum_{k_1,k_2}}
%  \times \left(\cos(k_1)\cos(k_2)+ \sin(k_1)\sin(k_2)\right)
 % \nonumber \\
  %  &= &
    \left(R_0(m,\lambda)\right)^2
    -\left(R_1(m,\lambda)\right)^2\nonumber \\
    && %\hphantom{2L^2\Bigl[}
    -\left(R_2^{\rm c}(m,\lambda) \right)^2
    + \left(R_2^{\rm s}(m,\lambda) \right)^2\nonumber \\
  \end{eqnarray}
with
\begin{eqnarray}
  R_0(m,\lambda) &= & \frac{1}{L} \sum_k e^{-\lambda E(k)} e^{{\rm i}2mk}
%  \nonumber\\
%  &=& \frac{1}{L}\Bigl(
%  e^{-\lambda E(0)}
%  +(-1)^me^{-\lambda E(\pi/2)} \nonumber \\
  %&&    +2 \sum_{0<k<\pi/2}e^{-\lambda E(k)}\cos(2mk)\Bigr)
  \nonumber   \; ,\\
    R_1(m,\lambda) &= & \frac{1}{L} \sum_k e^{-\lambda E(k)} e^{{\rm i}2mk}
    \frac{2Vn_a}{E(k)}\nonumber \; ,\\
%    &=&
%    \frac{1}{L}\Bigl(e^{-\lambda E(0)}\frac{2Vn_a}{E(0)}
 %   +(-1)^me^{-\lambda E(\pi/2)}\frac{2Vn_a}{E(\pi/2)}\nonumber \\
 %   && \hphantom{\frac{1}{L}\Bigl(}
 %   +2 \sum_{0<k<\pi/2}e^{-\lambda E(k)}\cos(2mk)\frac{2Vn_a}{E(k)}
 %   \Bigr) \nonumber   \; ,\\
 %   \nonumber \; ,\\
    R_2^{\rm c}(m,\lambda) &= & \frac{1}{L} \sum_k e^{-\lambda E(k)} e^{{\rm i}2mk}
    \frac{\widetilde{\epsilon}(k)}{E(k)} \cos(k) \nonumber \; ,\\
%    &=& \frac{1}{L}\Bigl(
%e^{-\lambda E(0)} \frac{\widetilde{\epsilon}(0)}{E(0)}\nonumber \\
%&& +2 \sum_{0<k<\pi/2}e^{-\lambda E(k)}\cos(2mk)
%\frac{\widetilde{\epsilon}(k)}{E(k)} \cos(k)\Bigr)
%\nonumber \; ,\\
%
   R_2^{\rm s}(m,\lambda) &= &
        \frac{1}{L} \sum_k e^{-\lambda E(k)} e^{{\rm i}2mk}
        \frac{\widetilde{\epsilon}(k)}{E(k)}
             {\rm i}\sin(k) \; . \nonumber \\
%             &=& -\frac{2}{L}
%             \sum_{0<k<\pi/2} e^{-\lambda E(k)} \sin(2mk)\sin(k)
%             \; .
\end{eqnarray}
All functions are real. This can be seen by separating the points $k=0$ and $k=\pi/2$,
and combining the terms for $k$ and $-k$.
In this way, the sums contain only $L/4+1$ summation terms.

Likewise, the contributions for odd lattice sites, $l= 2m+ 1$,
can be cast into the from
\begin{eqnarray}
  Q_{\rm odd}(2m+1,\lambda) &=&
  \frac{\tilde{Q}_1(2m+ 1,\lambda) -     \tilde{Q}_2(2m+ 1,\lambda)}{2L^2} \nonumber
  \\
%  2 \sum_{k_1,k_2}e^{-\lambda(E(k_1)+E(k_2))}
 % e^{{\rm i}(k_1+ k_2)(2m+ 1)}
  %\frac{\widetilde{\epsilon}(k_1)\widetilde{\epsilon}(k_2)}{E(k_1)E(k_2)}
  %\nonumber \\
%  &&-  2 \sum_{k_1,k_2}e^{-\lambda(E(k_1)+E(k_2))}
%  e^{{\rm i}(k_1+ k_2)2m}
%  \left(1- \frac{(2Vn_a)^2}{E(k_1)E(k_2)} \right)\nonumber \\
% && \hphantom{-  2 \sum_{k_1,k_2}}
%  \times \left(\cos(k_1)\cos(k_2)+ \sin(k_1)\sin(k_2)\right)
%  \nonumber \\
  %&= & 2L^2\Bigl[
 &=&   \left(R_3(m,\lambda)\right)^2\nonumber \\
&&    -\left(R_4^{\rm c}(m,\lambda) \right)^2
    + \left(R_4^{\rm s}(m,\lambda) \right)^2\nonumber \\
      && %\hphantom{2L^2\Bigl[}
  + \left(R_5^{\rm c}(m,\lambda) \right)^2
  - \left(R_5^{\rm s}(m,\lambda) \right)^2
  %\Bigr]
  \nonumber \\
  \end{eqnarray}
and
\begin{eqnarray}
\frac{  \tilde{Q}_3(2m+ 1,\lambda)}{(-{\rm i})2L^2} &= &
%  \sum_{k_1,k_2}e^{-\lambda(E(k_1)+E(k_2))}
%  e^{{\rm i}(k_1+ k_2)(2m+ 1)}
%  \left(\frac{2Vn_a}{E(k_1)}-\frac{2Vn_a}{E(k_2)}\right)
%  \nonumber \\
% && \hphantom{\sum_{k_1,k_2}}
%  \times \left(\sin(k_1)\cos(k_2)- \cos(k_1)\sin(k_2)\right)
%  \nonumber \\
  %
 %  &= & 2L^2(-{\rm i})\Bigl[
    R_5^{\rm s}(m,\lambda)R_4^{\rm c}(m,\lambda)
    - R_5^{\rm c}(m,\lambda)R_4^{\rm s}(m,\lambda)\nonumber\\
  \end{eqnarray}
with
\begin{eqnarray}
  R_3(m,\lambda) &= & \frac{1}{L} \sum_k
  e^{-\lambda E(k)} e^{{\rm i}(2m+ 1)k}
  \frac{\widetilde{\epsilon}(k)}{E(k)}
  \nonumber \; ,\\
%  &=& \frac{1}{L} \Bigl( e^{-\lambda E(0)} \frac{\widetilde{\epsilon}(0)}{E(0)}
%  \nonumber\\
%&&+ 2 \sum_{0<k<\pi/2} 
%    e^{-\lambda E(k)} \cos[(2m+ 1)k]
%  \frac{\widetilde{\epsilon}(k)}{E(k)}\Bigr)
%    \nonumber \; ,\\
    R_4^{\rm c}(m,\lambda) &= &
    \frac{1}{L} \sum_k e^{-\lambda E(k)} e^{{\rm i}(2m+ 1)k}  \cos(k)
    \nonumber \; ,\\
%    &=&
%    \frac{1}{L} \Bigl( e^{-\lambda E(0)} \nonumber\\ 
%&& +2 \sum_{0<k<\pi/2} e^{-\lambda E(k)} \cos[(2m+ 1)k] \cos(k)\Bigr)
%    \nonumber \; ,\\
%        R_4^{\rm s}(m,\lambda) &= &
%        \frac{1}{L} \sum_k e^{-\lambda E(k)} e^{{\rm i}(2m+ 1)k} {\rm i}\sin(k)
%        \nonumber \\
%        &=&     -\frac{1}{L} \Bigl( (-1)^m e^{-\lambda E(\pi/2)}\nonumber\\
%&&        +2 \sum_{0<k<\pi/2} e^{-\lambda E(k)} \sin[(2m+ 1)k] \sin(k)\Bigr)
%    \nonumber \; ,\\
 R_5^{\rm c}(m,\lambda) &= & \frac{1}{L} \sum_k e^{-\lambda E(k)} e^{{\rm i}(2m+ 1)k}
    \frac{2 V n_a}{E(k)}
    \cos(k) \nonumber \; , \\
%    &=&     \frac{1}{L} \Bigl( e^{-\lambda E(0)} \frac{2 V n_a}{E(0)}\nonumber \\
%&&    +2 \sum_{0<k<\pi/2} e^{-\lambda E(k)}     \frac{2 V n_a}{E(k)}
 %   \cos[(2m+ 1)k] \cos(k) \Bigr)
 %       \nonumber \; ,\\
        %
 R_5^{\rm s}(m,\lambda) &= &
        \frac{1}{L} \sum_k e^{-\lambda E(k)} e^{{\rm i}(2m+ 1)k}
        \frac{2Vn_a}{E(k)} {\rm i}\sin(k)\; .\nonumber 
        \\
%        &=&
%            -\frac{1}{L} \Bigl( (-1)^m e^{-\lambda E(\pi/2)}\nonumber \\
%&&            +2 \sum_{0<k<\pi/2} e^{-\lambda E(k)} \frac{2 V n_a}{E(k)}
%            \sin[(2m+ 1)k] \sin(k)\Bigr)\; .\nonumber
\end{eqnarray}
All functions are real, apart from $Q_3(2m+ 1,\lambda)$
that is purely imaginary.
This can be seen by separating the points $k=0$ and $k=\pi/2$,
and combining the terms for $k$ and $-k$.
In this way, the sums contain only $L/4+1$ summation terms.
Thus,
\begin{eqnarray}
  \delta e_0^{(2)} &= &
  -V^2 \int_0^{\infty} {\rm d }\lambda \nonumber \\
    && \biggl[ \sum_{m= 0}^{L/2-1}
\Bigl[\left[R_0(m,\lambda)\right]^2
  -\left[R_1(m,\lambda)\right]^2 \nonumber \\
  && \hphantom{\sum_{m= 0}^{L/2-1}\Bigl[}
    -\left[R_2^{\rm c}(m,\lambda) \right]^2
    + \left[R_2^{\rm s}(m,\lambda) \right]^2\Bigr]^2
\nonumber \\
    && %\hphantom{-\frac{V^2}{4 L^3}}
+\sum_{m= 0}^{L/2-1}\Bigl[
  \left[R_3(m,\lambda)\right]^2
  -\left[R_4^{\rm c}(m,\lambda) \right]^2 \nonumber\\
  && \hphantom{+\sum_{m= 0}^{L/2-1}\Bigl[}
    + \left[R_4^{\rm s}(m,\lambda) \right]^2
    + \left[R_5^{\rm c}(m,\lambda) \right]^2\nonumber\\
      && \hphantom{+\sum_{m= 0}^{L/2-1}\Bigl[}
  - \left[R_5^{\rm s}(m,\lambda) \right]^2\Bigr]^2\nonumber \\
&&
    - 4  \sum_{m= 0}^{L/2-1} \Bigl[
      R_5^{\rm s}(m,\lambda)R_4^{\rm c}(m,\lambda)
      \nonumber \\
      && \hphantom{- 4  \sum_{m= 0}^{L/2-1} \Bigl[}
    - R_5^{\rm c}(m,\lambda)R_4^{\rm s}(m,\lambda)\Bigr]^2\biggr]
      \, .
      \label{eq:twosumformulafinalnow}
\end{eqnarray}
This representation shows that only two sums need to be carried out,
one over momenta to calculate the functions $R_i^{\rm c,s}$ and another one
over the lattice sites. In addition, an integral must be carried out that
converges rapidly due to the exponentially decaying integrand.

To simplify matters, we use inversion symmetry which implies
that the sites $l$ and $L-l$ contribute equally.
This symmetry can also be inferred from the properties of the
functions $Q_i$.
%We find
%\begin{eqnarray}
%  \delta e_0^{(2)} &= &-V^2
%  \int_0^{\infty} {\rm d }\lambda
%\biggl[ \Bigl[\left[R_0(0,\lambda)\right]^2
%  -\left[R_1(0,\lambda)\right]^2
%    -\left[R_2^{\rm c}(0,\lambda) \right]^2
%    + \left[R_2^{\rm s}(0,\lambda) \right]^2\Bigr]^2 \nonumber \\
%&& 
%+ \Bigl[\left[R_0(L/4,\lambda)\right]^2
%  -\left[R_1(L/4,\lambda)\right]^2
%    -\left[R_2^{\rm c}(L/4,\lambda) \right]^2
%    + \left[R_2^{\rm s}(L/4,\lambda) \right]^2\Bigr]^2 \nonumber \\
%&& +2  \sum_{m= 1}^{L/4-1}
%\Bigl[\left[R_0(m,\lambda)\right]^2
%  -\left[R_1(m,\lambda)\right]^2
%    -\left[R_2^{\rm c}(m,\lambda) \right]^2
%    + \left[R_2^{\rm s}(m,\lambda) \right]^2\Bigr]^2
%\nonumber \\
%
%    && 
%    +2\sum_{m= 0}^{L/4-1}
%\Bigl[  \left[R_3(m,\lambda)\right]^2
%    -\left[R_4^{\rm c}(m,\lambda) \right]^2
%    + \left[R_4^{\rm s}(m,\lambda) \right]^2
%  + \left[R_5^{\rm c}(m,\lambda) \right]^2
%  - \left[R_5^{\rm s}(m,\lambda) \right]^2\Bigr]^2\nonumber \\
%&&
%  - 8  \sum_{m= 0}^{L/4-1} \Bigl[  R_5^{\rm s}(m,\lambda)R_4^{\rm c}(m,\lambda)
%    - R_5^{\rm c}(m,\lambda)R_4^{\rm s}(m,\lambda)\Bigr]^2
%  \biggr]\, .
%\label{eq:finaltwosumformula}
%\end{eqnarray}
Therefore, each lattice sum in eq.~(\ref{eq:twosumformulafinalnow})
contains only about $L/4$ terms.
For the complete set of system sizes
$L=\{8,16,32,64,128,256,512,1024\}$, a typical minimization for
fixed value of $V$ takes
a few hours on a notebook (Intel~i7 processor).

\subsubsection{Limit of strong coupling}

We address the
limit $V\gg 1$ where the Hartree-Fock energy and order parameters are
exact to leading order, see Sect.~IV C 2 of the main text.
To calculate the contribution from the second-order term, we thus may set
\begin{eqnarray}
  2VB_0 &=& 1 + {\cal O}\left(\frac{1}{V^2}\right) \; , \nonumber \\
  2Vn_a &= & 1- \frac{4}{V} + {\cal O}\left(\frac{1}{V^3}\right) \; ,
  \nonumber \\
  \widetilde{\epsilon}(k)&= & -4\cos(k)
  + {\cal O}\left(\frac{1}{V^2}\right) \; , \nonumber \\
  E(k)&= & V + \frac{4}{V}(2\cos^2(k)-1)
  + {\cal O}\left(\frac{1}{V^3}\right) \; , \nonumber \\
  \frac{2Vn_a}{E(k)} &= & 1-\frac{8\cos^2(k)}{V^2}
  + {\cal O}\left(\frac{1}{V^4}\right) \; , \nonumber \\
  \frac{\widetilde{\epsilon}(k)}{E(k)} &= & -\frac{4\cos(k)}{V}
  + {\cal O}\left(\frac{1}{V^3}\right) \; .
\end{eqnarray}
For the energy denominator in eq.~(\ref{eq:main21ndordergsnerergy})
we may thus approximate $E(k_1)+E(k_2)+E(p_1)+E(p_2)\approx 4 V$.

Therefore, keeping terms up to order $V^{-2}$, we approximate
\begin{eqnarray}
  R_0(m,0)&\approx& \frac{1}{2} \delta_{m,0}  \;, \nonumber \\
  R_1(m,0)&\approx& \left(\frac{1}{2}-\frac{2}{V^2}\right)
  \delta_{m,0} -\frac{1}{V^2}\delta_{m,1} \;, \nonumber \\
  R_2^{\rm c}(m,0)&\approx& -\frac{1}{V}  \delta_{m,0}
  -\frac{1}{2V}\delta_{m,1} \;, \nonumber \\
  R_2^{\rm s}(m,0)&\approx&   \frac{1}{2V}\delta_{m,1} \;, \nonumber \\
  R_3(m,0)&\approx& -\frac{1}{V} \delta_{m,0}  \;, \nonumber \\
  R_4^{\rm c}(m,0)&\approx& \frac{1}{4} \delta_{m,0}  \;, \nonumber \\
  R_4^{\rm s}(m,0)&\approx& -\frac{1}{4} \delta_{m,0}  \;, \nonumber \\
  R_5^{\rm c}(m,0)&\approx& \left(\frac{1}{4}-\frac{3}{2V^2}\right) \delta_{m,0}
  -\frac{1}{2V^2}\delta_{m,1} \;, \nonumber \\
  R_5^{\rm s}(m,0)&\approx& \left(-\frac{1}{4}+ \frac{1}{2V^2}\right)
  \delta_{m,0}
  +\frac{1}{2V^2}\delta_{m,1} \; .\nonumber \\
\end{eqnarray}
The excitations are all short-ranged so that the functions $R_i$ are
finite only for $m= 0,1$.
In eq.~(\ref{eq:twosumformulafinalnow}) we thus find
%\begin{eqnarray}
%\delta e_0^{(2)}(V\gg t) &\approx &
%  - \frac{V^2}{4V} \Biggl[
%    \left(\frac{1}{4}-\left(\frac{1}{2}-\frac{2}{V^2}\right)^2-\frac{1}{V^2}
%    \right)^2
%    \nonumber \\
%    && \hphantom{- \frac{V^2}{4V} \biggl[}
%      + 2\left( -\left(-\frac{1}{V^2}\right)^2
%      -\left(-\frac{1}{2V}\right)^2+ \left(\frac{1}{2V}\right)^2
%      \right)^2 \nonumber \\
%      && \hphantom{- \frac{V^2}{4V} \biggl[}
%        + 2\left(
%        \left(-\frac{1}{V}\right)^2
        %-\left(\frac{1}{4}\right)^2
        %+ \left(\frac{1}{4}\right)^2
  %      + \left(\frac{1}{4}-\frac{3}{2V^2}\right)^2
  %      -\left(-\frac{1}{4}+ \frac{1}{2V^2}\right)^2
  %             \right)^2 \nonumber \\
  %    && \hphantom{- \frac{V^2}{4V} \biggl[}
  %        +        2\left( \left(\frac{1}{2V^2}\right)^2
  %               -\left(\frac{1}{2V^2}\right)^2\right)^2
   %              \nonumber \\
   %                    && \hphantom{- \frac{V^2}{4V} \biggl[}
   %                -8 \left(\left(-\frac{1}{4}+ \frac{1}{2V^2}\right)\frac{1}{4}
  %-\left(\frac{1}{4}-\frac{3}{2V^2}\right)\left(-\frac{1}{4}\right)\right)^2
 % \Biggr]\nonumber\\
  %               &= & -\frac{1}{4V^3}\; ,
% \end{eqnarray}
\begin{equation}
  \delta e_0^{(2)}(V\gg t) \approx -\frac{1}{4V^3}
  \end{equation}
with corrections of the order $V^{-5}$.
Apparently, the second-order corrections remain finite for $V\gg t$,
\begin{eqnarray}
  e_0^{\rm HF}(V\gg t) &\approx & -\frac{1}{V}+ 2
  \frac{1}{V^3} \; ,\nonumber \\
    e_0^{\rm HF,2nd}(V\gg t) &\approx & -\frac{1}{V}+ \left(2-\frac{1}{4}\right)
  \frac{1}{V^3} \; , \nonumber\\
  e_0^{\rm exact}(V\gg t) &\approx & -\frac{1}{V}+ \frac{1}{V^3} \; .
  \label{appeq:2ndHFstrongcoupling}
\end{eqnarray}
The second order correction mildly improves the result
of the Hartree-Fock energy for
strong interactions.

\subsubsection{Vanishing charge-density wave order}

We can test our analytical expressions against the exact
second-order coefficient $e_0^{(2)}$ in eq.~(19) of the main text,
\begin{equation}
 e_0^{(2)}=  -\frac{2}{3 \pi^3} + \frac{1}{36 \pi}
 \approx -0.0126591 \; .
 \label{eq:e0secondanalyticnoVagain}
\end{equation}
When there is no charge-density wave order parameter, i.e., $n_a\equiv 0$,
$B_0= 1/\pi$, we may set $V= 0$ in eq.~(\ref{eq:twosumformulafinalnow}).
Since there is no symmetry breaking and $V= 0$, we see that $R_1\equiv 0$,
$R_5^{\rm c,s}\equiv 0$, and $\widetilde{\epsilon}(k)= \epsilon(k) =  -2t\cos(k)$.
In the thermodynamic limit we have
\begin{equation}
  \frac{1}{L}\sum_{-\pi/2\leq k<\pi/2} \to \int_{-\pi/2}^{\pi/2} \frac{{\rm d}k}{2\pi}
  \; , 
\end{equation}
and the integrals for the remaining functions $R_i$ are evaluated numerically
for $m\leq 50$ using {\sc Mathematica},\cite{Mathematica12APP}
\begin{eqnarray}
  e_0^{(2)} &\approx &
  -  \int_0^{\infty} {\rm d }\lambda
 \Bigl[\left[R_0(0,\lambda)\right]^2
    -\left[R_2^{\rm c}(0,\lambda) \right]^2
    + \left[R_2^{\rm s}(0,\lambda) \right]^2\Bigr]^2 \nonumber \\
&& +2  \sum_{m= 1}^{50}
\Bigl[\left[R_0(m,\lambda)\right]^2
    -\left[R_2^{\rm c}(m,\lambda) \right]^2
    + \left[R_2^{\rm s}(m,\lambda) \right]^2\Bigr]^2
\nonumber \\
    && 
    +2\sum_{m= 0}^{50}
\Bigl[  \left[R_3(m,\lambda)\right]^2
    -\left[R_4^{\rm c}(m,\lambda) \right]^2
    + \left[R_4^{\rm s}(m,\lambda) \right]^2\Bigr]^2\nonumber \\
&= & -0.0126587 %\;,
\end{eqnarray}
with relative deviations of the order $10^{-5}$ from the analytic
result~(\ref{eq:e0secondanalyticnoVagain}).
Note that the $m$th entries
in the lattice sums in eq.~(\ref{eq:twosumformulafinalnow})
decay proportional to $m^{-3}$.

The same result can be derived within seconds with an accuracy of $10^{-7}$
from the triple integrals in eq.~(\ref{eq:tripleintegral}),
\begin{equation}
  e_0^{(2)} \approx  -0.01265908184 \; .
\end{equation}

In the absence of a charge-density wave,
the ground-state energy as a function of $B_0$
is given by
\begin{eqnarray}
  e_0^{(2)}(B_0,0,V)&=& -\frac{2}{\pi}\left(1+B_0 V\right) +
  \left(\frac{1}{4}+B_0^2\right)V \nonumber \\
&&    + \left(-\frac{2}{3 \pi^3} + \frac{1}{36 \pi}\right)\frac{V^2}{1+VB_0}
\end{eqnarray}
because the Fock term just renormalizes the
dispersion relation, $\widetilde{\epsilon}(k)=(1+VB_0)\epsilon(k)$. 
The resulting minimization problem leads to a third-order equation
that can be solved analytically. For $V\lesssim 1.5$
we may use the approximation
%\begin{eqnarray}
\begin{equation}
  B_0(V\lesssim 1.5)\approx \frac{1}{\pi}
  -\frac{V^2 }{2}\frac{24-\pi^2}{36\pi(\pi+V)^2} \; .
  %\\
%  &&   - \frac{(24-\pi^2)^2V^5(\pi^2-5\pi V+15V^2)}{
%72 \pi^3 (36 \pi^5 -\pi^2 (24 - \pi^2) V^3 +
%   3 \pi (24 - \pi^2) V^4 - 6 (24 - \pi^2) V^5)
%  }\nonumber 
%\end{eqnarray}
\end{equation}

For moderately large~$V$, we find that $B_0$ is almost constant
so that the energy increases linearly with~$V$,
\begin{equation}
  e_0^{(2)}(B_0,0,4<V<10) \approx
-\frac{2}{\pi} +\gamma V +{\cal O}(V^2)
\end{equation}
and $\gamma\approx 1/4-1/\pi^2\approx 0.15$.
In the absence of the charge-density
wave order, the energy becomes positive around $V_+=5$.
Since the energy must go to zero for large interactions,
the CDW order must set in at some critical value $V_{\rm c}^{(2)}<V_+$.

\subsection{Occupation numbers}

The matrix element in eq.~(107) in the main text is readily calculated,
\begin{equation}
  \langle k_1,k_2;p_1,p_2  | \hat{n}_{s,\beta}| k_1,k_2;p_1,p_2
  \rangle = \delta_{k_1,s}+\delta_{k_2,s}
  \; .
\end{equation}
When we use the symmetry of the two-particle matrix element
in $k_1 \leftrightarrow k_2$, we readily see that
\begin{eqnarray}
  n_{s,\beta}&=&  \frac{1}{2L^2}  \sum_{p_1,p_2,k}
  \frac{V^2\left|A(s,k;p_1,p_2)\right|^2\delta_{s+k-p_1-p_2,0}   }{
    [E(s)+E(k)+E(p_1)+E(p_2)]^2}
  \nonumber \\
  && \hphantom{ \frac{1}{2L^2}  \sum_{p_1,p_2,k}}
  + \frac{V^2\left|B(s,k;p_1,p_2)\right|^2 }{
    [E(s)+E(k)+E(p_1)+E(p_2)]^2}\nonumber\\
  && \hphantom{ \frac{1}{2L^2}  \sum_{p_1,p_2,k}}
  \times \left(       \delta_{s+k-p_1-p_2,-\pi} + \delta_{s+k-p_1-p_2,\pi}\right)\;      .\nonumber \\
\end{eqnarray}
The triple sum reduces to a double sum when the Kronecker conditions
are taken into account, see Sect.~\ref{sec:gsenergy}.
Note that this expression is evaluated
at the optimal values for $B_0$ and $n_a$ so that the function is
evaluated only once.

The occupation numbers in second-order perturbation theory
can then be written as
\begin{eqnarray}
  n_{s,\beta}&=&  \frac{1}{2L^2} \sum_{s\leq p_1<\pi/2}
  \sum_{-\pi/2\leq p_2<\pi/2-p_1+s} \nonumber\\
&&  \frac{V^2\left|A(s,p_1+p_2-s;p_1,p_2)\right|^2}{
    [E(s)+E(p_1+p_2-s)+E(p_1)+E(p_2)]^2}\nonumber\\
  && + \frac{1}{2L^2} \sum_{-\pi/2\leq p_1<s}
  \sum_{-\pi/2+s-p_1\leq p_2<\pi/2}\nonumber\\
 && \frac{V^2\left|A(s,p_1+p_2-s;p_1,p_2)\right|^2}{
    [E(s)+E(p_1+p_2-s)+E(p_1)+E(p_2)]^2}\nonumber\\
&&+  \frac{1}{2L^2} \sum_{-\pi/2\leq p_1<s}
  \sum_{-\pi/2\leq p_2<-\pi/2+s-p_1} \nonumber\\
&&  \frac{V^2\left|B(s,p_1+p_2-s+\pi;p_1,p_2)\right|^2}{
    [E(s)+E(p_1+p_2-s+\pi)+E(p_1)+E(p_2)]^2} \nonumber \\
  &&+  \frac{1}{2L^2} \sum_{s< p_1<\pi/2}
  \sum_{\pi/2+s-p_1\leq p_2<\pi/2}\nonumber\\
&&  \frac{V^2\left|B(s,p_1+p_2-s-\pi;p_1,p_2)\right|^2}{
    [E(s)+E(p_1+p_2-s-\pi)+E(p_1)+E(p_2)]^2}\, .
    \nonumber \\\label{eq:nsbetasums}
\end{eqnarray}
This expression has to be evaluated for each $-\pi/2 \leq s<\pi/2$
using eqs.~(\ref{eq:finalAsq}) and~(\ref{eq:finalBsq}).

Note that the resulting momentum distribution is asymmetric
because we broke inversion symmetry on finite lattices by choosing
$k=-\pi/2$ occupied and $k=\pi/2$ unoccupied.
Therefore, the momentum distribution must be symmetrized.
In the region $-\pi/2\leq s \leq \pi/2$ we set 
\begin{equation}
  \bar{n}_{\beta,s}= \left\{
  \begin{array}{rcl}
    n_{\beta,-\pi/2} & \hbox{for} & s=-\pi/2 \; , \\
    \left(n_{\beta,s}+n_{\beta,-s}\right)/2&\hbox{for} &
    -\pi/2<s<\pi/2\; ,\\
    n_{\beta,-\pi/2} & \hbox{for} & s=\pi/2 \; .
    \end{array}
  \right.
\end{equation}
Here, $s=2\pi m_s/L$ with $m_s=-L/4,-L/4+1,\ldots,L/4$.

It is possible to express the momentum distribution using a single
sum plus an integral. This is not necessary because
the expression~(\ref{eq:nsbetasums})
  involves only a double sum for each momentum value
so that there are no memory problems even at $L=1000$.

In the thermodynamic limit, we find that the first two terms
in eq.~(\ref{eq:nsbetasums})
map onto each other by inverting the momentum,
$n_{s,\beta}^{(1)}
= n_{-s,\beta}^{(2)}$. Likewise,
$n_{s,\beta}^{(3)}= n_{-s,\beta}^{(4)}$ so that
the momentum distribution
is inversion symmetric,
$n_{s,\beta}= n_{-s,\beta}$.
Thus we find,
\begin{eqnarray}
  n_{s,\beta}&=& n_{s,\beta}^{(1)}+ n_{-s,\beta}^{(1)}  + n_{s,\beta}^{(4)}
+   n_{-s,\beta}^{(4)} \; , \nonumber \\[6pt]
  n_{s,\beta}^{(1)}&= & \frac{V^2}{2} \int_s^{\pi/2} \frac{{\rm d}p_1}{2\pi}
  \int_{-\pi/2}^{\pi/2-p_1+ s} \frac{{\rm d}p_2}{2\pi}\nonumber\\
&&\frac{\left|A(s,p_1+p_2-s;p_1,p_2)\right|^2}{
    [E(s)+E(p_1+p_2-s)+E(p_1)+E(p_2)]^2}  \; , \nonumber \\[6pt]
  n_{s,\beta}^{(4)}&= & \frac{V^2}{2} \int_s^{\pi/2} \frac{{\rm d}p_1}{2\pi}
  \int_{\pi/2-p_1+ s}^{\pi/2} \frac{{\rm d}p_2}{2\pi}\nonumber \\
 && \frac{\left|B(s,p_1+p_2-s-\pi;p_1,p_2)\right|^2}{
    [E(s)+E(p_1+p_2-s-\pi)+E(p_1)+E(p_2)]^2}\, .
  \nonumber\\
\end{eqnarray}
On an i7 notebook, the momentum distributions are done in minutes.

Of particular interest is the density of particles
in the upper Hartree-Fock band,
\begin{equation}
  n_{\beta}(V)
  = \frac{1}{L} \sum_s n_{\beta,s}= \int_{-\pi/2}^{\pi/2}
  \frac{{\rm d}s}{2\pi}  n_{\beta,s} \leq \frac{1}{2} \; .
\end{equation}
Since Hartree-Fock theory becomes exact for $V\to 0$ and for $V\to \infty$,
we have
\begin{eqnarray}
  n_{\beta}(V\to 0) &\to & 0 \; , \nonumber \\
  n_{\beta}(V\to \infty) &\to & 0 \; ,
\end{eqnarray}
so that $n_{\beta}(V)$ has (at least) one maximum. We expect that
the physical properties change at the maximum,
i.e., $V_{\rm max}= V_{\rm c}$. Therefore, $n_{\beta}(V)$ can be used to monitor
the charge-density wave transition.

%\bibliography{tVmodel}% Produces the bibliography via BibTeX.

\begin{thebibliography}{56}%
\makeatletter
\providecommand \@ifxundefined [1]{%
 \@ifx{#1\undefined}
}%
\providecommand \@ifnum [1]{%
 \ifnum #1\expandafter \@firstoftwo
 \else \expandafter \@secondoftwo
 \fi
}%
\providecommand \@ifx [1]{%
 \ifx #1\expandafter \@firstoftwo
 \else \expandafter \@secondoftwo
 \fi
}%
\providecommand \natexlab [1]{#1}%
\providecommand \enquote  [1]{``#1''}%
\providecommand \bibnamefont  [1]{#1}%
\providecommand \bibfnamefont [1]{#1}%
\providecommand \citenamefont [1]{#1}%
\providecommand \href@noop [0]{\@secondoftwo}%
\providecommand \href [0]{\begingroup \@sanitize@url \@href}%
\providecommand \@href[1]{\@@startlink{#1}\@@href}%
\providecommand \@@href[1]{\endgroup#1\@@endlink}%
\providecommand \@sanitize@url [0]{\catcode `\\12\catcode `\$12\catcode
  `\&12\catcode `\#12\catcode `\^12\catcode `\_12\catcode `\%12\relax}%
\providecommand \@@startlink[1]{}%
\providecommand \@@endlink[0]{}%
\providecommand \url  [0]{\begingroup\@sanitize@url \@url }%
\providecommand \@url [1]{\endgroup\@href {#1}{\urlprefix }}%
\providecommand \urlprefix  [0]{URL }%
\providecommand \Eprint [0]{\href }%
\providecommand \doibase [0]{http://dx.doi.org/}%
\providecommand \selectlanguage [0]{\@gobble}%
\providecommand \bibinfo  [0]{\@secondoftwo}%
\providecommand \bibfield  [0]{\@secondoftwo}%
\providecommand \translation [1]{[#1]}%
\providecommand \BibitemOpen [0]{}%
\providecommand \bibitemStop [0]{}%
\providecommand \bibitemNoStop [0]{.\EOS\space}%
\providecommand \EOS [0]{\spacefactor3000\relax}%
\providecommand \BibitemShut  [1]{\csname bibitem#1\endcsname}%
\let\auto@bib@innerbib\@empty
%</preamble>
\bibitem [{\citenamefont {Bethe}(1931)}]{Bethe1931}%
  \BibitemOpen
  \bibfield  {author} {\bibinfo {author} {\bibfnamefont {H.}~\bibnamefont
  {Bethe}},\ }\href@noop {} {\bibfield  {journal} {\bibinfo  {journal}
  {Zeitschrift f\"ur Physik}\ }\textbf {\bibinfo {volume} {71}},\ \bibinfo
  {pages} {205} (\bibinfo {year} {1931})}\BibitemShut {NoStop}%
\bibitem [{\citenamefont {Orbach}(1958)}]{Orbach}%
  \BibitemOpen
  \bibfield  {author} {\bibinfo {author} {\bibfnamefont {R.}~\bibnamefont
  {Orbach}},\ }\href@noop {} {\bibfield  {journal} {\bibinfo  {journal} {Phys.
  Rev.}\ }\textbf {\bibinfo {volume} {112}},\ \bibinfo {pages} {309} (\bibinfo
  {year} {1958})}\BibitemShut {NoStop}%
\bibitem [{\citenamefont {Yang}\ and\ \citenamefont
  {Yang}(1966{\natexlab{a}})}]{PhysRev.150.321}%
  \BibitemOpen
  \bibfield  {author} {\bibinfo {author} {\bibfnamefont {C.~N.}\ \bibnamefont
  {Yang}}\ and\ \bibinfo {author} {\bibfnamefont {C.~P.}\ \bibnamefont
  {Yang}},\ }\href@noop {} {\bibfield  {journal} {\bibinfo  {journal} {Phys.
  Rev.}\ }\textbf {\bibinfo {volume} {150}},\ \bibinfo {pages} {321} (\bibinfo
  {year} {1966}{\natexlab{a}})}\BibitemShut {NoStop}%
\bibitem [{\citenamefont {Yang}\ and\ \citenamefont
  {Yang}(1966{\natexlab{b}})}]{PhysRev.150.327}%
  \BibitemOpen
  \bibfield  {author} {\bibinfo {author} {\bibfnamefont {C.~N.}\ \bibnamefont
  {Yang}}\ and\ \bibinfo {author} {\bibfnamefont {C.~P.}\ \bibnamefont
  {Yang}},\ }\href@noop {} {\bibfield  {journal} {\bibinfo  {journal} {Phys.
  Rev.}\ }\textbf {\bibinfo {volume} {150}},\ \bibinfo {pages} {327} (\bibinfo
  {year} {1966}{\natexlab{b}})}\BibitemShut {NoStop}%
\bibitem [{\citenamefont {Johnson}\ \emph {et~al.}(1973)\citenamefont
  {Johnson}, \citenamefont {Krinsky},\ and\ \citenamefont
  {McCoy}}]{PhysRevA.8.2526}%
  \BibitemOpen
  \bibfield  {author} {\bibinfo {author} {\bibfnamefont {J.~D.}\ \bibnamefont
  {Johnson}}, \bibinfo {author} {\bibfnamefont {S.}~\bibnamefont {Krinsky}}, \
  and\ \bibinfo {author} {\bibfnamefont {B.~M.}\ \bibnamefont {McCoy}},\
  }\href@noop {} {\bibfield  {journal} {\bibinfo  {journal} {Phys. Rev. A}\
  }\textbf {\bibinfo {volume} {8}},\ \bibinfo {pages} {2526} (\bibinfo {year}
  {1973})}\BibitemShut {NoStop}%
\bibitem [{\citenamefont {Babelon}\ \emph {et~al.}(1983)\citenamefont
  {Babelon}, \citenamefont {{de Vega}},\ and\ \citenamefont
  {Viallet}}]{BABELON198313}%
  \BibitemOpen
  \bibfield  {author} {\bibinfo {author} {\bibfnamefont {O.}~\bibnamefont
  {Babelon}}, \bibinfo {author} {\bibfnamefont {H.~J.}\ \bibnamefont {{de
  Vega}}}, \ and\ \bibinfo {author} {\bibfnamefont {C.~M.}\ \bibnamefont
  {Viallet}},\ }\href@noop {} {\bibfield  {journal} {\bibinfo  {journal}
  {Nucl.\ Phys.\ B}\ }\textbf {\bibinfo {volume} {220}},\ \bibinfo {pages} {13}
  (\bibinfo {year} {1983})}\BibitemShut {NoStop}%
\bibitem [{\citenamefont {Woynarovich}(1982)}]{Woynarovich82}%
  \BibitemOpen
  \bibfield  {author} {\bibinfo {author} {\bibfnamefont {F.}~\bibnamefont
  {Woynarovich}},\ }\href@noop {} {\bibfield  {journal} {\bibinfo  {journal}
  {Journal of Physics A: Mathematical and General}\ }\textbf {\bibinfo {volume}
  {15}},\ \bibinfo {pages} {2985} (\bibinfo {year} {1982})}\BibitemShut
  {NoStop}%
\bibitem [{\citenamefont {Virosztek}\ and\ \citenamefont
  {Woynarovich}(1984)}]{ViWo84}%
  \BibitemOpen
  \bibfield  {author} {\bibinfo {author} {\bibfnamefont {A.}~\bibnamefont
  {Virosztek}}\ and\ \bibinfo {author} {\bibfnamefont {F.}~\bibnamefont
  {Woynarovich}},\ }\href@noop {} {\bibfield  {journal} {\bibinfo  {journal}
  {Journal of Physics A: Mathematical and General}\ }\textbf {\bibinfo {volume}
  {17}},\ \bibinfo {pages} {3029} (\bibinfo {year} {1984})}\BibitemShut
  {NoStop}%
\bibitem [{\citenamefont {Baxter}(1973)}]{Baxter1973}%
  \BibitemOpen
  \bibfield  {author} {\bibinfo {author} {\bibfnamefont {R.~J.}\ \bibnamefont
  {Baxter}},\ }\href@noop {} {\bibfield  {journal} {\bibinfo  {journal}
  {Journal of Statistical Physics}\ }\textbf {\bibinfo {volume} {9}},\ \bibinfo
  {pages} {145} (\bibinfo {year} {1973})}\BibitemShut {NoStop}%
\bibitem [{\citenamefont {Izergin}\ \emph {et~al.}(1999)\citenamefont
  {Izergin}, \citenamefont {Kitanine}, \citenamefont {Maillet},\ and\
  \citenamefont {Terras}}]{IZERGIN1999679}%
  \BibitemOpen
  \bibfield  {author} {\bibinfo {author} {\bibfnamefont {A.~G.}\ \bibnamefont
  {Izergin}}, \bibinfo {author} {\bibfnamefont {N.}~\bibnamefont {Kitanine}},
  \bibinfo {author} {\bibfnamefont {J.~M.}\ \bibnamefont {Maillet}}, \ and\
  \bibinfo {author} {\bibfnamefont {V.}~\bibnamefont {Terras}},\ }\href@noop {}
  {\bibfield  {journal} {\bibinfo  {journal} {Nucl.\ Phys.\ B}\ }\textbf
  {\bibinfo {volume} {554}},\ \bibinfo {pages} {679} (\bibinfo {year}
  {1999})}\BibitemShut {NoStop}%
\bibitem [{\citenamefont {Kl\"umper}(1993)}]{Kluemper93}%
  \BibitemOpen
  \bibfield  {author} {\bibinfo {author} {\bibfnamefont {A.}~\bibnamefont
  {Kl\"umper}},\ }\href@noop {} {\bibfield  {journal} {\bibinfo  {journal}
  {Zeitschrift f\"ur Physik B: Condensed Matter}\ }\textbf {\bibinfo {volume}
  {91}},\ \bibinfo {pages} {507} (\bibinfo {year} {1993})}\BibitemShut
  {NoStop}%
\bibitem [{\citenamefont {Jimbo}\ \emph {et~al.}(1992)\citenamefont {Jimbo},
  \citenamefont {Miki}, \citenamefont {Miwa},\ and\ \citenamefont
  {Nakayashiki}}]{JMMN92}%
  \BibitemOpen
  \bibfield  {author} {\bibinfo {author} {\bibfnamefont {M.}~\bibnamefont
  {Jimbo}}, \bibinfo {author} {\bibfnamefont {K.}~\bibnamefont {Miki}},
  \bibinfo {author} {\bibfnamefont {T.}~\bibnamefont {Miwa}}, \ and\ \bibinfo
  {author} {\bibfnamefont {A.}~\bibnamefont {Nakayashiki}},\ }\href@noop {}
  {\bibfield  {journal} {\bibinfo  {journal} {Physics Letters A}\ }\textbf
  {\bibinfo {volume} {168}},\ \bibinfo {pages} {256} (\bibinfo {year}
  {1992})}\BibitemShut {NoStop}%
\bibitem [{\citenamefont {Jimbo}\ and\ \citenamefont {Miwa}(1996)}]{JiMi96}%
  \BibitemOpen
  \bibfield  {author} {\bibinfo {author} {\bibfnamefont {M.}~\bibnamefont
  {Jimbo}}\ and\ \bibinfo {author} {\bibfnamefont {T.}~\bibnamefont {Miwa}},\
  }\href@noop {} {\bibfield  {journal} {\bibinfo  {journal} {Journal of Physics
  A: Mathematical and General}\ }\textbf {\bibinfo {volume} {29}},\ \bibinfo
  {pages} {2923} (\bibinfo {year} {1996})}\BibitemShut {NoStop}%
\bibitem [{\citenamefont {Kitanine}\ \emph {et~al.}(2000)\citenamefont
  {Kitanine}, \citenamefont {Maillet},\ and\ \citenamefont {Terras}}]{KMT00}%
  \BibitemOpen
  \bibfield  {author} {\bibinfo {author} {\bibfnamefont {N.}~\bibnamefont
  {Kitanine}}, \bibinfo {author} {\bibfnamefont {J.~M.}\ \bibnamefont
  {Maillet}}, \ and\ \bibinfo {author} {\bibfnamefont {V.}~\bibnamefont
  {Terras}},\ }\href@noop {} {\bibfield  {journal} {\bibinfo  {journal}
  {Nuclear Physics B}\ }\textbf {\bibinfo {volume} {567}},\ \bibinfo {pages}
  {554} (\bibinfo {year} {2000})}\BibitemShut {NoStop}%
\bibitem [{\citenamefont {G\"ohmann}\ \emph {et~al.}(2005)\citenamefont
  {G\"ohmann}, \citenamefont {Kl\"umper},\ and\ \citenamefont {Seel}}]{GKS05}%
  \BibitemOpen
  \bibfield  {author} {\bibinfo {author} {\bibfnamefont {F.}~\bibnamefont
  {G\"ohmann}}, \bibinfo {author} {\bibfnamefont {A.}~\bibnamefont
  {Kl\"umper}}, \ and\ \bibinfo {author} {\bibfnamefont {A.}~\bibnamefont
  {Seel}},\ }\href@noop {} {\bibfield  {journal} {\bibinfo  {journal} {Journal
  of Physics A: Mathematical and General}\ }\textbf {\bibinfo {volume} {38}},\
  \bibinfo {pages} {1833} (\bibinfo {year} {2005})}\BibitemShut {NoStop}%
\bibitem [{\citenamefont {Boos}\ \emph {et~al.}(2009)\citenamefont {Boos},
  \citenamefont {Jimbo}, \citenamefont {Miwa}, \citenamefont {Smirnov},\ and\
  \citenamefont {{Ta\-ke\-yama}}}]{BJMST08a}%
  \BibitemOpen
  \bibfield  {author} {\bibinfo {author} {\bibfnamefont {H.}~\bibnamefont
  {Boos}}, \bibinfo {author} {\bibfnamefont {M.}~\bibnamefont {Jimbo}},
  \bibinfo {author} {\bibfnamefont {T.}~\bibnamefont {Miwa}}, \bibinfo {author}
  {\bibfnamefont {F.}~\bibnamefont {Smirnov}}, \ and\ \bibinfo {author}
  {\bibfnamefont {Y.}~\bibnamefont {{Ta\-ke\-yama}}},\ }\href@noop {}
  {\bibfield  {journal} {\bibinfo  {journal} {Communications in Mathematical
  Physics}\ }\textbf {\bibinfo {volume} {286}},\ \bibinfo {pages} {875}
  (\bibinfo {year} {2009})}\BibitemShut {NoStop}%
\bibitem [{\citenamefont {Jimbo}\ \emph {et~al.}(2009)\citenamefont {Jimbo},
  \citenamefont {Miwa},\ and\ \citenamefont {Smirnov}}]{JMS09}%
  \BibitemOpen
  \bibfield  {author} {\bibinfo {author} {\bibfnamefont {M.}~\bibnamefont
  {Jimbo}}, \bibinfo {author} {\bibfnamefont {T.}~\bibnamefont {Miwa}}, \ and\
  \bibinfo {author} {\bibfnamefont {F.}~\bibnamefont {Smirnov}},\ }\href@noop
  {} {\bibfield  {journal} {\bibinfo  {journal} {Journal of Physics A:
  Mathematical and General}\ }\textbf {\bibinfo {volume} {42}},\ \bibinfo
  {pages} {304018} (\bibinfo {year} {2009})}\BibitemShut {NoStop}%
\bibitem [{\citenamefont {Boos}\ and\ \citenamefont
  {G\"ohmann}(2009)}]{BoGo09}%
  \BibitemOpen
  \bibfield  {author} {\bibinfo {author} {\bibfnamefont {H.}~\bibnamefont
  {Boos}}\ and\ \bibinfo {author} {\bibfnamefont {F.}~\bibnamefont
  {G\"ohmann}},\ }\href@noop {} {\bibfield  {journal} {\bibinfo  {journal}
  {Journal of Physics A: Mathematical and General}\ }\textbf {\bibinfo {volume}
  {42}},\ \bibinfo {pages} {315001} (\bibinfo {year} {2009})}\BibitemShut
  {NoStop}%
\bibitem [{\citenamefont {Bortz}\ and\ \citenamefont
  {G\"ohmann}(2005)}]{BortzGoehmann2005}%
  \BibitemOpen
  \bibfield  {author} {\bibinfo {author} {\bibfnamefont {M.}~\bibnamefont
  {Bortz}}\ and\ \bibinfo {author} {\bibfnamefont {F.}~\bibnamefont
  {G\"ohmann}},\ }\href@noop {} {\bibfield  {journal} {\bibinfo  {journal}
  {Eur. Phys. J. B}\ }\textbf {\bibinfo {volume} {46}},\ \bibinfo {pages} {399}
  (\bibinfo {year} {2005})}\BibitemShut {NoStop}%
\bibitem [{\citenamefont {Damerau}\ \emph {et~al.}(2007)\citenamefont
  {Damerau}, \citenamefont {G\"ohmann}, \citenamefont {Hasenclever},\ and\
  \citenamefont {{Kl\"um- \newline per}}}]{Damerauetal2007}%
  \BibitemOpen
  \bibfield  {author} {\bibinfo {author} {\bibfnamefont {J.}~\bibnamefont
  {Damerau}}, \bibinfo {author} {\bibfnamefont {F.}~\bibnamefont {G\"ohmann}},
  \bibinfo {author} {\bibfnamefont {N.}~\bibnamefont {Hasenclever}}, \ and\
  \bibinfo {author} {\bibfnamefont {A.}~\bibnamefont {{Kl\"um- \newline
  per}}},\ }\href@noop {} {\bibfield  {journal} {\bibinfo  {journal} {J.\
  Phys.\ A}\ }\textbf {\bibinfo {volume} {40}},\ \bibinfo {pages} {4439}
  (\bibinfo {year} {2007})}\BibitemShut {NoStop}%
\bibitem [{\citenamefont {Boos}\ \emph {et~al.}(2008)\citenamefont {Boos},
  \citenamefont {Damerau}, \citenamefont {G\"ohmann}, \citenamefont
  {Kl\"umper}, \citenamefont {{Su\-zu\-ki}},\ and\ \citenamefont
  {Wei{\ss}e}}]{Boosetal2008}%
  \BibitemOpen
  \bibfield  {author} {\bibinfo {author} {\bibfnamefont {H.~E.}\ \bibnamefont
  {Boos}}, \bibinfo {author} {\bibfnamefont {J.}~\bibnamefont {Damerau}},
  \bibinfo {author} {\bibfnamefont {F.}~\bibnamefont {G\"ohmann}}, \bibinfo
  {author} {\bibfnamefont {A.}~\bibnamefont {Kl\"umper}}, \bibinfo {author}
  {\bibfnamefont {J.}~\bibnamefont {{Su\-zu\-ki}}}, \ and\ \bibinfo {author}
  {\bibfnamefont {A.}~\bibnamefont {Wei{\ss}e}},\ }\href@noop {} {\bibfield
  {journal} {\bibinfo  {journal} {Journal of Statistical Mechanics: Theory and
  Experiment}\ }\textbf {\bibinfo {volume} {2008}},\ \bibinfo {pages} {P08010}
  (\bibinfo {year} {2008})}\BibitemShut {NoStop}%
\bibitem [{\citenamefont {Babenko}\ \emph {et~al.}(2021)\citenamefont
  {Babenko}, \citenamefont {G\"ohmann}, \citenamefont {Kozlowski},
  \citenamefont {Sirker},\ and\ \citenamefont
  {Suzuki}}]{PhysRevLett.126.210602}%
  \BibitemOpen
  \bibfield  {author} {\bibinfo {author} {\bibfnamefont {C.}~\bibnamefont
  {Babenko}}, \bibinfo {author} {\bibfnamefont {F.}~\bibnamefont {G\"ohmann}},
  \bibinfo {author} {\bibfnamefont {K.~K.}\ \bibnamefont {Kozlowski}}, \bibinfo
  {author} {\bibfnamefont {J.}~\bibnamefont {Sirker}}, \ and\ \bibinfo {author}
  {\bibfnamefont {J.}~\bibnamefont {Suzuki}},\ }\href@noop {} {\bibfield
  {journal} {\bibinfo  {journal} {Phys. Rev. Lett.}\ }\textbf {\bibinfo
  {volume} {126}},\ \bibinfo {pages} {210602} (\bibinfo {year}
  {2021})}\BibitemShut {NoStop}%
\bibitem [{\citenamefont {G\"{o}hmann}\ \emph {et~al.}(2022)\citenamefont
  {G\"{o}hmann}, \citenamefont {Kozlowski}, \citenamefont {Sirker},\ and\
  \citenamefont {Suzuki}}]{Goehmannetal2021}%
  \BibitemOpen
  \bibfield  {author} {\bibinfo {author} {\bibfnamefont {F.}~\bibnamefont
  {G\"{o}hmann}}, \bibinfo {author} {\bibfnamefont {K.~K.}\ \bibnamefont
  {Kozlowski}}, \bibinfo {author} {\bibfnamefont {J.}~\bibnamefont {Sirker}}, \
  and\ \bibinfo {author} {\bibfnamefont {J.}~\bibnamefont {Suzuki}},\
  }\href@noop {} {\bibfield  {journal} {\bibinfo  {journal} {SciPost Phys.}\
  }\textbf {\bibinfo {volume} {12}},\ \bibinfo {pages} {158} (\bibinfo {year}
  {2022})}\BibitemShut {NoStop}%
\bibitem [{\citenamefont {Jordan}\ and\ \citenamefont
  {Wigner}(1928)}]{JordanWigner1928}%
  \BibitemOpen
  \bibfield  {author} {\bibinfo {author} {\bibfnamefont {P.}~\bibnamefont
  {Jordan}}\ and\ \bibinfo {author} {\bibfnamefont {E.}~\bibnamefont
  {Wigner}},\ }\href@noop {} {\bibfield  {journal} {\bibinfo  {journal}
  {Zeit\-schrift f\"ur Phy\-sik}\ }\textbf {\bibinfo {volume} {47}},\ \bibinfo
  {pages} {631} (\bibinfo {year} {1928})}\BibitemShut {NoStop}%
\bibitem [{\citenamefont {Di~Castro}\ and\ \citenamefont
  {Metzner}(1991)}]{PhysRevLett.67.3852}%
  \BibitemOpen
  \bibfield  {author} {\bibinfo {author} {\bibfnamefont {C.}~\bibnamefont
  {Di~Castro}}\ and\ \bibinfo {author} {\bibfnamefont {W.}~\bibnamefont
  {Metzner}},\ }\href@noop {} {\bibfield  {journal} {\bibinfo  {journal} {Phys.
  Rev. Lett.}\ }\textbf {\bibinfo {volume} {67}},\ \bibinfo {pages} {3852}
  (\bibinfo {year} {1991})}\BibitemShut {NoStop}%
\bibitem [{\citenamefont {Shankar}(1994)}]{RevModPhys.66.129}%
  \BibitemOpen
  \bibfield  {author} {\bibinfo {author} {\bibfnamefont {R.}~\bibnamefont
  {Shankar}},\ }\href@noop {} {\bibfield  {journal} {\bibinfo  {journal} {Rev.
  Mod. Phys.}\ }\textbf {\bibinfo {volume} {66}},\ \bibinfo {pages} {129}
  (\bibinfo {year} {1994})}\BibitemShut {NoStop}%
\bibitem [{\citenamefont {Giamarchi}(2004)}]{Thierrybook}%
  \BibitemOpen
  \bibfield  {author} {\bibinfo {author} {\bibfnamefont {T.}~\bibnamefont
  {Giamarchi}},\ }\href@noop {} {\emph {\bibinfo {title} {{Quantum Physics in
  One Dimension}}}}\ (\bibinfo  {publisher} {Cla\-ren\-don Press},\ \bibinfo
  {address} {Oxford},\ \bibinfo {year} {2004})\BibitemShut {NoStop}%
\bibitem [{\citenamefont {Kosterlitz}\ and\ \citenamefont
  {Thouless}(1973)}]{KTtransition}%
  \BibitemOpen
  \bibfield  {author} {\bibinfo {author} {\bibfnamefont {J.~M.}\ \bibnamefont
  {Kosterlitz}}\ and\ \bibinfo {author} {\bibfnamefont {D.~J.}\ \bibnamefont
  {Thouless}},\ }\href@noop {} {\bibfield  {journal} {\bibinfo  {journal}
  {Journal of Physics C: Solid State Physics}\ }\textbf {\bibinfo {volume}
  {6}},\ \bibinfo {pages} {1181} (\bibinfo {year} {1973})}\BibitemShut
  {NoStop}%
\bibitem [{\citenamefont {K\"uhner}\ \emph {et~al.}(2000)\citenamefont
  {K\"uhner}, \citenamefont {White},\ and\ \citenamefont
  {Monien}}]{PhysRevB.61.12474}%
  \BibitemOpen
  \bibfield  {author} {\bibinfo {author} {\bibfnamefont {T.~D.}\ \bibnamefont
  {K\"uhner}}, \bibinfo {author} {\bibfnamefont {S.~R.}\ \bibnamefont {White}},
  \ and\ \bibinfo {author} {\bibfnamefont {H.}~\bibnamefont {Monien}},\
  }\href@noop {} {\bibfield  {journal} {\bibinfo  {journal} {Phys. Rev. B}\
  }\textbf {\bibinfo {volume} {61}},\ \bibinfo {pages} {12474} (\bibinfo {year}
  {2000})}\BibitemShut {NoStop}%
\bibitem [{\citenamefont {Montenegro-Filho}\ \emph {et~al.}(2020)\citenamefont
  {Montenegro-Filho}, \citenamefont {Matias},\ and\ \citenamefont
  {Coutinho-Filho}}]{PhysRevB.102.035137}%
  \BibitemOpen
  \bibfield  {author} {\bibinfo {author} {\bibfnamefont {R.~R.}\ \bibnamefont
  {Montenegro-Filho}}, \bibinfo {author} {\bibfnamefont {F.~S.}\ \bibnamefont
  {Matias}}, \ and\ \bibinfo {author} {\bibfnamefont {M.~D.}\ \bibnamefont
  {Coutinho-Filho}},\ }\href@noop {} {\bibfield  {journal} {\bibinfo  {journal}
  {Phys. Rev. B}\ }\textbf {\bibinfo {volume} {102}},\ \bibinfo {pages}
  {035137} (\bibinfo {year} {2020})}\BibitemShut {NoStop}%
\bibitem [{\citenamefont {Georges}\ and\ \citenamefont
  {Yedidia}(1991)}]{PhysRevB.43.3475}%
  \BibitemOpen
  \bibfield  {author} {\bibinfo {author} {\bibfnamefont {A.}~\bibnamefont
  {Georges}}\ and\ \bibinfo {author} {\bibfnamefont {J.~S.}\ \bibnamefont
  {Yedidia}},\ }\href@noop {} {\bibfield  {journal} {\bibinfo  {journal} {Phys.
  Rev. B}\ }\textbf {\bibinfo {volume} {43}},\ \bibinfo {pages} {3475}
  (\bibinfo {year} {1991})}\BibitemShut {NoStop}%
\bibitem [{\citenamefont {van Dongen}(1994)}]{PhysRevB.50.14016}%
  \BibitemOpen
  \bibfield  {author} {\bibinfo {author} {\bibfnamefont {P.~G.~J.}\
  \bibnamefont {van Dongen}},\ }\href@noop {} {\bibfield  {journal} {\bibinfo
  {journal} {Phys. Rev. B}\ }\textbf {\bibinfo {volume} {50}},\ \bibinfo
  {pages} {14016} (\bibinfo {year} {1994})}\BibitemShut {NoStop}%
\bibitem [{\citenamefont {White}(1992)}]{White-1992b}%
  \BibitemOpen
  \bibfield  {author} {\bibinfo {author} {\bibfnamefont {S.~R.}\ \bibnamefont
  {White}},\ }\href@noop {} {\bibfield  {journal} {\bibinfo  {journal} {Phys.
  Rev. Lett.}\ }\textbf {\bibinfo {volume} {69}},\ \bibinfo {pages} {2863}
  (\bibinfo {year} {1992})}\BibitemShut {NoStop}%
\bibitem [{\citenamefont {White}(1993)}]{White-1993}%
  \BibitemOpen
  \bibfield  {author} {\bibinfo {author} {\bibfnamefont {S.~R.}\ \bibnamefont
  {White}},\ }\href@noop {} {\bibfield  {journal} {\bibinfo  {journal} {Phys.
  Rev. B}\ }\textbf {\bibinfo {volume} {48}},\ \bibinfo {pages} {10345}
  (\bibinfo {year} {1993})}\BibitemShut {NoStop}%
\bibitem [{\citenamefont {Schollw\"ock}(2005)}]{Schollwock-2005}%
  \BibitemOpen
  \bibfield  {author} {\bibinfo {author} {\bibfnamefont {U.}~\bibnamefont
  {Schollw\"ock}},\ }\href@noop {} {\bibfield  {journal} {\bibinfo  {journal}
  {Rev. Mod. Phys.}\ }\textbf {\bibinfo {volume} {77}},\ \bibinfo {pages} {259}
  (\bibinfo {year} {2005})}\BibitemShut {NoStop}%
\bibitem [{sup()}]{suppmat}%
  \BibitemOpen
  \href@noop {} {}\bibinfo {note} {{See Supplemental Material at [URL will be
  inserted by publisher] for details on extrapolation schemes, series
  expansions of Bethe Ansatz results, and Hartree-Fock
  calculations.}}\BibitemShut {Stop}%
\bibitem [{\citenamefont {Banerjee}\ and\ \citenamefont
  {Wilkerson}(2017)}]{BanerjeeWilkerson}%
  \BibitemOpen
  \bibfield  {author} {\bibinfo {author} {\bibfnamefont {S.}~\bibnamefont
  {Banerjee}}\ and\ \bibinfo {author} {\bibfnamefont {B.}~\bibnamefont
  {Wilkerson}},\ }\href@noop {} {\bibfield  {journal} {\bibinfo  {journal}
  {International Journal of Number Theory}\ }\textbf {\bibinfo {volume} {13}},\
  \bibinfo {pages} {2097} (\bibinfo {year} {2017})}\BibitemShut {NoStop}%
\bibitem [{\citenamefont {Garoufalidis}\ and\ \citenamefont
  {Zagier}(2021)}]{GaroufalidisZagier2021}%
  \BibitemOpen
  \bibfield  {author} {\bibinfo {author} {\bibfnamefont {S.}~\bibnamefont
  {Garoufalidis}}\ and\ \bibinfo {author} {\bibfnamefont {D.}~\bibnamefont
  {Zagier}},\ }\href@noop {} {\bibfield  {journal} {\bibinfo  {journal} {The
  Ramanujan Journal}\ }\textbf {\bibinfo {volume} {55}},\ \bibinfo {pages}
  {219} (\bibinfo {year} {2021})}\BibitemShut {NoStop}%
\bibitem [{\citenamefont {Lieb}\ \emph {et~al.}(1961)\citenamefont {Lieb},
  \citenamefont {Schultz},\ and\ \citenamefont {Mattis}}]{LIEB1961407}%
  \BibitemOpen
  \bibfield  {author} {\bibinfo {author} {\bibfnamefont {E.}~\bibnamefont
  {Lieb}}, \bibinfo {author} {\bibfnamefont {T.}~\bibnamefont {Schultz}}, \
  and\ \bibinfo {author} {\bibfnamefont {D.}~\bibnamefont {Mattis}},\
  }\href@noop {} {\bibfield  {journal} {\bibinfo  {journal} {Annals of
  Physics}\ }\textbf {\bibinfo {volume} {16}},\ \bibinfo {pages} {407}
  (\bibinfo {year} {1961})}\BibitemShut {NoStop}%
\bibitem [{\citenamefont {G\"ohmann}(2022)}]{Goehmannprivatecomm}%
  \BibitemOpen
  \bibfield  {author} {\bibinfo {author} {\bibfnamefont {F.}~\bibnamefont
  {G\"ohmann}},\ }\href@noop {} {}\bibinfo {howpublished} {private
  communication} (\bibinfo {year} {2022})\BibitemShut {NoStop}%
\bibitem [{\citenamefont {Hellmann}(2015)}]{Hellmann2}%
  \BibitemOpen
  \bibfield  {author} {\bibinfo {author} {\bibfnamefont {H.}~\bibnamefont
  {Hellmann}},\ }\href@noop {} {\emph {\bibinfo {title} {{Einf\"uhrung in die
  Quantenchemie}}}}\ (\bibinfo  {publisher} {Springer},\ \bibinfo {address}
  {Berlin},\ \bibinfo {year} {2015})\BibitemShut {NoStop}%
\bibitem [{\citenamefont {Feynman}(1939)}]{Feynman}%
  \BibitemOpen
  \bibfield  {author} {\bibinfo {author} {\bibfnamefont {R.}~\bibnamefont
  {Feynman}},\ }\href@noop {} {\bibfield  {journal} {\bibinfo  {journal} {Phys.
  Rev.}\ }\textbf {\bibinfo {volume} {56}},\ \bibinfo {pages} {340} (\bibinfo
  {year} {1939})}\BibitemShut {NoStop}%
\bibitem [{\citenamefont {Schulz}(1990)}]{PhysRevLett.64.2831}%
  \BibitemOpen
  \bibfield  {author} {\bibinfo {author} {\bibfnamefont {H.~J.}\ \bibnamefont
  {Schulz}},\ }\href@noop {} {\bibfield  {journal} {\bibinfo  {journal} {Phys.
  Rev. Lett.}\ }\textbf {\bibinfo {volume} {64}},\ \bibinfo {pages} {2831}
  (\bibinfo {year} {1990})}\BibitemShut {NoStop}%
\bibitem [{\citenamefont {Karrasch}\ and\ \citenamefont
  {Moore}(2012)}]{PhysRevB.86.155156}%
  \BibitemOpen
  \bibfield  {author} {\bibinfo {author} {\bibfnamefont {C.}~\bibnamefont
  {Karrasch}}\ and\ \bibinfo {author} {\bibfnamefont {J.~E.}\ \bibnamefont
  {Moore}},\ }\href@noop {} {\bibfield  {journal} {\bibinfo  {journal} {Phys.
  Rev. B}\ }\textbf {\bibinfo {volume} {86}},\ \bibinfo {pages} {155156}
  (\bibinfo {year} {2012})}\BibitemShut {NoStop}%
\bibitem [{\citenamefont {Ejima}\ \emph {et~al.}(2005)\citenamefont {Ejima},
  \citenamefont {Gebhard},\ and\ \citenamefont {Nishimoto}}]{Ejima2005}%
  \BibitemOpen
  \bibfield  {author} {\bibinfo {author} {\bibfnamefont {S.}~\bibnamefont
  {Ejima}}, \bibinfo {author} {\bibfnamefont {F.}~\bibnamefont {Gebhard}}, \
  and\ \bibinfo {author} {\bibfnamefont {S.}~\bibnamefont {Nishimoto}},\
  }\href@noop {} {\bibfield  {journal} {\bibinfo  {journal} {Europhysics
  Letters ({EPL})}\ }\textbf {\bibinfo {volume} {70}},\ \bibinfo {pages} {492}
  (\bibinfo {year} {2005})}\BibitemShut {NoStop}%
\bibitem [{\citenamefont {Giamarchi}\ and\ \citenamefont
  {Schulz}(1989)}]{PhysRevB.39.4620}%
  \BibitemOpen
  \bibfield  {author} {\bibinfo {author} {\bibfnamefont {T.}~\bibnamefont
  {Giamarchi}}\ and\ \bibinfo {author} {\bibfnamefont {H.~J.}\ \bibnamefont
  {Schulz}},\ }\href@noop {} {\bibfield  {journal} {\bibinfo  {journal} {Phys.
  Rev. B}\ }\textbf {\bibinfo {volume} {39}},\ \bibinfo {pages} {4620}
  (\bibinfo {year} {1989})}\BibitemShut {NoStop}%
\bibitem [{\citenamefont {{Wolfram Research{,} Inc.}}(2021)}]{Mathematica12}%
  \BibitemOpen
  \bibfield  {author} {\bibinfo {author} {\bibnamefont {{Wolfram Research{,}
  Inc.}}},\ }\href@noop {} {\emph {\bibinfo {title} {Mathematica, {V}ersion
  12.3}}}\ (\bibinfo  {publisher} {Wolfram Research{,} Inc.},\ \bibinfo
  {address} {Champaign, IL},\ \bibinfo {year} {2021})\BibitemShut {NoStop}%
\bibitem [{\citenamefont {Legeza}\ \emph {et~al.}(2003)\citenamefont {Legeza},
  \citenamefont {R\"oder},\ and\ \citenamefont {Hess}}]{legeza2003}%
  \BibitemOpen
  \bibfield  {author} {\bibinfo {author} {\bibfnamefont {{\"O}.}~\bibnamefont
  {Legeza}}, \bibinfo {author} {\bibfnamefont {J.}~\bibnamefont {R\"oder}}, \
  and\ \bibinfo {author} {\bibfnamefont {B.~A.}\ \bibnamefont {Hess}},\
  }\href@noop {} {\bibfield  {journal} {\bibinfo  {journal} {Phys. Rev. B}\
  }\textbf {\bibinfo {volume} {67}},\ \bibinfo {pages} {125114} (\bibinfo
  {year} {2003})}\BibitemShut {NoStop}%
\bibitem [{\citenamefont {Legeza}\ and\ \citenamefont
  {S\'olyom}(2004)}]{Legeza2004}%
  \BibitemOpen
  \bibfield  {author} {\bibinfo {author} {\bibfnamefont {{\"O}.}~\bibnamefont
  {Legeza}}\ and\ \bibinfo {author} {\bibfnamefont {J.}~\bibnamefont
  {S\'olyom}},\ }\href@noop {} {\bibfield  {journal} {\bibinfo  {journal}
  {Phys. Rev. B}\ }\textbf {\bibinfo {volume} {70}},\ \bibinfo {pages} {205118}
  (\bibinfo {year} {2004})}\BibitemShut {NoStop}%
\bibitem [{\citenamefont {Woynarovich}\ and\ \citenamefont
  {Eckle}(1987)}]{WoynarovichEckle}%
  \BibitemOpen
  \bibfield  {author} {\bibinfo {author} {\bibfnamefont {F.}~\bibnamefont
  {Woynarovich}}\ and\ \bibinfo {author} {\bibfnamefont {H.-P.}\ \bibnamefont
  {Eckle}},\ }\href@noop {} {\bibfield  {journal} {\bibinfo  {journal} {Journal
  of Physics A: Mathematical and General}\ }\textbf {\bibinfo {volume} {20}},\
  \bibinfo {pages} {L97} (\bibinfo {year} {1987})}\BibitemShut {NoStop}%
\bibitem [{\citenamefont {Affleck}\ \emph {et~al.}(1989)\citenamefont
  {Affleck}, \citenamefont {Gepner}, \citenamefont {Schulz},\ and\
  \citenamefont {Ziman}}]{Affleck_1989}%
  \BibitemOpen
  \bibfield  {author} {\bibinfo {author} {\bibfnamefont {I.}~\bibnamefont
  {Affleck}}, \bibinfo {author} {\bibfnamefont {D.}~\bibnamefont {Gepner}},
  \bibinfo {author} {\bibfnamefont {H.~J.}\ \bibnamefont {Schulz}}, \ and\
  \bibinfo {author} {\bibfnamefont {T.}~\bibnamefont {Ziman}},\ }\href@noop {}
  {\bibfield  {journal} {\bibinfo  {journal} {Journal of Physics A:
  Mathematical and General}\ }\textbf {\bibinfo {volume} {22}},\ \bibinfo
  {pages} {511} (\bibinfo {year} {1989})}\BibitemShut {NoStop}%
\bibitem [{\citenamefont {Rutkevich}(2020)}]{Rutkevich2021}%
  \BibitemOpen
  \bibfield  {author} {\bibinfo {author} {\bibfnamefont {S.~B.}\ \bibnamefont
  {Rutkevich}},\ }\href@noop {} {\bibfield  {journal} {\bibinfo  {journal}
  {Phys. Rev. E}\ }\textbf {\bibinfo {volume} {101}},\ \bibinfo {pages}
  {032115} (\bibinfo {year} {2020})}\BibitemShut {NoStop}%
\bibitem [{\citenamefont {Gebhard}\ and\ \citenamefont
  {Legeza}(2021)}]{PhysRevB.104.245118}%
  \BibitemOpen
  \bibfield  {author} {\bibinfo {author} {\bibfnamefont {F.}~\bibnamefont
  {Gebhard}}\ and\ \bibinfo {author} {\bibfnamefont {{\"O}.}~\bibnamefont
  {Legeza}},\ }\href@noop {} {\bibfield  {journal} {\bibinfo  {journal} {Phys.
  Rev. B}\ }\textbf {\bibinfo {volume} {104}},\ \bibinfo {pages} {245118}
  (\bibinfo {year} {2021})}\BibitemShut {NoStop}%
\bibitem [{\citenamefont {Hamer}(1985)}]{Hamer_1985}%
  \BibitemOpen
  \bibfield  {author} {\bibinfo {author} {\bibfnamefont {C.~J.}\ \bibnamefont
  {Hamer}},\ }\href@noop {} {\bibfield  {journal} {\bibinfo  {journal} {Journal
  of Physics A: Mathematical and General}\ }\textbf {\bibinfo {volume} {18}},\
  \bibinfo {pages} {L1133} (\bibinfo {year} {1985})}\BibitemShut {NoStop}%
\bibitem [{\citenamefont {Mishra}\ \emph {et~al.}(2011)\citenamefont {Mishra},
  \citenamefont {Carrasquilla},\ and\ \citenamefont
  {Rigol}}]{PhysRevB.84.115135}%
  \BibitemOpen
  \bibfield  {author} {\bibinfo {author} {\bibfnamefont {T.}~\bibnamefont
  {Mishra}}, \bibinfo {author} {\bibfnamefont {J.}~\bibnamefont
  {Carrasquilla}}, \ and\ \bibinfo {author} {\bibfnamefont {M.}~\bibnamefont
  {Rigol}},\ }\href@noop {} {\bibfield  {journal} {\bibinfo  {journal} {Phys.
  Rev. B}\ }\textbf {\bibinfo {volume} {84}},\ \bibinfo {pages} {115135}
  (\bibinfo {year} {2011})}\BibitemShut {NoStop}%
\bibitem [{\citenamefont {Carrasquilla}\ \emph {et~al.}(2013)\citenamefont
  {Carrasquilla}, \citenamefont {Manmana},\ and\ \citenamefont
  {Rigol}}]{PhysRevA.87.043606}%
  \BibitemOpen
  \bibfield  {author} {\bibinfo {author} {\bibfnamefont {J.}~\bibnamefont
  {Carrasquilla}}, \bibinfo {author} {\bibfnamefont {S.~R.}\ \bibnamefont
  {Manmana}}, \ and\ \bibinfo {author} {\bibfnamefont {M.}~\bibnamefont
  {Rigol}},\ }\href@noop {} {\bibfield  {journal} {\bibinfo  {journal} {Phys.
  Rev. A}\ }\textbf {\bibinfo {volume} {87}},\ \bibinfo {pages} {043606}
  (\bibinfo {year} {2013})}\BibitemShut {NoStop}%
\end{thebibliography}

\begin{thebibliography}{8}%
\makeatletter
\providecommand \@ifxundefined [1]{%
 \@ifx{#1\undefined}
}%
\providecommand \@ifnum [1]{%
 \ifnum #1\expandafter \@firstoftwo
 \else \expandafter \@secondoftwo
 \fi
}%
\providecommand \@ifx [1]{%
 \ifx #1\expandafter \@firstoftwo
 \else \expandafter \@secondoftwo
 \fi
}%
\providecommand \natexlab [1]{#1}%
\providecommand \enquote  [1]{``#1''}%
\providecommand \bibnamefont  [1]{#1}%
\providecommand \bibfnamefont [1]{#1}%
\providecommand \citenamefont [1]{#1}%
\providecommand \href@noop [0]{\@secondoftwo}%
\providecommand \href [0]{\begingroup \@sanitize@url \@href}%
\providecommand \@href[1]{\@@startlink{#1}\@@href}%
\providecommand \@@href[1]{\endgroup#1\@@endlink}%
\providecommand \@sanitize@url [0]{\catcode `\\12\catcode `\$12\catcode
  `\&12\catcode `\#12\catcode `\^12\catcode `\_12\catcode `\%12\relax}%
\providecommand \@@startlink[1]{}%
\providecommand \@@endlink[0]{}%
\providecommand \url  [0]{\begingroup\@sanitize@url \@url }%
\providecommand \@url [1]{\endgroup\@href {#1}{\urlprefix }}%
\providecommand \urlprefix  [0]{URL }%
\providecommand \Eprint [0]{\href }%
\providecommand \doibase [0]{http://dx.doi.org/}%
\providecommand \selectlanguage [0]{\@gobble}%
\providecommand \bibinfo  [0]{\@secondoftwo}%
\providecommand \bibfield  [0]{\@secondoftwo}%
\providecommand \translation [1]{[#1]}%
\providecommand \BibitemOpen [0]{}%
\providecommand \bibitemStop [0]{}%
\providecommand \bibitemNoStop [0]{.\EOS\space}%
\providecommand \EOS [0]{\spacefactor3000\relax}%
\providecommand \BibitemShut  [1]{\csname bibitem#1\endcsname}%
\let\auto@bib@innerbib\@empty
%</preamble>
\bibitem [{\citenamefont {Mishra}\ \emph {et~al.}(2011)\citenamefont {Mishra},
  \citenamefont {Carrasquilla},\ and\ \citenamefont
  {Rigol}}]{PhysRevB.84.115135APP}%
  \BibitemOpen
  \bibfield  {author} {\bibinfo {author} {\bibfnamefont {T.}~\bibnamefont
  {Mishra}}, \bibinfo {author} {\bibfnamefont {J.}~\bibnamefont
  {Carrasquilla}}, \ and\ \bibinfo {author} {\bibfnamefont {M.}~\bibnamefont
  {Rigol}},\ }\href@noop {} {\bibfield  {journal} {\bibinfo  {journal} {Phys.
  Rev. B}\ }\textbf {\bibinfo {volume} {84}},\ \bibinfo {pages} {115135}
  (\bibinfo {year} {2011})}\BibitemShut {NoStop}%
\bibitem [{\citenamefont {Carrasquilla}\ \emph {et~al.}(2013)\citenamefont
  {Carrasquilla}, \citenamefont {Manmana},\ and\ \citenamefont
  {Rigol}}]{PhysRevA.87.043606APP}%
  \BibitemOpen
  \bibfield  {author} {\bibinfo {author} {\bibfnamefont {J.}~\bibnamefont
  {Carrasquilla}}, \bibinfo {author} {\bibfnamefont {S.~R.}\ \bibnamefont
  {Manmana}}, \ and\ \bibinfo {author} {\bibfnamefont {M.}~\bibnamefont
  {Rigol}},\ }\href@noop {} {\bibfield  {journal} {\bibinfo  {journal} {Phys.
  Rev. A}\ }\textbf {\bibinfo {volume} {87}},\ \bibinfo {pages} {043606}
  (\bibinfo {year} {2013})}\BibitemShut {NoStop}%
\bibitem [{\citenamefont {{Wolfram Research{,} Inc.}}(2021)}]{Mathematica12APP}%
  \BibitemOpen
  \bibfield  {author} {\bibinfo {author} {\bibnamefont {{Wolfram Research{,}
  Inc.}}},\ }\href@noop {} {\emph {\bibinfo {title} {Mathematica, {V}ersion
  12.3}}}\ (\bibinfo  {publisher} {Wolfram Research{,} Inc.},\ \bibinfo
  {address} {Champaign, IL},\ \bibinfo {year} {2021})\BibitemShut {NoStop}%
\bibitem [{\citenamefont {G\"ohmann}(2022)}]{GoehmannprivatecommAPP}%
  \BibitemOpen
  \bibfield  {author} {\bibinfo {author} {\bibfnamefont {F.}~\bibnamefont
  {G\"ohmann}},\ }\href@noop {} {}\bibinfo {howpublished} {private
  communication} (\bibinfo {year} {2022})\BibitemShut {NoStop}%
\bibitem [{\citenamefont {Yang}\ and\ \citenamefont
  {Yang}(1966{\natexlab{a}})}]{PhysRev.150.321APP}%
  \BibitemOpen
  \bibfield  {author} {\bibinfo {author} {\bibfnamefont {C.~N.}\ \bibnamefont
  {Yang}}\ and\ \bibinfo {author} {\bibfnamefont {C.~P.}\ \bibnamefont
  {Yang}},\ }\href@noop {} {\bibfield  {journal} {\bibinfo  {journal} {Phys.
  Rev.}\ }\textbf {\bibinfo {volume} {150}},\ \bibinfo {pages} {321} (\bibinfo
  {year} {1966}{\natexlab{a}})}\BibitemShut {NoStop}%
\bibitem [{\citenamefont {Yang}\ and\ \citenamefont
  {Yang}(1966{\natexlab{b}})}]{PhysRev.150.327APP}%
  \BibitemOpen
  \bibfield  {author} {\bibinfo {author} {\bibfnamefont {C.~N.}\ \bibnamefont
  {Yang}}\ and\ \bibinfo {author} {\bibfnamefont {C.~P.}\ \bibnamefont
  {Yang}},\ }\href@noop {} {\bibfield  {journal} {\bibinfo  {journal} {Phys.
  Rev.}\ }\textbf {\bibinfo {volume} {150}},\ \bibinfo {pages} {327} (\bibinfo
  {year} {1966}{\natexlab{b}})}\BibitemShut {NoStop}%
\bibitem [{\citenamefont {Banerjee}\ and\ \citenamefont
  {Wilkerson}(2017)}]{BanerjeeWilkersonAPP}%
  \BibitemOpen
  \bibfield  {author} {\bibinfo {author} {\bibfnamefont {S.}~\bibnamefont
  {Banerjee}}\ and\ \bibinfo {author} {\bibfnamefont {B.}~\bibnamefont
  {Wilkerson}},\ }\href@noop {} {\bibfield  {journal} {\bibinfo  {journal}
  {International Journal of Number Theory}\ }\textbf {\bibinfo {volume} {13}},\
  \bibinfo {pages} {2097} (\bibinfo {year} {2017})}\BibitemShut {NoStop}%
\bibitem [{\citenamefont {Garoufalidis}\ and\ \citenamefont
  {Zagier}(2021)}]{GaroufalidisZagier2021APP}%
  \BibitemOpen
  \bibfield  {author} {\bibinfo {author} {\bibfnamefont {S.}~\bibnamefont
  {Garoufalidis}}\ and\ \bibinfo {author} {\bibfnamefont {D.}~\bibnamefont
  {Zagier}},\ }\href@noop {} {\bibfield  {journal} {\bibinfo  {journal} {The
  Ramanujan Journal}\ }\textbf {\bibinfo {volume} {55}},\ \bibinfo {pages}
  {219} (\bibinfo {year} {2021})}\BibitemShut {NoStop}%
\end{thebibliography}


\end{document}